\documentclass[journal]{IEEEtran}

\usepackage{hyperref}
\usepackage{graphicx}
\usepackage[utf8]{inputenc}
\usepackage{amssymb}                  % add symbole 
\usepackage{amsmath}                  % add align
\usepackage{color}
\usepackage[numbers]{natbib}
\usepackage{multirow}		
 \usepackage{comment}
\usepackage{caption}
\usepackage{subcaption}
\usepackage{url}
\usepackage{breqn}

\newcommand\sophie{\textcolor{black}}
 \newcommand\offline{\textcolor{black}}
 
\DeclareCaptionFormat{cont}{#1 (cont.)#2#3\par}

\begin{document}

% \begin{frontmatter}
\title{Gaussian Anamorphosis for Ensemble Kalman Filter Analysis of SAR-Derived Wet Surface Ratio Observations}

% \address[cerfacs]{CERFACS, 31057 Toulouse Cedex 1, France,}
% \address[ceci]{CECI, CNRS UMR 5318/CERFACS, 31057 Toulouse Cedex 1, France}
% \address[cnes]{CNES, 31401 Toulouse Cedex 9, France}

% \author[cerfacs,ceci]{Thanh Huy Nguyen\corref{mycorrespondingauthor}}
% \cortext[mycorrespondingauthor]{Corresponding author}
% \ead{thnguyen@cerfacs.fr}

% \author[cerfacs]{Andrea Piacentini}
	
% \author[cerfacs,ceci]{Sophie Ricci}

% \author[cnes]{Gwendoline Blanchet}
% \author[cnes]{Raquel Rodriquez-Suquet}

\author{Thanh~Huy~Nguyen,~\IEEEmembership{Member,~IEEE,}
        Sophie~Ricci,
        Andrea~Piacentini,
        Ehouarn~Simon,
        Raquel~Rodriguez-Suquet,
        and Santiago~Peña~Luque% <-this % stops a space
\thanks{%Manuscript received December 1, 2022; revised January 1, 2023. 
This work was supported in part by the Centre National d'Études Spatiales (CNES) and in part by the Centre Européen de Recherche et de Formation Avancée en Calcul Scientifique (CERFACS) within the framework of the Space for Climate Observatory (SCO). (Corresponding author: Thanh Huy Nguyen.)}
\thanks{T.H. Nguyen and S. Ricci are with the Centre Européen de Recherche et de Formation Avancée en Calcul Scientifique (CERFACS), 31057 Toulouse Cedex 1, France, and also with the CECI Laboratory, CERFACS/CNRS UMR 5318, 31057 Toulouse Cedex 1, France (e-mail: thnguyen@cerfacs.fr; ricci@cerfacs.fr).}
\thanks{A. Piacentini is with the Centre Européen de Recherche et de Formation Avancée en Calcul Scientifique (CERFACS), 31057 Toulouse Cedex 1, France
(piacentini.palm@gmail.com).}% <-this % stops a space
\thanks{E. Simon is with the Institut National Polytechnique de Toulouse (Toulouse INP) and also with the Institut de Recherche en Informatique de Toulouse (IRIT), Toulouse, France (ehouarn.simon@toulouse-inp.fr).}
\thanks{R. Rodriquez Suquet and S. Peña Luque are with the Centre National d'Études Spatiales (CNES), 31401 Toulouse Cedex 9, France (e-mail: raquel.rodriguezsuquet@cnes.fr; santiago.penaluque@cnes.fr).}% <-this % stops a space
}

% \author{\thanks{}}

\maketitle

\begin{abstract}
Flood simulation and forecast capability have been greatly improved thanks to advances in data assimilation (DA) strategies incorporating various types of observations; many are derived from spatial Earth Observation. This paper focuses on the assimilation of 2D flood observations derived from Synthetic Aperture Radar (SAR) images acquired during a flood event with a dual state-parameter Ensemble Kalman Filter (EnKF). Binary wet/dry maps are here expressed in terms of wet surface ratios (WSR) over a number of subdomains of the floodplain. This ratio is further assimilated jointly with in-situ water-level observations to improve the flow dynamics within the floodplain. However, the non-Gaussianity of the observation errors associated with SAR-derived measurements break a major hypothesis for the application of the EnKF, thus jeopardizing the optimality of the filter analysis. The novelty of this paper lies in the treatment of the non-Gaussianity of the SAR-derived WSR observations with a Gaussian anamorphosis process (GA).  This DA strategy was validated and applied over the Garonne Marmandaise catchment (South-west of France) represented with the TELEMAC-2D hydrodynamic model, first in a twin experiment and then for a major flood event that occurred in January-February 2021. It was shown that assimilating SAR-derived WSR observations, in complement to the in-situ water-level observations significantly improves the representation of the flood dynamics. Also, the GA transformation brings further improvement to the DA analysis, while not being a critical component in the DA strategy. This study heralds a reliable solution for flood forecasting over poorly gauged catchments thanks to available remote-sensing datasets.
\end{abstract}


% \end{frontmatter}

% \linenumbers

% \clearpage
% \tableofcontents

% %%%% debut macro %%%%
% \newenvironment{changemargin}[2]{\begin{list}{}{%
% \setlength{\topsep}{0pt}%
% \setlength{\leftmargin}{0pt}%
% \setlength{\rightmargin}{0pt}%
% \setlength{\listparindent}{\parindent}%
% \setlength{\itemindent}{\parindent}%
% \setlength{\parsep}{0pt plus 1pt}%
% \addtolength{\leftmargin}{#1}%
% \addtolength{\rightmargin}{#2}%
% }\item }{\end{list}}
% %%%% fin macro %%%%	

% \include{1_introduction}
% \include{2_material}
% \include{3_methods}
% \include{4_experimental_settings}
% \include{5_results_new}
% \include{6_conclusion_acknowledgements}

\section{Introduction}
In 2021, the Emergency Event Database EM-DAT recorded 432 disastrous events related to natural hazards worldwide, among which flooding dominates with 223 occurrences \cite{EMDAT2021}. %These events affected more than 100 million people and accounted for an economic loss of 74 billion USD. 
Globally, flooding alone is responsible for approximately 40\% of all natural disasters \cite{shah2018}, and as many as 1.47 billion people---nearly 20\% of the world population---are directly exposed to flood risks \cite{rentschler2020people}. 
%\cite{}.
% Several strategies can be considered for flood mitigation, especially working on river, agricultural and urban infrastructures. 
As such, flood monitoring and prediction is crucial in terms of cost-to-benefit ratio. 
The forecast mode is essential for civil and industry protection services, while the hindcast mode allows for damage assessments \cite{jamali2018rapid,pinter2017} and flood defense design studies \cite{shah2018,begg2018}. Early warning and emergency management systems rely on the combination of dense and reliable observing network with numerical models possessing robust forecasting capabilities. 
 
\subsection{Remote-sensing flood observations}
While hydrologic and hydraulic numerical model\sophie{s} play an indispensable role in forecast capability, their efficiency is limited by the uncertainties inherently existing in their input data. Such uncertain data includes rainfall, discharge inflow, and geometry of the river and the floodplain, namely bathymetric and topographic errors from utilized Digital Elevation Models (DEM), as well as hydraulic parameter errors due to the calibration of friction coefficients. In this context, DA has emerged as an efficient tool in hydrology to reduce these uncertainties, by combining numerical model outputs with various observations from in-situ gauge measurements and/or from satellite Earth Observations. Indeed, the increasing volume of data from space missions has provided more actors involved in flood management with heterogeneous and relevant satellite data, namely altimetry (e.g. TOPEX/POSEIDON, Jason-1/2/3, SARAL/AltiKa, Sentinel-3, Sentinel-6/Jason-CS, SWOT), optical (e.g. SPOT, LANDSAT-7/8/9, MODIS/VIIRS, Pléiades, Sentinel-2) and Synthetic Aperture Radar (SAR) (e.g. Sentinel-1, TerraSAR-X/TanDEM-X, COSMO-SkyMed, ALOS-2 PALSAR-2, RADARSAT-1/2, ENVISAT ASAR, RISAT-1).
 
A classical DA approach  stands in the assimilation of water surface elevation (WSE) data, either from in-situ gauge measurements, from altimetry satellites, or retrieved from remote-sensing (RS) images using flood edge location information combined with complementary DEM data. %\citet{Dasgupta2021review} provided an updated review on the assimilation of remote-sensing (RS) data with hydraulic models in the purpose of improving flood inundation forecasts. 
Satellite SAR data is particularly advantageous as it allows an all-weather day-and-night global-coverage imagery of continental water, depicted by low backscatter (BS) values resulted from the specular reflection of the incident radar pulses \cite{martinis2015flood}. 
The assimilation of RS-derived WSE is typically convenient as it deals with a diagnostic variable of the model, yet it can suffer from the lack of precision of high-resolution topographic data as noted in various studies  \cite{matgen2011towards,giustarini2011assimilating,hostache2009water,garcia2015satellite}. Nevertheless, many research works \sophie{have proposed} the assimilation of RS-derived WSE as summarized in \cite[Table 1]{revilla2016integrating}. Most commonly used strategies are similar to that of \citet{giustarini2011assimilating}: flood edges are identified on SAR images and  integrated with an available DEM to derive the WSE on the floodplain, which are then compared with and/or assimilated to the WSE simulated by 1D or 2D hydrodynamic model to sequentially update the model state and parameters. 
 %% From Hostache 2018 : Many studies on data assimilation into hydraulic models or forecasting systems integrate synthetic, in situ or remote sensing-derived observations of water levels. For example, see Table 1 in Revilla-Romero et al. (2016), and also Neal et al. (2007), Matgen et al. (2010), Hostache et al. (2010), Giustarini et al. (2011), Yoon et al. (2012), Andreadis and Schumann (2014), García-Pintado et al. (2015), Hostache et al. (2015), and Xu et al. (2017).
 
The need to retrieve WSE from flood extents can be avoided with direct assimilation of SAR-derived flood probability maps or flood extent maps. The assimilation of surface water extents has been presented in large-scale hydrology and in catchment-scale hydrodynamic by various approaches.
In \cite{revilla2016integrating}, daily surface water extents from the so-called Global Flood Detection System are assimilated with an Ensemble Kalman Filter (EnKF) that relies on the random perturbations of the precipitation---input to the distributed hydrological rainfall-runoff LISFLOOD model \cite{van2010lisflood}---with a focus on Africa and South America catchments. 
%In the context of large-scale hydrology, the surface water extents are here derived from low-resolution ($0.1\degree \times 0.1\degree$) passive microwave remote sensors (merged product provided from GFDS using TRMM and AMSR-E sensors, \url{http://www.gdacs.org/flooddetection/}). 
%Since water bodies have lower emissivity than land, the changes in temperature brightness---which was derived from low-resolution passive microwave remote sensors---can be associated with flooding while allowing to estimate the portion of within-pixel water and land over a $0.09\degree \times 0.09\degree$ raster grid. 
At daily time steps, the innovations of streamflow volumes (i.e. the differences between simulated and observed ones) are computed. They are then used by an EnKF algorithm with a state-augmented strategy to correct and update the simulated groundwater levels in the catchment, instead of the simulated streamflow levels, in order to improve streamflow forecast\sophie{s}. It was shown that the assimilation of RS-derived surface water extents greatly improves flood peak forecasting in terms of timing and volume for slow-motion ungauged catchments. 
\citet{lai2014variational} proposed a 4D-Var (four-dimensional variational) DA scheme implemented on top of a 2D Shallow Water model; differentiated with an automatic differentiation tool called TAPENADE \cite{hascoet:inria-00069880}; to assimilate flood extent observations derived from MODIS (Moderate-resolution Imaging Spectroradiometer) \sophie{in order} to correct roughness parameter\sophie{s} over \sophie{the floodplain}. The assimilation of flood extent data was shown to be suitable for improving flood modeling in the floodplain or similar areas with slowly-varying bed slopes.%The flood extent is related to the water level through a wet/dry status matrix devised from the observation and the simulation and prescribed with a level of certainty for each pixel. 

% From Hostache et al. 2018 papier - Pour mémoire  -  Promising results have already been obtained by assimilating flood extents derived from satellite images into a hydraulic model (Lai et al., 2014) and into a forecasting system (Revilla-Romero et al., 2016). The study by Lai et al. (2014) was based on 4DVAR (variational) data assimilation, with no objective toward real-time forecasting; they showed that the assimilation of a flood extent map derived from MODIS data (250-m spatial resolution) allows for the optimization of a lumped friction parameter. Revilla-Romero et al. (2016) used the ensemble Kalman filter to assimilate low resolution (0.1∘ × 0.1∘) satellite-derived flood extents based on the Global Flood Detection System (http://www.gdacs.org/flooddetection/) into a global forecasting system composed of a hydrological model and a routing function, with an objective toward real-time forecasting; their study over 101 stations in Africa and South America shows that flood extent assimilation improves simulated streamflow at the majority of stream gauges, especially at the gauges with poorest skill scores on open-loop runs.
%
While both \cite{revilla2016integrating} and \cite{lai2014variational} strategies \sophie{rely} on the expression of flood extents as a function of the model state, other research works %from \cite{cooper2019observation} and \cite{hostache2018near} 
\sophie{propose} a more direct use of SAR observations.
% Copy Paste de la partie Cooper from our AGU paper
\citet{cooper2019observation} proposed an observation operator that directly takes into account synthetical SAR BS values as observations, in an Observing System Simulation Experiment (OSSE) framework, in order to circumvent the needed flood extent mapping and flood pixel-wise probability estimation processes.
%However, this approach has only been implemented with synthetical SAR images within the scope of a twin experiment. 
It relies on the assumption that SAR images must yield distinct distributions of wet and dry BS values, which may not hold for real SAR data \sophie{that may}  require \sophie{further} treatments, such as hierarchical split-based approach \cite{chini2017hierarchical}. 
\citet{hostache2018near} presents the assimilation of ENVISAT ASAR-derived flood probability maps using a Particle Filter (PF) approach with a sequential importance sampling into a coupled hydrologic-hydraulic model. 
%Several follow-up papers focus on the improvement of this strategy [37, 38]. 
As detailed in \cite{giustarini2016probabilistic}, such a probabilistic flood map represents the probability of an observed BS value to correspond to a flood pixel, assuming that its prior probability to be flooded or non-flooded are two Gaussian probability density functions (PDF). 
%\sophie{The PF framework used in \cite{matgen2011towards,giustarini2011assimilating,hostache2018near,Dasgupta2020,DiMauro2021} offers the key advantage of relaxing the assumption that observation errors are Gaussian, and allows to propagating a non-Gaussian distribution through non-linear hydrologic and hydrodynamic models  \cite{moradkhani2008hydrologic}. This makes it better suited for a DA of probabilistic flood maps than the more widely used EnKF \cite{Neal2007,garcia2015satellite,revilla2016integrating} or variational approaches \cite{lai2014variational}.} %When such assumptions no longer hold, the variational or Kalman filter analysis will provide a variance-minimizing result that would not be the optimal estimate.

%Unlike the widely used EnKF \cite{burgers1988, evensen1994sequential}, which simplifies the recursive estimation by assuming a Gaussian distribution for both the model and the observation error structure, the PF relaxes the need for restrictive assumptions regarding the shape of the probability density functions and can easily manage the propagation of a non-Gaussian distribution through nonlinear hydrologic and hydrodynamic models \cite{moradkhani2008hydrologic}.

\subsection{Dealing with non-Gaussianity in DA}
The non-Gaussianity characteritics of SAR-derived observation errors need to be properly accounted for in the framework of DA. Indeed, the optimality of the KF and variational analysis relies on the Gaussian assumption for the background and observation errors, as well as on the linearity of the observation operator that relates the control and the observation spaces \cite{asch2016data}. When these assumptions no longer hold, the KF or variational analysis can still be used but they are suboptimal. %This may also be the case when dealing with positive or bounded variables \cite{beal2010}, \cite{doron2011}. 
As such, when the Gaussianity assumptions are strongly violated, a pre-processing step is necessary.

%\citet{schumann2007deriving} showed that the PDFs of ENVISAT ASAR-derived water level observations are non-Gaussian at many river cross-sections. \modif{However, the spatial coverage offered by RS is such that it is not necessary to consider all measurements derived from an image.} Therefore, 
For instance, \citet{Neal2009} proposed an adaptive sampling method that only assimilates the measurements that did not fail \sophie{the} normality test. A classical approach in Numerical Weather Prediction (NWP), yet still questionable for extreme events, 
consists in rejecting outlier observations with a Quality Control (QC) procedure applied on the innovation (also called \textit{misfit}), %between observation and background model equivalent, 
assuming that the non-Gaussianity is entirely attributable to observation errors, as these observations are statistically unlikely and their assimilation may lead to a spurious analysis \cite{isaksen2017, bonavita2012}. The simplest approach is thus to assume that the remaining data is correct and follows a Gaussian distribution, then to apply classical variational DA algorithms. In the context of an operational NWP, the \textquotedblleft Gaussian plus flat\textquotedblright~distribution is often used as a refinement with a gray zone between correct data and grossly erroneous data \cite{anderson1999variational,Tavolato2015}. In addition, \citet{Tavolato2015} found that a Huber norm (i.e. Gaussian distribution in the center of the distribution and exponential distribution at the tails) was the most suitable distribution to describe the statistics of the innovations, assuming that the majority of the outliers cannot be considered as gross errors and that they may provide some relevant information. The observation cost function with QC based on the Huber norm relaxes the rejection threshold for large misfits. This allows to keep observations with large innovations in the analysis, which is particularly beneficial for \sophie{the representation of} extreme events.  

The non-Gaussianity of the control and/or observation errors can be handled using a DA algorithm that does not require Gaussianity assumptions. For instance, \sophie{the} PF works with the entire probability function, instead of focusing on the first and second moments of the statistics like KF and variational algorithm \sophie{do}. A considerable number of studies advocate for such a solution, working with a PF or with a Bayesian approach to assimilate SAR-derived observations.
Indeed, the PF framework used in \cite{matgen2011towards,giustarini2011assimilating,hostache2018near,Dasgupta2020,DiMauro2021} offers the key advantage of relaxing the assumption that observation errors are Gaussian, and allows to propagating a non-Gaussian distribution through non-linear hydrologic and hydrodynamic models  \cite{moradkhani2008hydrologic}. This makes it better suited for a DA of probabilistic flood maps than the more widely used EnKF \cite{garcia2015satellite,revilla2016integrating,Neal2007} or variational approaches \cite{lai2014variational}.
In addition, in the context of an OSSE, \citet{Dasgupta2021network} analyzed the impacts of the characteristics of the RS observing network with a PF that assimilates SAR-based probabilistic flood maps into the LISFLOOD-FP hydrodynamic model for a synthetical set-up of a flood event in the Clarence catchment (Australia). They asserted that the location and timing of the SAR images is more important than the revisit interval for flood forecast accuracy. As a follow-up study of \cite{hostache2018near}, \citet{DiMauro2021} introduced an enhanced PF algorithm with a tempering coefficient that depends on the size of the desired effective ensemble size after the assimilation. This aims to inflate the posterior probability and avoid degeneracy of a PF that assimilates SAR-derived probabilistic flood maps. %\modif{This strategies succeeds in improving prediction of streamflow and water elevation, it also increases the Critical Success Index (CSI) score for the predicted flood extents.}

\subsection{Gaussian anamorphosis}
A different strategy stands in transforming the distribution of the SAR-derived observations into a Gaussian distribution, compatible with KF-based DA algorithms. This alleviates the need for advanced processes on the PF implementation to avoid ensemble collapse. Such a strategy is reported in the literature as Gaussian Anamorphosis (GA), also known as normal-score transform. GA is a pre-processing step that maps the control and/or observation variables onto a transformed space where the Gaussianity assumption is better fulfilled. %This transformation either results from a selection on the control and/or observations variables, or from the parametric or non-parametric transformation of the control and/or observation spaces with an anamorphosis function.
GA was proposed by \citet{bertino2003} and it has been investigated in different works \cite{zhou2011}, \cite{brankart2012}, \cite{simon2009}, \cite{simon2012}.
%\cite{zhou2011}, \cite{brankart2012} from Amezcua2014.
In most studies, it was applied to state analysis, as opposed to model parameter analysis (which is performed in this present work). It involves transforming the state variables and observations into new variables with Gaussian features, over which the DA analysis is computed. The inverse of the GA transform must then be used to remap the analysis result back onto the original space.

\citet{bertino2003} presented the application of the EnKF to transformed Gaussian---or \textit{anamorphosed}---state variables, with the integration of the anamorphosis in the update step of the analysis for an ocean model. In order to deal with the large dimension of the model state, it is assumed that the distribution of the state variables at different locations are identical, thus a homogeneous anamorphosis function is chosen all over the spatial domain. This has been assessed, in OSSE mode, for a simplified 1D ecological model. Two versions of \sophie{the} EnKF are compared, assuming either a Gaussian or a log-normal distribution for the errors on the synthetical measurements and the ecological model variables. The ordinary (or \textit{plain}) version of the EnKF  leads to the negative values for the nutriment, phytoplankton and herbivore concentrations (i.e. the parameters considered in \cite{bertino2003}), which should be truncated, thus resulting in repetitive biases and corrections for these variables that lead to undesired artificial spring blooms. The novel GA version of the EnKF with the log-normal transform reduces the spurious \textquotedblleft false starts\textquotedblright~of spring blooms and leads to more realistic ecological cycles.
In continuity of this work, \citet{simon2009} presents the application of such a non-Gaussian extension of the EnKF to perform model state estimations for a 3D coupled ocean physical-ecosystem model that present non-Gaussian and positive variables. They demonstrated, within an OSSE, that the assimilation of anamorphosed synthetical chlorophyll-$a$ surface concentration data %(with a log-normal prescribed error)
presents a slight advantage in effectiveness compared to the plain EnKF with a simple post-processing of negative values. This advocates for the use of GA when dealing with similar RS observations. \citet{simon2012} considers dual state-parameter estimation (\cite{moradkhani2005dual})
%(\cite{anderson2001}, \cite{moradkhani2005dual}, \cite{annan2005}) 
in OSSE, with an EnKF for a simplified 1D ocean ecosystem model (continuing from \cite{bertino2003}) that presents non-Gaussian and positive variables. It was shown that GA overcomes the limits of \sophie{the classical} EnKF when dealing with positive variables.

\citet{beal2010} also showed that GA leads to an improved estimation of \sophie{the} 3D state of a coupled physical-biogeochemical ocean model with respect to a classical EnKF approach. \citet{schoniger2012parameter} promoted the use of GA in the field of hydrogeology for the correction of the subsurface hydrologic state. Indeed, the state vector of a groundwater flow numerical model is typically non-Gaussian in the presence of strong spatial heterogeneity of the hydraulic conductivity field. Drawdown, pressure head, and concentration state are rendered Gaussian with a non-parametric function and the discretized field of log-conductivity, which are parameters to the model, is estimated sequentially with DA. The parameter-only EnKF scheme obliterates the need for a back transformation step on updated transformed state variables; it also guarantees that the simulated state is coherent with the updated parameters in the sense of the model's governing equation\sophie{s}. The merits of GA for parameter estimation with EnKF were demonstrated in OSSE assimilating synthetical drawdown measurements. Also in hydraulic tomography, the authors of \cite{zhou2012pattern,li2012groundwater,xu2013power} describe a normal-score transformed (also named GA) EnKF to generate log-conductivity realizations which are conditional to log-conductivity and/or transient piezometric head data.

\citet{amezcua2014} studied the merits of a joint transformation for the state and observation variables that yields joint Gaussianity in the transformed space for the analysis step of the EnKF. Indeed, when mapping the model state space onto a transformed space with Gaussian properties, the observation operator may become non-linear, thus raising a new issue for the filter. The joint GA transformed is compared to a transformation in the state variable and/or observation space, for the univariate case. This work was carried out over a simple case in which the Bayesian posterior can be obtained analytically. It was demonstrated that, in spite of the various GA strategies, the optimality of the analysis is not reached but the joint transformation outperforms the other strategies as its solution is closer to that of the Bayesian solution. 

\subsection{Scope of the article}
The overall objective of the article is to reduce comprehensively the uncertainties in the model parameters, forcing and hydraulic state, and consequently improve the overall flood re-analysis and forecast capability, especially in the floodplain,  by assimilating relevant SAR-derived \offline{flood} observations. This article presents as the subsequent study of \cite{nguyenagu2022,nguyen2022hic,nguyen2022tuc}
that proposes a DA approach to accommodate 2D SAR-derived observations alongside with in-situ water level (WL) time-series within an EnKF framework for the TELEMAC-2D\footnote{\url{www.opentelemac.org}} (T2D) hydrodynamic model \cite{hervouet2007hydrodynamics} of the Garonne Marmandaise catchment (South-west France). %, with a dual state-parameter estimation implementation. 
Sentinel-1 SAR-derived flood extent \offline{maps} are expressed in terms of wet surface ratios (WSR) computed as the ratio between the number of wet pixels detected on SAR-derived flood extent maps over the total number of pixels in a subdomain of the floodplain. 
% Faut il garder ca ? Est ce qu'on se sert du threshold sur les backscatter values dans FloodML  ?
%In this context, the SAR backscatter signal is post-processed into a wet/dry pixel image and the speckle error accumulates with  in the wet/dry classification. Indeed, the radiometric distributions of land and water bodies overlap and the identification of a threshold on the backscatter value is difficult \cite{hostache2009water}. 

The novelty of this article stands in the treatment of the non-Gaussianity of such SAR-derived WSR observations. A dual state-parameter DA strategy \cite{moradkhani2005dual} is implemented to reduce the uncertainties in the parameters associated with friction coefficients and upstream forcing. The control vector is augmented with a WL state correction that is uniform over a limited number of subdomains in the floodplain. Here the EnKF algorithm is favored and implemented as it allows to stochastically estimate the covariance matrices between the model inputs/parameters and its outputs, without formulating the tangent linear of the hydrodynamic model \sophie{with respect to its parameters}, under the assumption that the errors in the control vector are properly described by a Gaussian distribution.
As this property is violated when considering SAR-derived WSR observations, the GA strategy described in \cite{simon2012} is implemented here. As opposed to previously cited studies, our observation variables may be strictly null (i.e. dry area) or strictly equal to 1 (i.e. totally flooded area). As a consequence, particular efforts were made on the treatment of these extreme values to ensure that the anamorphosis function remains bijective. 

The remainder of the article is organized as follows. Section~\ref{Data} presents the data that are used in this study. The DA strategy is detailed in Section~\ref{Methods} with a focus on anamorphosis. Section~\ref{ExpeSett} presents the experimental settings for the OSSE and real DA experiments. Section~\ref{results} discusses the merits of the DA and GA strategies for OSSE and real experiments with assessments in the control and observation spaces. Conclusion and perspectives are finally given in Section~\ref{ConclusionPerspectives}.

% CE QU'ON VEUT CONCLURE : 
%Assimiler des SAR derived WSR into EnKF avec anamoprhose pour se débarasser de la non Gaussianité. On montre que l'usage de EnKF avec anamorphose donne de bon résulatt, mais que si on ne fait pas d'anamorphose, c'est pas trs grave, on  a déjà de bons résultats (papier AGU). 

\section{Data}
\label{Data}

\subsection{The Garonne Marmandaise catchment and Observations}

\begin{figure*}[t]
\centering
    \includegraphics[width=0.85\linewidth]{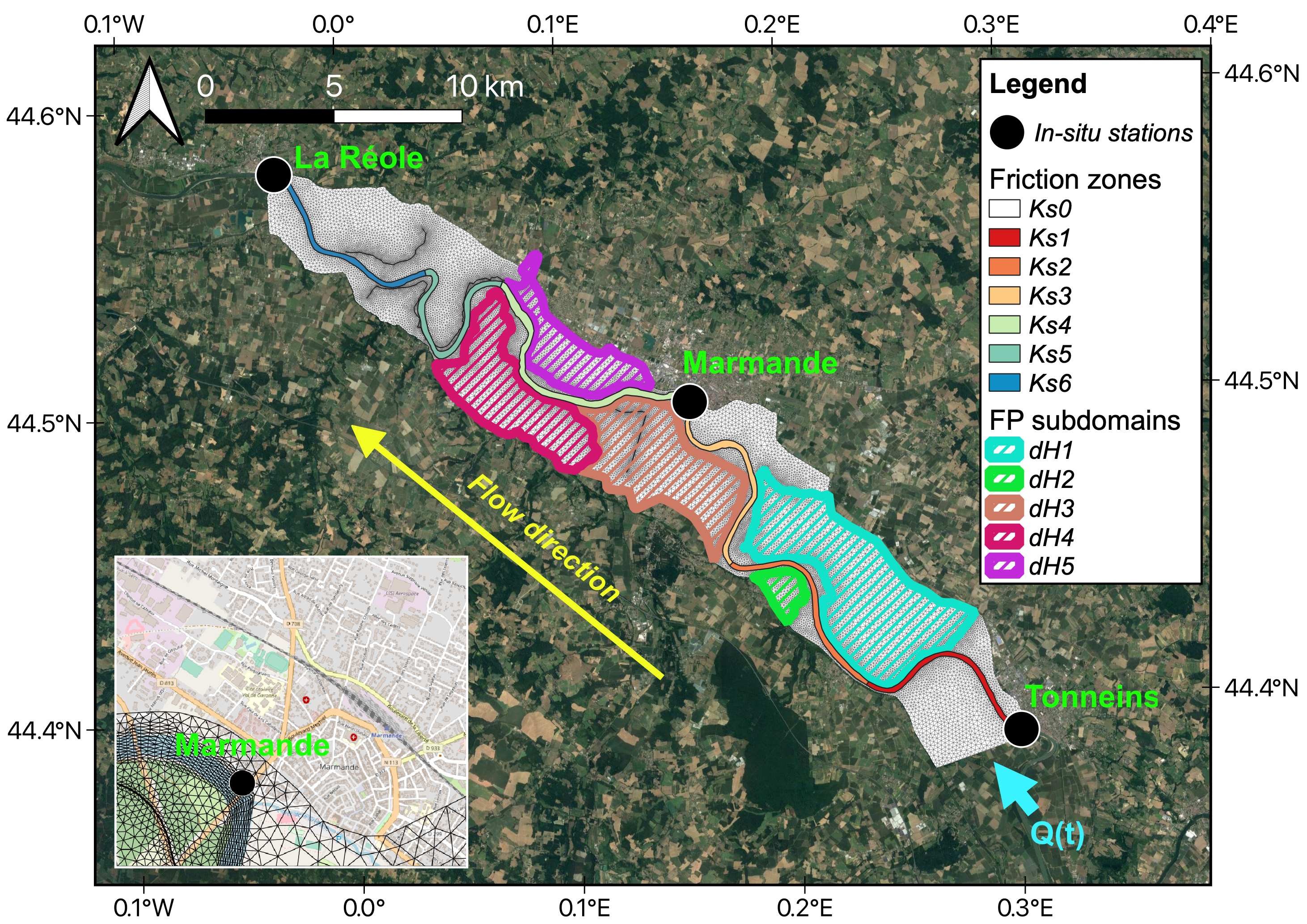}
     \caption{T2D Garonne Marmandaise domain. The VigiCrue observing stations are indicated as black circles.
     The different river friction zones are indicated as colored segments of the Garonne River. The floodplain is divided into five subdomains that are hatched in different colors. The inset figure shows the urban area of Marmande nearby its namesake gauge station. (Note. From \cite{nguyenagu2022} by Nguyen \textit{et al.} (2022), \textit{Water Resources Research}, 58, e2022WR033155. CC BY-NC.)}
     \label{fig:study_area}
\end{figure*}

\begin{figure}[h]
    \centering
    \includegraphics[width=\linewidth]{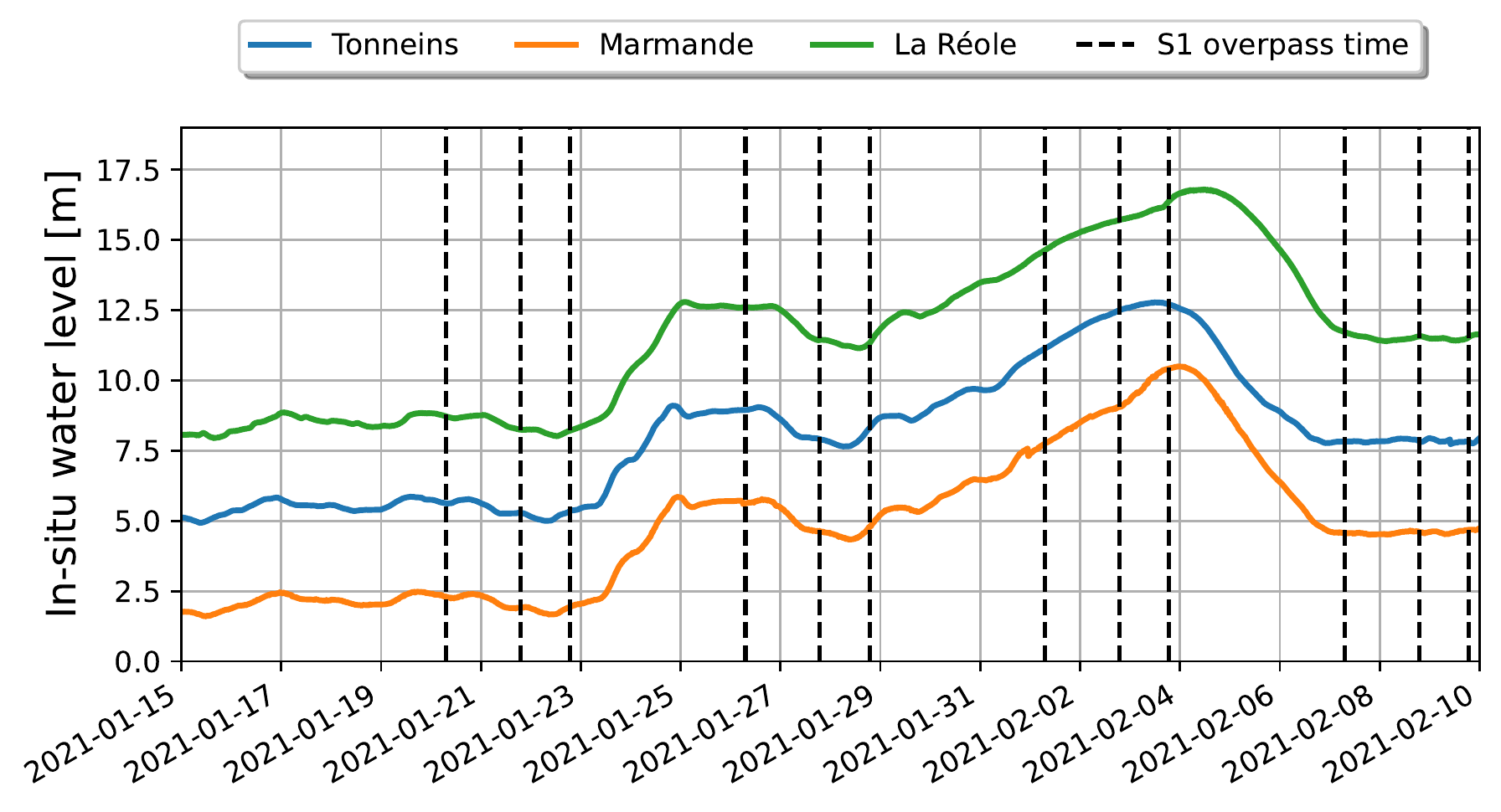}
    \caption{Water level $H$ time-series at Tonneins (blue), Marmande (orange) and La Réole (green) for 2021 flood event. Vertical dashed lines indicate S1 overpass times.} %Horizontal dash-dotted lines represent the highest risk level at each observing stations.}
    \label{fig:Htvigicrue2021}
\end{figure}

The study is carried out over a reach of the Garonne \sophie{R}iver near Marmande for the major flood event that occurred in January-February 2021. The hydrodynamic numerical model T2D is used to simulate and predict the WL (denoted by $H$ [$m$]) and velocity (denoted by $u$ and $v$ [$m.s^{-1}$] for the two horizontal components) from which flood risk can be assessed. The study case as well as the hydrodynamic model T2D that was set up to represent the dynamics of the flow are fully described in~\cite{nguyenagu2022,NguyenTGRS2022}. The major sources of uncertainties \sophie{are} associated with friction coefficients in the riverbed and in the floodplain, the upstream forcing, as well as with the hydraulic state in subdomains of the floodplain.
The friction coefficients include six coefficients in the riverbed ($K_{s_1}$ to $K_{s_6}$) and one in the floodplain $K_{s_0}$, the upstream forcing is corrected by a multiplicative factor $\mu$, whereas the hydraulic state in subdomains \sophie{of the floodplain} is corrected with a (spatially) uniform WL \sophie{additive increment} $\delta H_1$ to $\delta H_5$ over five subdomains, as illustrated in Figure~\ref{fig:study_area}. 
The a priori values for these aforementioned random variables---their PDF are assumed to be Gaussian---are presented  in \cite{nguyenagu2022}.
The 2021 flood event is of a 20-year return period.
Figure~\ref{fig:Htvigicrue2021} depicts available in-situ WL observations, measured every 15 minutes at Tonneins (blue), Marmande (orange) and La Réole (green). %Vertical dashed lines represent the overpass time of Sentinel-1 satellites providing the SAR images. 
The study is carried over a period of 25 days between 2021-01-16 and 2021-02-10 that capture the flood and the recess phases. %All of the time-varying plots in this article are made in local time UTC +01:00.

Sentinel-1 (S1) %is the first satellite series of the Copernicus program 
\cite{torres2012gmes}, carrying a C-band SAR system, with a central frequency of 5.405 GHz. In this work, the Interferometric Wide (IW) mode is used with 250-km-wide swath and a ground resolution of approximately 20$\times$22 m, further  resampled, reprojected and distributed at 10$\times$10 m for the Ground Range Detected (GRD) products. %In order to improve the revisit time, 
S1 operates as a constellation of two polar-orbiting identical satellites, launched respectively on 2014-04-03 and 2016-04-26, allowing a six-day repeat cycle, until 2021-12-23. %Future satelites in this program Sentinel-1C and Sentinel-1D to be operational in the upcoming years will further enable flood monitoring and mapping even better.
The S1 products are used %leveraged as the predominant data source 
to produce binary water maps with the FloodML software based on a Random Forest Machine Learning algorithm %Machine Learning algorithms developed by CNES and CLS in the framework of the FloodML project 
\cite{2020AGUFMIN041..09H,kettig}. More details can be found in \cite{NguyenTGRS2022,NguyenTUC2021,pena2021sentinel}.
%The specifications of the flood extent mapping method applied to S1 images are detailed in 

The 2021 flood event was observed by twelve S1 images, represented by vertical dashed lines in Figure~\ref{fig:Htvigicrue2021}. The flood peak was reached on 2021-02-04 and it exceeded the highest threshold level for flood risk at Marmande
%---horizontal dash-dotted orange line in Figure~\ref{fig:Htvigicrue2021}---
set out by the French national flood forecasting center (SCHAPI) in collaboration with the departmental prefect. %It should be noted that for the S1 images from the ascending orbit 132, a small part of the downstream area (including La Réole) is a no-data area as it is out of range from the acquisition.
The \sophie{validation of the results} is also performed using independent observations from relevant high water marks (HWM) dataset. It is a collaborative dataset of collected flood marks\footnote{\url{https://www.reperesdecrues.developpement-durable.gouv.fr/}} maintained by the local flood forecast services (SPCs) or by the flood risk prevention and management service (GEMAPI) in the floodplain. In the aftermath of the 2021 flood event, 178 HWM observations were collected (accessed in December 2021). It is worth-noting that, due to high cloud cover during this flood event, no Sentinel-2 optical image acquired during the 2021 flood event provides proper observations.

%POur les erreurs d'obs : 
%Questions sur les erreurs des données SAR qu'on utilise
%\sophie{Quelle est la nature de l'erreur sur l'information issue du Random Forest de FloodML (wet/dry pixel) ? 
%Ne faut il pas aussi considérer les erreurs de classification entre pixel dry/wet et la non sépération claire des 2 distributions  ? voir Figure 3 du  papier Hostache 2009 + voir Cooper et al. 2019, p 2542, premier paragraph par exemple. } 

%\sophie{Préciser la taille du vecteur d'observation, on en a besoin pour l'explicati de la gneration de l'anamoprhose (sampling)  : il faut que je regarde l'annexe de Simon2009 pour comprendre concrètement combien d'obs on utilise sur un cycle : on a 5 zones de WSR, avec disons 2 passages de S1 par cycle d'assim, et 75 membres. Ca fait 5 * 2 * 75 éléments pour construire l'histogramme... est ce critique ?}

\subsection{Imperfection in SAR backscatter information}
SAR images, like all active coherent imaging systems, suffer from an inherent error called \textit{speckle}. It is caused by constructive and destructive interference of coherent waves reflected by the many elementary scatterers contained within an imaged resolution cell. It should be noted that, a radiometric and a geometrical correction \cite{blacknell1989geometric,small2011flattening} are classically applied as pre-processing on raw SAR data, so that most of the remaining errors in SAR images lies in speckle errors.
In the detected SAR images (intensity or amplitude), the speckle is usually described as a non-Gaussian multiplicative noise. It can be considered as a random variable whose power and magnitude, respectively, follow a negative exponential and a Rayleigh distribution \cite{xie2002statistical}. The inevitable presence of speckles in SAR images makes the image interpretations and analyses particularly difficult. However, their undesirable effects can be reduced with various filtering methods \cite{kuan1985adaptive,lopes1990adaptive,lee1994}. 
In addition, multilooking approach also reduces speckle but at the expense of the image spatial resolution. While avoiding such a loss of resolution, adaptive filters also allow for a significant reduction of speckle while better preserving resolution but they alter the statistical properties of the image. Temporal multilooking comes as an interesting solution when image time-series are available. Yet, none of these methods is capable of a complete removal of all speckle in the image \cite{deledalle2017mulog,abergel2018subpixellic}.
Therefore, the non-Gaussianity of SAR-derived BS  observations, more specifically due to speckle, raises major difficulties for image post-processing. As presented in \cite{xie2002sar}, a logarithmic transform can be applied to convert the signal with multiplicative noise into one with additive noise, easier to be treated by analyzing and standard image processing techniques. Once this transform is achieved, the statistical properties of the log-transformed multilook speckle noise are described for further use in SAR image post-processing. 

\subsection{Error in SAR-derived flood observations}
As aforementioned, wet pixels on SAR images exhibit low BS since most of the incident radar pulses are specularly reflected away upon arrival at the water surfaces as opposed to dry pixels that exhibit high BS values. While this properties favors the use of S1 images to detect inundated areas, the variability of water roughness and speckle effect may come as a limitation, especially in urban environment or vegetated areas \cite{martinis2015backscatter, pierdicca2018flood}. 

%the detection of inundated areas is straightforward on SAR images, with several exceptions, e.g. in urban environment, vegetated areas, or when facing variability of water roughness and speckle effect. 
%Indeed, mis-detection of flooded vegetation areas (i.e. partially submerged vegetation) mainly occurs because some SAR signals (e.g. C-band and X-band)  being caught in volume scattering from the canopy and cannot reach the water surfaces beneath vegetation, or some others related to multiple-bounce effects between the tree trunks and the submerging water surfaces \cite{martinis2015backscatter, pierdicca2018flood}. Similar situations could also occur in urban areas due to the complexity of the landscape geometry, e.g. shadow, layover, and highly reflective scatterers. 

%During a flood event, SAR images provide a clear distinction between flooded and non-flooded areas. %Pixels in flooded or other wet areas such as lakes and rivers have low backscatter values and appear as dark areas on SAR images; dry areas have higher backscatter values, and dry pixels therefore appear brighter. 
A number of techniques exist for separating pixels into wet and dry areas based on BS values. \sophie{Most} methods include thresholding \cite{henry2006envisat} with varying levels of user interpretation (as compared in \cite{brown2016progress}), region growing and clustering \cite{horritt2001}, and change detection \cite{hostache2012change}. These techniques can be used to provide observational information for \sophie{DA} frameworks but they have also been used for flood mapping and monitoring (e.g. \cite{matgen2011towards,brown2016progress}) \sophie{as well as} for validation and calibration of inundation models \cite{mason2009flood,dibaldassarre2009technique,wood2016calibration}. 
However, uncertainty in flood extent mapping from SAR images, originates from both the input images and the performed classification algorithm. As a result, resulting classification overall accuracy of flooded areas varies considerably and only in rare cases, exceeds 90\% \cite{Schumann2012}.

\begin{figure*}[t]
    \centering
    \includegraphics[width=0.8\textwidth]{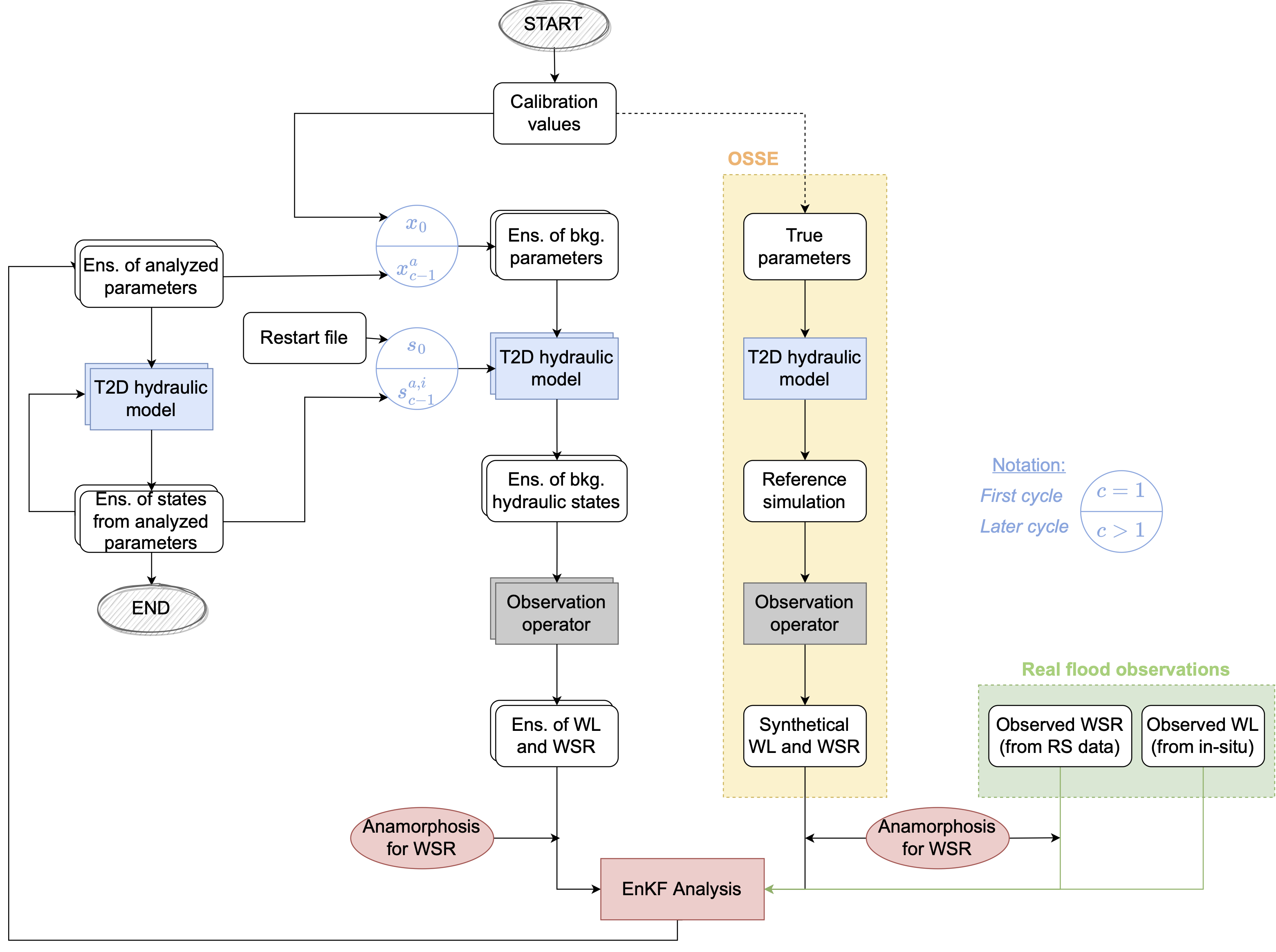}
    \caption{Diagram for DA strategy in OSSE and real DA experiments.}
    \label{fig:ExpSetWorkflow}
\end{figure*}

\section{Methods}
\label{Methods}

\subsection{Workflow for the DA algorithm}\label{subsect:EnKFctl}

The general framework for DA experimental settings is depicted in  Figure~\ref{fig:ExpSetWorkflow}. Each box represents a step of the DA algorithm, with a multi-layering aspect indicating ensemble steps. The different ensemble inputs, outputs, and variables are represented in white boxes, the observation operators in gray boxes, the T2D model simulations in blue boxes, and the steps conducted in the transformed Gaussian space are represented in red boxes. %\sophie{Maybe we change color for boxes dedicated to the observation operator ? and leave blue for T2D simulations only.} 
The real observations are shown in the green block, and the specific steps to generate synthetical observations for the OSSE are gathered in the yellow block on the right-hand side. 
The blue circles indicate different input choices for the first cycle and for the subsequent ones.

%\subsection{Description of the control vector}\label{subsect:EnKFctl}
The control vector ${\bf x}$ for the EnKF DA algorithm, has a size up to $n=13$ depending on the DA experimental settings reported in Table~\ref{tab:runs} (in Section~\ref{DAExpeSett}). At most, it gathers seven friction coefficients $K_{s_k}$ with $k\in [0,6]$, one multiplicative parameter $\mu$ to modify the time-varying upstream BC $Q(t)$, and five state corrective variables $\delta H_k$ with $k\in [1,5]$ over the floodplain subdomains. The friction coefficients and multiplicative coefficient for the discharge are constant over a DA cycle and vary from one DA cycle to another.
%The implemented DA algorithm consists in a cycled stochastic EnKF, where the control vector denoted by ${\bf x}$ is composed of $n=13$ parameters. The control vector gathers seven friction coefficients $K_{s_k}$ with $k\in [0,6]$, one multiplicative parameter $\mu$ to modify the time-varying upstream BC $Q(t)$, and five state corrective variables $\delta H_k$ with $k\in [1,5]$ over the floodplain subdomains. 
%The friction and multiplicative coefficient for the discharge are constant over a DA cycle and vary from one DA cycle to another. Different DA experiments are carried out in the present paper, depending on the definition including all or part of this list of coefficients and variables, as reported by Table~\ref{tab:runs} (in Section~\ref{DAExpeSett}). 

\begin{figure}[!h]
\centering
    \includegraphics[width=\linewidth]{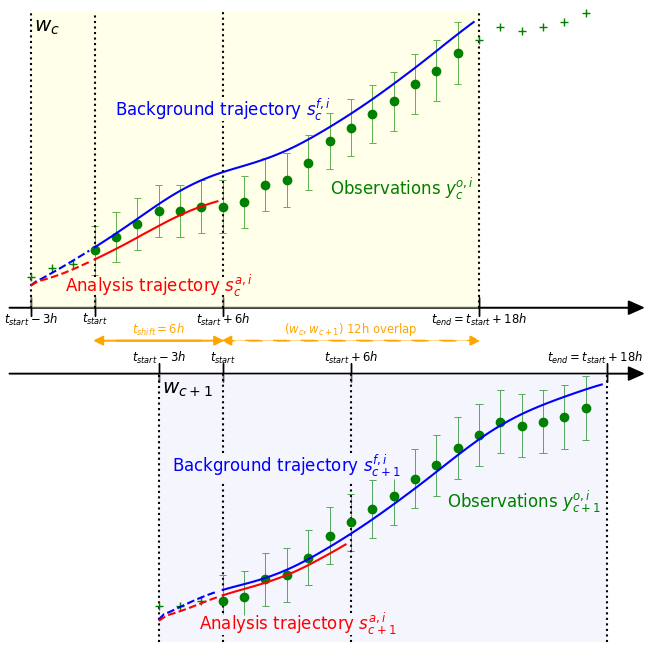}
     \caption{Schematic for the DA cycling, illustrated for $W_c$ and $W_{c+1}$, with background trajectory (blue) and analysis trajectory (red). The trajectory are represented by solid lines during its cycle when compared with the observations (green circles) and by dashed lines for the spin-up periods before the cycles. %Green pluses represent the observations outside the respective cycle (thus not assimilated). 
     The observations errors is represented with a green vertical bar.}
     \label{fig:cycling}
\end{figure}

Each DA cycle $c$ covers a time window, denoted by $W_c$ over $T = [t_{start},~t_{end}]$ of duration $18$ hours over which $n_{obs, c}$ observations are assimilated, as illustrated in Figure~\ref{fig:cycling}. Over each DA cycle $W_c$, a first simulation with the direct model is carried out---i.e. the background trajectory plotted in blue---and each observation is compared with its model equivalent at the respective observation time over $T$. The observations assimilated over $W_c$ are represented as green circles, whereas the observations that are not assimilated over $W_c$ are depicted as green pluses. The resulting misfit vector is used for the cycle analysis that provides a correction to the control vector. Then, the corrected control vector is used to carry out an analysis trajectory plotted in red, thus providing a coherent updated analyzed hydraulic state. The analysis trajectories are issued over a $9$-hour window starting at $t_{start}-3~hr$ in order to allow the hydraulic state to become coherent with the updated parameters and forcing by $t_{start}$, and ending at $t_{start}+6~hr$. This provides the final analyzed hydraulic state over $[t_{start}, ~t_{start}+6~hr]$ for $W_c$. The cycling of the DA algorithm then consists in sliding the time window of a period $t_{shift}=6$ hours so that the cycles $W_c$ and $W_{c+1}$ overlap for 12 hours. It should be noted that for the very first DA cycle $W_1$ of an event, a 24-hour period of direct simulation (or \textit{restart}) is achieved before comparing the model outputs to the observations in order to limit the impact of the initial condition.

%As mentioned in previous works \cite{nguyenagu2022,NguyenTGRS2022}, it could be argued that this DA algorithm is more similar to a smoother than a filter as it operates over a sliding time window. Yet, as the control vector is composed of model parameters and corrections that are assumed constant over the assimilation window (as opposed to the model state), the smoothing resumes to a filtering scheme. 

\subsection{Description of the EnKF forecast step}\label{subsect:EnKFfctstep}

The description of the EnKF forecast is detailed in \cite{nguyenagu2022,NguyenTGRS2022}. 
In the following, $i \in [1, N_e]$ indicates the member index within an ensemble of size $N_e$.  ${\bf x}^{f,i}_c$ (respectively,  ${\bf x}^{a,i}_c$) stands for the forecast (respectively, analysis) control vector for member $i$ over DA cycle $c$.  The EnKF forecast step consists in the propagation in time, over $W_c$, of the $N_e$ control and model state vectors. These steps are represented in the middle branch of Figure~\ref{fig:ExpSetWorkflow}. The background hydraulic state for each member ${\mathbf{s}}^{f,i}_{c}$ results from the integration of the hydrodynamic model ${\cal{M}}_c$: ${\mathbb{R}}^n \rightarrow {\mathbb{R}}^m$  from the control space to the model state (of dimension $m$) over cycle $c$: 
\begin{equation}
{\mathbf{s}}^{f,i}_{c} = {\cal{M}}_{c}\left({\bf s}^{a,i}_{c-1},{\bf x}^{f,i}_{c}\right),
\label{eq:stateforecast}
\end{equation}
where ${\bf s}^{a,i}_{c-1}$ is a restart file saved from the previous analysis at cycle $c-1$. As aforementioned, in order to avoid inconsistencies between the state and the analyzed set of parameters at $t_{start}$, a short spin-up integration is run on the three hours preceding $t_{start}$.
The equivalent of the control vector in the observation space  ${\bf y}^{f,i}_c$ is computed with the observation operator ${{\cal{H}}_c}$: ${\mathbb{R}}^m \rightarrow {\mathbb{R}}^{n_{obs}}$: 
\begin{equation}
{\mathbf{y}}^{f,i}_{c} = {{\cal{H}}_c}\left({\mathbf{s}}^{f,i}_{c}\right).
\label{eq:ctlequivobs}
\end{equation}

The observation vector is noted $\mathbf{y}^{o,i}_c$, it gathers in-situ WL and S1-derived WSR observations for cycle $c$. The in-situ WL subpart of $\mathbf{y}^{o,i}_c$ is noted $\mathbf{y}^{o,i}_{c, \mathrm{H}}$, and the WSR subpart of $\mathbf{y}^{o,i}_c$ is noted $\mathbf{y}^{o,i}_{c, \mathrm{WSR}}$. 
The equivalent of $\mathbf{y}^{o,i}_c$ results from Eq.~\ref{eq:ctlequivobs} that extracts simulated WL at locations and time of in-situ measurements $\mathbf{y}^{o,i}_{c, \mathrm{H}}$ and compute a wet/dry pixel mask from the computed WL simulated 2D field at S1 overpass times in order to compute WSR over each subdomains of the floodplain.

\subsection{Anamorphosis in the observation space}\label{subsect:GAstep}
%\subsection{Anamorphosis in the observation space for EnKF analysis}\label{subsect:GAstep}
The step of the DA that relates to the Gaussian anamorphosis are shown in red in Figure~\ref{fig:ExpSetWorkflow}. The model equivalent of WSR observations is a subset of ${\mathbf{y}}^{f,i}_{c}$ and is noted ${\mathbf{y}}^{f,i}_{c,\mathrm{WSR}}$ in the following. Similarly to the observations ${\mathbf{y}}^{o}_{c,\mathrm{WSR}}$, the model equivalent ${\mathbf{y}}^{f,i}_{c,\mathrm{WSR}}$ follows a non-Gaussian distribution and is bounded within $[0,1]$. In order to prevent the loss of optimality of the EnKF, the GA is applied in the observation space (also called physical space in the following) \offline{on} the model equivalents ${\mathbf{y}}^{f,i}_{c,\mathrm{WSR}}$ and \offline{on} the observations ${\bf y}^{o,i}_{c,\mathrm{WSR}}$. % (same notation for synthetical and real observations).

\begin{figure}[h]
  \centering
  \begin{subfigure}[b]{0.525\textwidth}
    \centering
    \includegraphics[width=\linewidth]{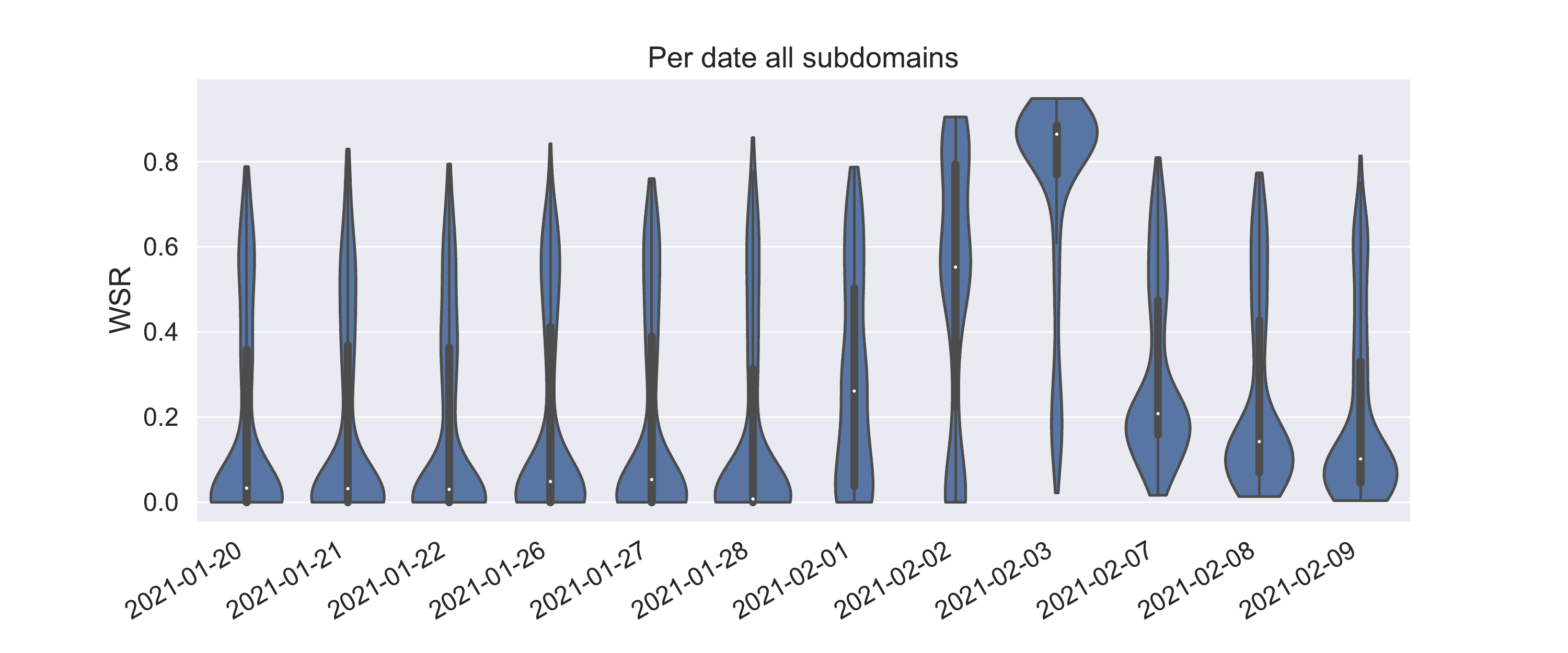}
    \caption{per date}
    \label{fig:WSR_violinplots_all_zones}
  \end{subfigure}
  %\hfill
  
  \begin{subfigure}[b]{0.525\textwidth}
    \centering
    \includegraphics[height=3.75cm]{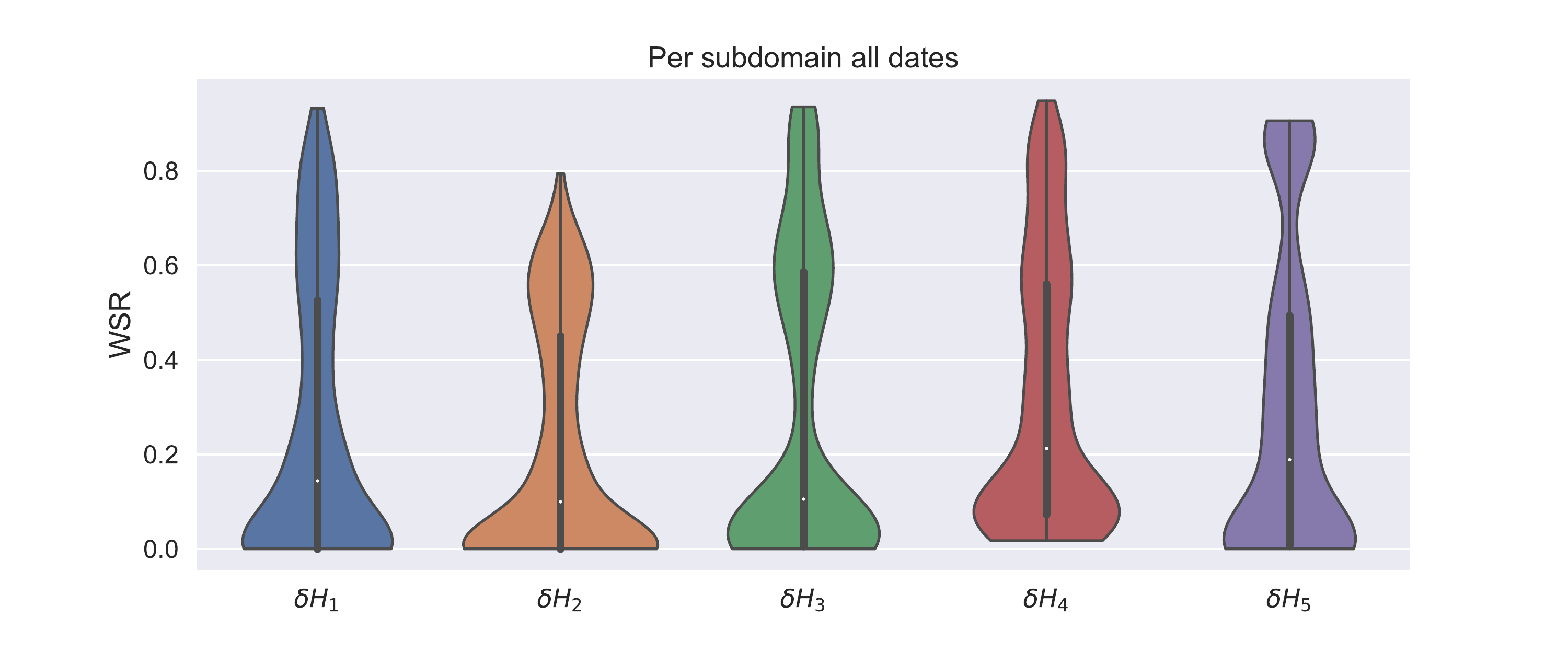}
    \caption{per subdomain}\label{fig:WSR_violinplots_all_days}
  \end{subfigure}
  \caption{Violin plots computed considering the model equivalents ${\mathbf{y}}^{f,i}_{c,\mathrm{WSR}}$ of SAR-derived WSR observations during the 2021 flood event, (a) generated over all five subdomains of the floodplain, at each S1 overpass time, and (b) generated over 2021 flood event, for each of the five subdomains.}
  \label{fig:WSR_violinplots_all_zones_all_days}
\end{figure}

Figure~\ref{fig:WSR_violinplots_all_zones_all_days} depicts the violin plots of WSR values by aggregating the model equivalents of all SAR-derived WSR observations over the five subdomains of the floodplain from all 75 members of the ensemble, for each S1 overpass time in panel (\textit{a}) and by aggregating all WSR model equivalents over all the times of S1 for each floodplain subdomain in panel (\textit{b}), over the entire 2021 flood event. In this Figure, the WSR model equivalents originate from the members of the ensemble of experiment IDA that sequentially assimilates in-situ WL observations at Tonneins, Marmande and La Réole \cite{NguyenTGRS2022}. %\modif{Our previous work \cite{NguyenTGRS2022} showed that IDA results in the floodplain are similar to that of an ensemble run without assimilation since the in-situ observations do not allow to account for errors in the floodplain.} \sophie{I think we can remove this sentence}
It appears that, for most dates before the rising limb (2021-02-02) and after the flood peak (2021-02-03), null and small values of WSR  prevail as indicated by the flat bottom of the violin plots, whereas higher values of WSR (close to 1) are reached near the flood peak as indicated by the larger flat top of the violin plots. It is worth-noting that for all dates and all floodplain subdomains, the distribution of WSR model equivalents is clearly non-Gaussian.

For a particular cycle $c$, the anamorphosis function $\phi_c$ is a non-linear bijective function, that maps WSR physical space to a Gaussian space:
\begin{equation}
    \tilde{{\bf y}}^{o,i}_{c, \mathrm{WSR}} = \phi_c \left({\bf y}^{o,i}_{c, \mathrm{WSR}} \right), \; \tilde{{\bf y}}^{f,i}_{c, \mathrm{WSR}} = \phi_c \left({\bf y}^{f,i}_{c, \mathrm{WSR}} \right).
\end{equation}

This comes down to redefining the observation operator ${\cal{H}}_c$ as $\tilde{{\cal{H}}}_c$ that now maps the hydraulic state variable ${\mathbf{s}}^{f,i}_{c}$ onto the transformed space:
\begin{equation}
{\tilde{\mathbf{y}}}^{f,i}_{c} = {\tilde{{\cal{H}}}_c}\left({\mathbf{s}}^{f,i}_{c}\right) = \phi_c \left({{\cal{H}}}_c\left({\mathbf{s}}^{f,i}_{c}\right) \right)
%\label{eq:ctlequivobs}
\end{equation}
where ${{\tilde{\cal{H}}}_c}$: ${\mathbb{R}}^m \rightarrow {\mathbb{R}}^{n_{obs}}$ selects, extracts and eventually interpolates model outputs at times and locations of the in-situ WL observations $\mathbf{y}^o_{c, \mathrm{H}}$, whereas it selects, extracts and applies the anamorphosis function $\phi_c$ at times and locations of WSR observations ${\bf y}^{o,i}_{c,\mathrm{WSR}}$, over $W_c$. This corresponds to prescribing an identity function for the anamorphosis of the in-situ WL observations and $\phi_c$ for that of the WSR observations.

The anamorphosis function $\phi_c$ is devised from the empirical marginal distribution of the variable in the observation space. For that purpose, all observations in time and space over $W_c$ are taken into account simultaneously. The algorithm to build the anamorphosis function is fully described in \cite[Appendix A]{simon2009}. It involves three steps: 
\begin{enumerate}
    \item Construction of the empirical anamorphosis function based on the marginal distribution of the non-Gaussian variable;
    \item Interpolation of the empirical piecewise function to allow a bijective function;
    \item Definition of the tails of the function necessary to process extremity values.
\end{enumerate}
% Complete details of this construction can be found in \cite[Appendix A]{simon2009}. 
The first step of the process is widely used in the geostatistical studies and thoroughly detailed in \cite{chiles1999geostatistics}. It maps the sampled values in the physical space onto the Gaussian space. The second and third steps are intended to ensure a bijective function, as presented in \cite{simon2009}. 
Three different approaches to select relevant samples in the physical space are presented in \cite{simon2012} in the context joint dual state-parameter estimation:  
(\textit{i}) a static approach working with existing realizations of the variables to transform; (\textit{ii}) a dynamic approach: the sample is populated by the members of the forecast ensemble at the time of the analysis; (\textit{iii}) an hybrid approach: combining both approaches which was shown to be more suitable for the problem of combined state-parameter estimation with non-linear models. %\modif{In the present work, since the size of the observation vector is \sophie{small} compared to that of the ensemble, the dynamic approach was favored.}
In order to build an anamorphosis that is coherent with the flood behavior, the dynamic approach is favored. Nevertheless, it should be noted that the computational cost of the anamorphosis function construction and application is considerably low, given that the size of the ensemble and the observation vectors are small.

% \begin{comment}
\begin{figure*}[h]
\centering
\begin{subfigure}[b]{0.32\textwidth}
\centering\includegraphics[width=\linewidth]{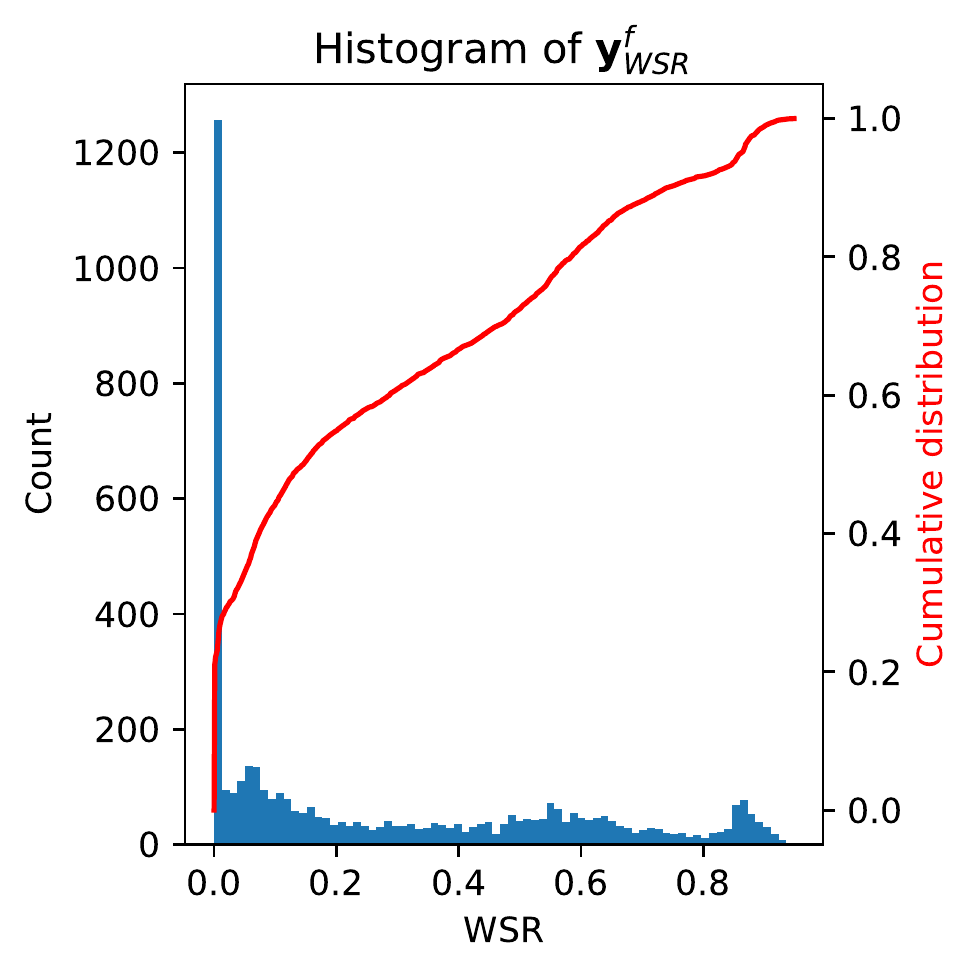}\caption{}\label{fig:bijective_a}
\end{subfigure}\hfill
\begin{subfigure}[b]{0.32\textwidth}
\centering\includegraphics[width=\linewidth]{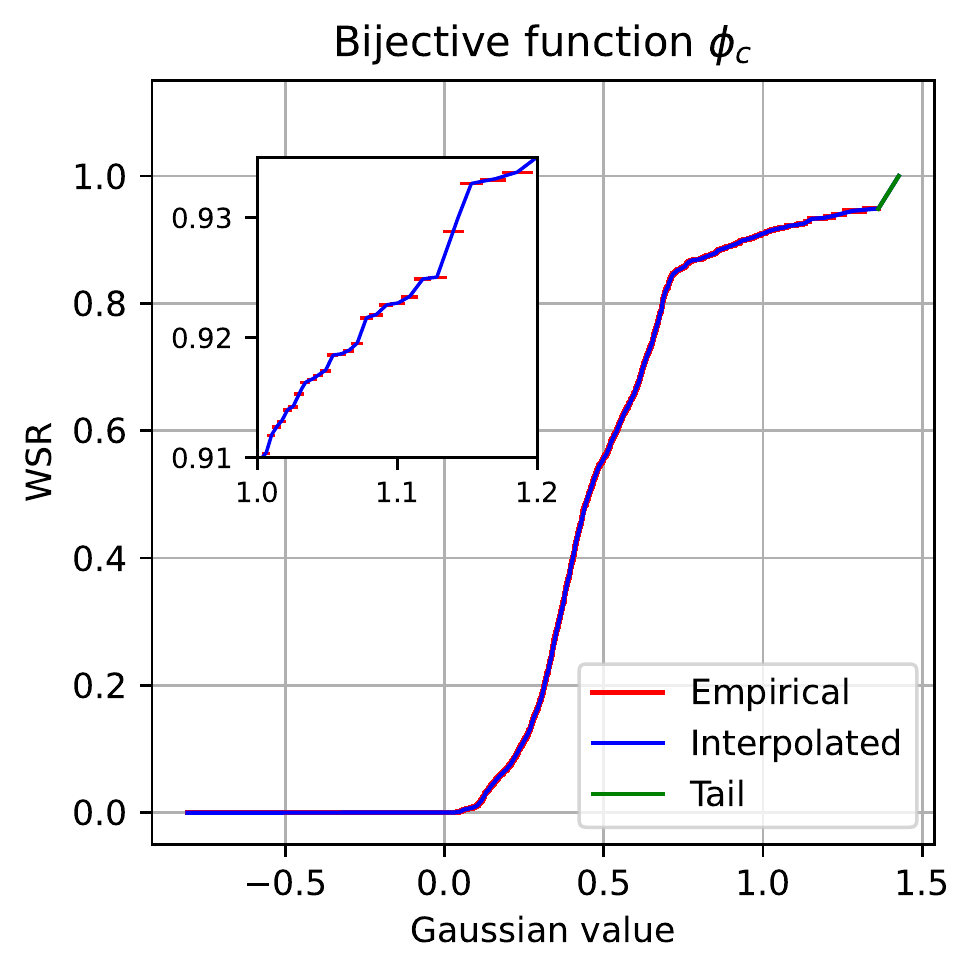}\caption{}\label{fig:bijective_b}
\end{subfigure}\hfill
\begin{subfigure}[b]{0.32\textwidth}
\centering\includegraphics[width=\linewidth]{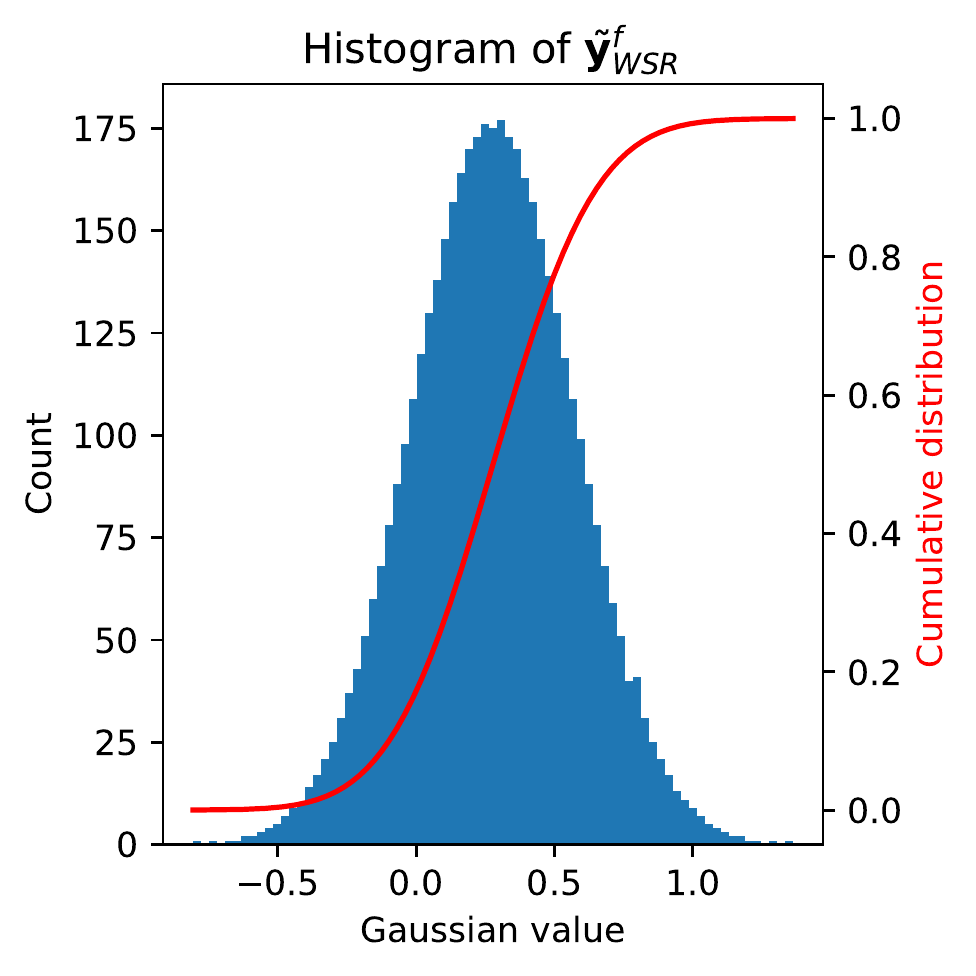}\caption{}\label{fig:bijective_c}
\end{subfigure}
\caption{(a) Histogram of model equivalent ${\mathbf{y}}^{f,i}_{c,\mathrm{WSR}}$ aggregating all subdomains and all dates; (b) Bijective function to transform from the physical space ($y$-axis) to the Gaussian space ($x$-axis); (c) Histogram of the transformed model equivalent $\tilde{\mathbf{y}}^{f,i}_{c,\mathrm{WSR}}$.}
\label{fig:bijective}
\end{figure*}
% \end{comment}

The anamorphosis is illustrated in Figure~\ref{fig:bijective} computed by aggregating all WSR model equivalents. The histogram of ${\bf y}^{f,i}_{c, \mathrm{WSR}}$ is shown in Figure~\ref{fig:bijective_a}, the devised anamorphosis function $\phi^c$ is plotted in Figure~\ref{fig:bijective_b} and the histogram of transformed WSR model equivalents $\tilde{{\bf y}}^{f,i}_{c, \mathrm{WSR}}$ is depicted in Figure~\ref{fig:bijective_c}. For both Figure~\ref{fig:bijective_a} and Figure~\ref{fig:bijective_c}, the cumulative distributions that were used for the function construction are shown in red. In contrast to the eco-biological variables treated in \cite{simon2009,simon2012}, WSR values are likely to reach the bounds of their domain definition, as WSR is strictly equal to 0 for dry areas and to 1 for flooded areas. As a consequence, the tails of the anamorphosis function requires a special treatment in this work to build a bijective function, to ensure a non-null dispersion among the ensemble members and artificially distinguish those all equal to 0 and 1 values---illustrated by the flat bottoms and tops in the violin plots in Figure~\ref{fig:WSR_violinplots_all_zones_all_days}. %Indeed, a linear resampling of the zeros is applied to convert them to an ordered list of small values bounded between $10^{-15}$ and $5\times 10^{-4}$. 
Indeed, a uniform random noise of small magnitude bounded between $10^{-15}$ and $5\times 10^{-4}$ is added to these zeros. These bound values are considered as small with respect to the WSR measurements that lie in $[0, 1]$, while still significant regarding the numerical precision of our numerical schemes. Similar strategy is also applied for the values close to 1. In the case where the anamorphosis function does not reach the 0 and 1 bounds, the function is extrapolated to cover the whole possible domain of WSR value (green segment of the bijective function shown in Figure~\ref{fig:bijective_b}).

\begin{figure}[h]
  \centering
  \includegraphics[width=0.6\linewidth]{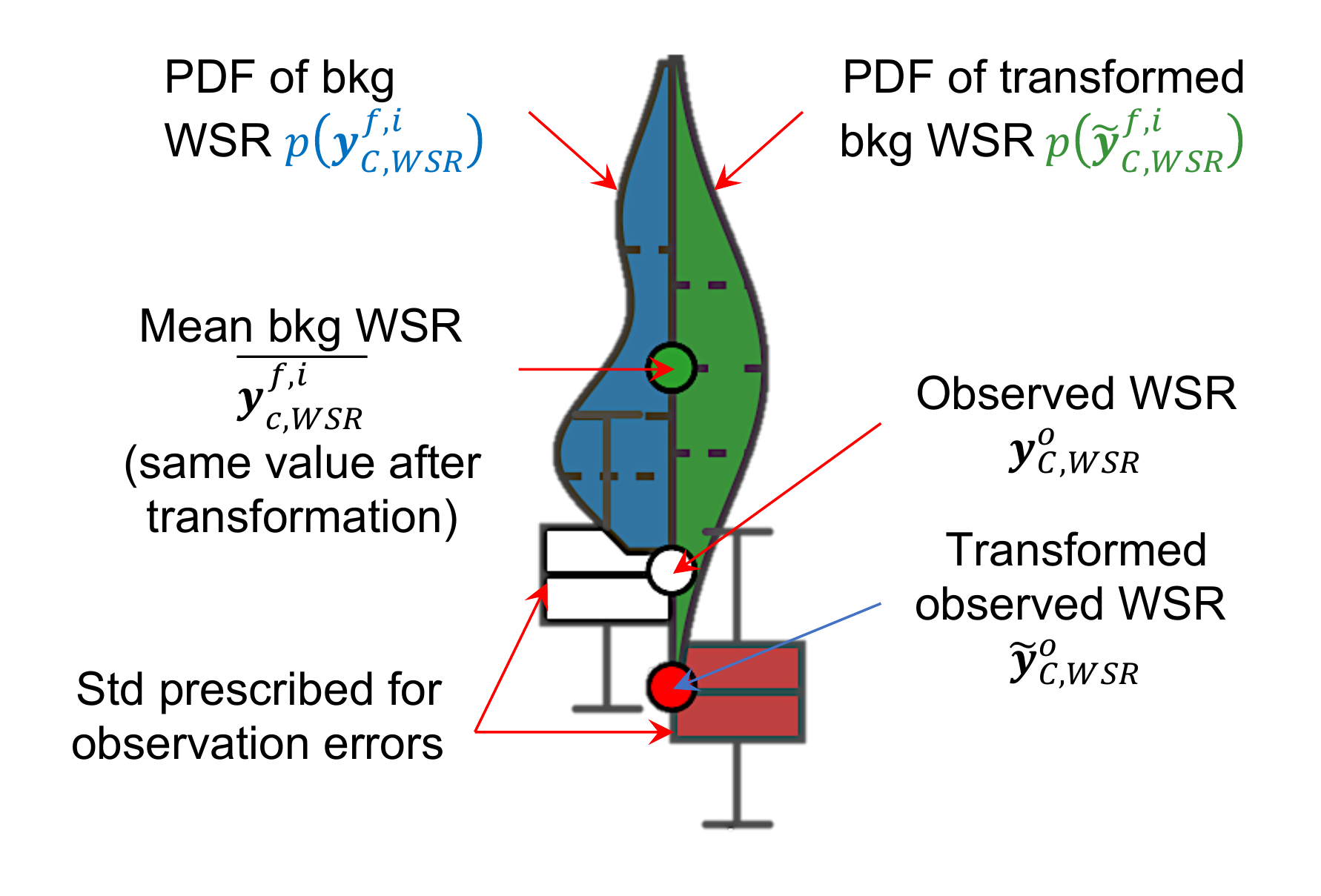}
  \caption{Split-violin plot representing the GA process.}
  \label{fig:WSR_violinplots}
\end{figure}

A composite illustration of the GA carried out on one subdomain (e.g. $\delta H_2$) at the flood peak (cycle $W_C$) is shown in Figure~\ref{fig:WSR_violinplots}. This split-violin plot represents the PDFs of the model equivalents for the WSR observations before GA ${\mathbf{y}}^{f,i}_{C,\mathrm{WSR}}$ (represented by the blue area) and after the GA $\tilde{\mathbf{y}}^{f,i}_{C,\mathrm{WSR}}$ (represented by the green area), using all WSR values from 75 members for the subdomain. 
% \sophie{dire combine d'obs on traite, sur combien d'info on fait la pdf}.
It appears, on this example, that the non-Gaussian PDF has been well morphed into a Gaussian PDF. The transformation is carried out centered around the mean of the background WSR values, represented by the green circle, that overlaps with the mean of the transformed distribution (blue circle). The observed WSR value ${\mathbf{y}}^{o}_{C,\mathrm{WSR}}$ in the physical space (respectively, transformed space) is represented with a white (respectively, red) dot. The prescribed standard deviation for the observation error is indicated by the white (respectively, red) box plot in the physical (respectively, transformed) space. 
% \sophie{introduire l'idée des 2 solutions pour la std}.
The observation error standard deviation is preserved in the transformed space in this work. %A different approach based on the anamorphosis of each perturbed observations for the stochastic version of the EnKF was tested, but it did not bring any improvement to the DA results.
An alternative approach using the standard deviation computed based on the transformed variables (performed by \cite{simon2009}) was also tested, but it did not bring any improvement to the DA results. 
Such a representation extended to all dates and all subdomains of the floodplain is shown in Section~\ref{ssec:res_GA}.

% \clearpage
\subsection{Description of the EnKF analysis step in the anamorphosed space}\label{subsect:EnKFanastep}

The EnKF analysis step usually stands in the update of the control ${\bf x}_c^{a,i}$ and \offline{the} associated model state vector ${\bf s}_c^{a,i}$, here achieved in the anamorphosed space and represented by a red rectangle in Figure~\ref{fig:ExpSetWorkflow}. This differs from the classical EnKF analysis \cite{nguyenagu2022,NguyenTGRS2022}, as the computation of the innovations and the covariance matrices are achieved in the transformed space using the transformed observation operator ${{\tilde{\cal{H}}}_c}$:
\begin{equation}
{\bf x}_c^{a,i} = {\bf x}^{f,i}_{c} + \mathbf{K}_{c} \left(\tilde{\bf y}^{o,i}_{c} - \tilde{\bf y}^{f,i}_{c}\right).
\label{eq:ctlana}
\end{equation}

The Kalman gain is further computed from covariance matrices stochastically estimated within the ensemble, considering anamorphosed observation vectors  $\tilde{\bf y}^{f,i}$  in place of ${\bf y}^{f,i}_{c}$.
\begin{equation}
	\mathbf{K}_c = \mathbf{P}^{\bf{x},\tilde{\bf{y}}}_c {\left[ \mathbf{P}^{\tilde{\bf{y}},\tilde{\bf{y}}}_c + \mathbf{R}_{c} \right]}^{-1}.
	\label{eq:EnKF_ana_Klambda_gain_chap12}
\end{equation}
%$\mathbf{P}^{\bf{x},\tilde{\bf{y}}}_{c}$  is the covariance matrix between the error in the control vector and the error in $\tilde{\mathbf{y}}^{f}_{c}$. 
%$\mathbf{P}^{\tilde{\bf{y}},\tilde{\bf{y}}}_{c}$ is the covariance matrix of the error in the background state equivalent in the transformed observation space $\tilde{\mathbf{y}}^{f}_{c}$
\sophie{comment equations here to be consistent with the amount of observation given in the forescast step paragraph, and stays different from WRR section}
\begin{comment}
where
\begin{equation}
	\mathbf{P}^{\bf{x},\tilde{\bf{y}}}_{c} = \frac{1}{N_e - 1} \mathbf{X}_c^\top \tilde{\mathbf{Y}}_c \in \mathbb{R}^{n \times n_{obs}}
	\label{eq:Pxy}
\end{equation}
\begin{equation}
	\mathbf{P}^{\tilde{\bf{y}},\tilde{\bf{y}}_{c}} = \frac{1}{N_e - 1} \tilde{\mathbf{Y}}_c^\top \tilde{\mathbf{Y}}_c \in \mathbb{R}^{n_{obs} \times n_{obs}}
	\label{eq:Pyy}
\end{equation}
with:
\begin{equation}
	\mathbf{X}_c = \left[ {\bf x}^{f,1}_c- \overline{{\bf x}^f_c}, \cdots, {\bf x}^{f,N_e}_{c} -\overline{{\bf x}^f_c}\right] \in \mathbb{R}^{n \times N_e}
\end{equation}
\begin{equation}
	\tilde{\mathbf{Y}}_c = \left[ {\tilde{\bf y }}^{f,1}_c- \overline{\tilde{\bf y}^f_c}, \cdots, \; \tilde{{\bf y}}^{f,N_e}_{c} - \overline{\tilde{{\bf y}}^f_c}\right] \in \mathbb{R}^{n_{obs} \times N_e}
	\label{eq:ens_anomaly_matrix}
\end{equation}
and 
\begin{equation}
	\overline{{\bf x}^f_c} = \frac{1}{N_e} \sum\limits_{\substack{i = 1}}^{N_e} {{\bf x}^{f,i}_c} \in \mathbb{R}^{n}
	\label{eq:ens_mean_x}
\end{equation}
\begin{equation}
	\overline{\tilde{{\bf y}}^f_c} = \frac{1}{N_e} \sum\limits_{\substack{i = 1}}^{N_e} {\tilde{{\bf y}}^{f,i}_c} \in \mathbb{R}^{n_{obs}}.
	\label{eq:ens_mean_y}
\end{equation}
${\bf R}_c= \sigma_{obs,c}^2 {\mathbf I}_{n_{obs}}$ is the observation error covariance matrix, where ${\mathbf I}_{n_{obs}}$ is the ${n_{obs} \times n_{obs}}$ identity matrix. 
\end{comment}

\sophie{Then, similar to Equation~\eqref{eq:stateforecast},  the hydrodynamic state ${\bf s}^{a,i}_c$ associated with each analyzed control vector ${\bf x}^{a,i}_c$  results from the integration of the hydrodynamic model ${\cal{M}}_c$ with the updated parameters in $\underline{\underline{\bf x}}{}^{a,i}_{c}$: 
\begin{equation}
	{\mathbf{s}}^{a,i}_{c} = {\cal{M}}_{c}\left({\bf s}^{a,i}_{c-1},\underline{\underline{\bf x}}{}^{a,i}_{c}\right),
	\label{eq:stateanalyzed}
\end{equation}
where $\underline{\underline{\bf x}}{}^{a,i}_{c}$ gathers $(K_{s_k})^{a,i}_{c}$, $\mu^{a,i}_c$ and the state correction in the floodplain $\delta H^{a,i}_k$ over $W_c$, starting from the same initial condition as each background simulation within the ensemble. In should be noted that, in order to preserve a smooth WL field, the mean of the analysis for $\overline{\delta H_k^a}$ computed within the ensemble is considered, instead of individual member values. }

\section{Experimental setting}
\label{ExpeSett}

\subsection{Generation of synthetical observations for OSSE}
 
 %\begin{figure}[!ht]
 %\centering
 %    \includegraphics[width=0.8\linewidth]{fig/truth_params2.pdf}
 %     \caption{True values of the control vector for the reference simulation over the synthetical 2021 event in OSSE. First panel: $K_{s_0}$; second panel: $K_{s_k}$ with $k\in [1,6]$; third panel: $\delta H_k$ with $k\in [1,5]$; last panel: $\mu$ (left y-axis, dashed curve) and $Q(t)$ (right y-axis, solid curve).} %These color codes are identical to those of Figure~\ref{fig:study_area}.}%The S1 overpass times are indicated as vertical black dashed lines.} 
 %     \label{fig:OSSE_truth}
 %\end{figure}

The OSSE experimental setting here is similar to that described in \cite{nguyenagu2022,nguyen2022hic}. It is represented on the right branch in Figure~\ref{fig:ExpSetWorkflow}, framed in \sophie{a} yellow block. The reference simulation (further denoted as {\it True}) is a deterministic simulation with a selected set of time-varying parameters for friction and inflow. 
The true value for  $K_{s_k}$ with $k\in [0,6]$, $\mu$ and $\delta H_k$ with $k\in [1,5]$ are set from the results of a previous DA experiment on the real 2021 flood event where in-situ WL observations were assimilated, and then a WL correction is added in the floodplain to account for T2D inability to empty the floodplain. 
\begin{comment}
\modif{In Figure~\ref{fig:OSSE_truth}, the time-varying true friction coefficient for the floodplain $K_{s_0}$ is plotted in black on the top panel and the true friction coefficients $K_{s_k}$ with $k\in [1,6]$ for the river bed are depicted on the second panel. %The color codes correspond to the ones used in Figure \ref{fig:study_area}.
%\sophie{The true value for the multiplicative correction $\mu$ is represented  by a solid cyan curve (left y-axis).}
The true values for state correction $\delta H$  were set up with negative cosine curves for the three first groups of three consecutive S1 observations (from the beginning of the event until the flood peak), and a constant water removal of $-18$ cm over the flood recession period. They are shown on the third panel in Figure~\ref{fig:OSSE_truth}. For the sake of consistency, the color codes for $K_{s_k}$ with $k\in [1,6]$ and for $\delta H_k$ with $k\in [1,5]$ are identical to their effective areas depicted in Figure~\ref{fig:study_area}. 
Lastly, the time-series inflow discharge $Q(t)$ for the OSSE is represented on the bottom panel. The true multiplicative correction $\mu$, represented by a dashed cyan curve (left $y$-axis), applied on the original BC (right $y$-axis), was also issued from a previous DA experiments and is added a small cosine function as perturbation.}
\end{comment}
%
Synthetical in-situ and SAR-derived flood extent observations are generated using the observation operator described in Section~\ref{Methods} applied to the reference simulation outputs. These observations are further assimilated to retrieve the model parameters from the reference simulation, starting from a priori values.  
%The reference simulation is used to provide synthetic observations at the in-situ and S1 observation times from the real 2021 event, using the observation operator described in Section~\ref{Methods}. 
%This stands in the extraction of the true WL values at all observing times and locations, first to generate synthetical in-situ WL observations, and second to extract the wet/dry pixels from synthetical flood maps for the WSR computation. % \modif{An additional observation error is then added to the WL and WSR values.} %\sophie{préciser la nature de ce bruit. c'est un bruit gaussien de std 0.1m pour WL je suppose, mais combien pour WSR ?} 
%The synthetical in-situ and WSR observations are then assimilated in a DA experiment, with a priori settings that differ from the truth. The OSSE experiments aim at assessing the performance of the DA method involving both types of observations (in-situ WL and S1-derived WSR), especially its capacity to retrieve the true parameters (forcing upstream data, friction coefficients and hydraulic state correction). 

\subsection{Configuration for DA experiments and metric assessment}
\label{DAExpeSett}

\begin{table*}[t]
	\centering
	\caption{Settings for OSSE and real experiments.}
	\label{tab:runs}
	\begin{tabular}{cccc}
		\hline
		Name of the     & Observing & Ensemble & Control \\ 
		experiment  & network &  size $N_e$ & vector \\ \hline
		FR  & No assim & 1 & - \\ %\hline
		IDA$_{osse}$/IDA  & In-situ WL & 75 & $K_{s_{[0:6]}}, \mu $ \\ %\hline
% 		IWDA$_{(OSSE)}$/IWDA & Yes & In-situ WL and WSR & 75 & $K_{s_{[0:6]}}, \mu $ \\ %\hline
		IHDA$_{osse}$/IHDA  & In-situ WL and WSR & 75 & $K_{s_{[0:6]}}, \mu, \delta H_{[1:5]}$ \\
		IGDA$_{osse}$/IGDA  & In-situ WL and WSR & 75 & $K_{s_{[0:6]}}, \mu, \delta H_{[1:5]}$ \\\hline
	\end{tabular}
\end{table*}

Table~\ref{tab:runs} presents the experimental settings for the free run and for the DA simulations that are devised with different control vector and observing network. 
%In both OSSE and real event experiments, one  deterministic control run (named {\it Free run}, without assimilation or open-loop, noted FR in the following), and three DA experiments were carried out  with different configurations regarding the types of observations that are assimilated and the active components of the control vector, as detailed in Table~\ref{tab:runs}. 
It should be noted that FR is not represented in Figure~\ref{fig:ExpSetWorkflow} as this simulation does not involve DA and is used only to assess the merits of DA. For both OSSE and real experiments, FR is devised using the observed upstream time-series as forcing, as well as the friction coefficients that result from calibration. There is no state correction in the floodplain in FR.
The proposed EnKF approach carried out by DA experiments is represented in Figure~\ref{fig:ExpSetWorkflow}, where the middle branch represents the forecast step and the left branch is the analysis. The real DA experiments use the same forcing and friction coefficient settings as FR at the first \sophie{DA} cycle, and then involve a sequential correction over the next cycles. They assimilate real in-situ WL and WSR observations, represented as green boxes in Figure~\ref{fig:ExpSetWorkflow}. 

The observing network is composed of in-situ WL observations at the three Vigicrue stations Tonneins, Marmande and La Réole, every 15 minutes,  eventually completed with WSR values computed over the five floodplain zones at S1 overpass times. 
%Then, two settings for the control vector are handled, the first one gathers all seven friction coefficients and the inflow multiplicative coefficient, whereas the second one is extended with the water state correction in the floodplain. With these configurations, three DA experiments are devised and named IDA, IHDA and IGDA. 
IDA experiment only assimilates in-situ WL observations and the control vector is limited to friction coefficients $K_{s_k}$ with $k\in [0,6]$ and the inflow multiplicative coefficient $\mu$. The control vector for IHDA is extended to include $\delta H_k$ with $k\in [1,5]$. It assimilates the same in-situ WL observations \sophie{like IDA, as well as} WSR observations. Lastly, experiment IGDA is at the core of the present work. It is similar to IHDA but the anamorphosis is applied on the  vectors expressed onto the observation space in order to alleviate for the violation of the Gaussian assumption for WSR observations. The analysis for IGDA is thus carried out in the transformed Gaussian space.

The observation error (Section~\ref{subsect:GAstep}) is set proportional to the value of the in-situ WL observations $\mathbf{y}^{o}_{c, \mathrm{H}}$, while it is prescribed as a scalar fixed value for the WSR observations $\mathbf{y}^{o}_{c, \mathrm{WSR}}$.  As such, the standard deviation of the in-situ WL error is fixed to $\tau=15\%$ so that $\sigma_{obs,c, \mathrm{H}} = \tau \mathbf{y}^{o}_{c,\mathrm{H}}$. On the other hand, the standard deviation of the SAR-derived WSR observations is fixed to $0.1$ (and up to $0.2$ depending on how early the observation time within each $W_c$). These values stem from the assessment of the FloodML algorithm that provides wet/dry classification results (validated on five test sites all over the world) with an overall accuracy of $86.86\%$ \cite{NguyenTGRS2022}. All DA experiments where carried out using $N_e=75$ members. In the following, the subscript $osse$ is used in the experiment name to distinguish the OSSE \sophie{mode} from the real mode. 

The metric employed for 1D assessment is formulated as the root-mean-square error (RMSE) of the WL times-series computed with respect to in-situ WL observations (synthetical or real\sophie{)}, over time, at the three observing stations. The metric for 2D assessment is the Critical Success Index ($\mathrm{CSI}$), also known as Intersection-over-Union ($\mathrm{IoU}$) for RS-based object segmentation tasks, which formulates the fit between a simulated 2D flood extent and an observed flood extent (synthetical or real) derived from S1 observations and FloodML algorithm \cite{kettig}. $\mathrm{CSI}$ varies from 0\% when no common area (i.e. no agreement) is found between the simulated and the observed flood extents, and reach their maximal value of 100\% when the prediction perfectly fits the observed flood extents. More details on the formulation of these metrics can be found in \cite{NguyenTGRS2022}.

\section{Results}
\label{results}

The impact of the GA on the distribution of the WSR model equivalent is shown in subsection~\ref{ssec:res_GA}. The results of the DA experiments 
% when assimilating WSR data in the floodplain which are complementary to in-situ WL data in the river bed 
are shown in the control space (subsection~\ref{sect:res_control}) and the observation space (subsection~\ref{sect:res_observation_space}) for both OSSE and real experiments, using RMSE computed at the observing stations and CSI computed over the whole domain.

\subsection{Result of GA}\label{ssec:res_GA}
\begin{figure*}[!t]
    \centering
    \begin{subfigure}[b]{0.495\textwidth}
    \centering\includegraphics[trim=0 0 5.11cm 0,clip,width=\linewidth]{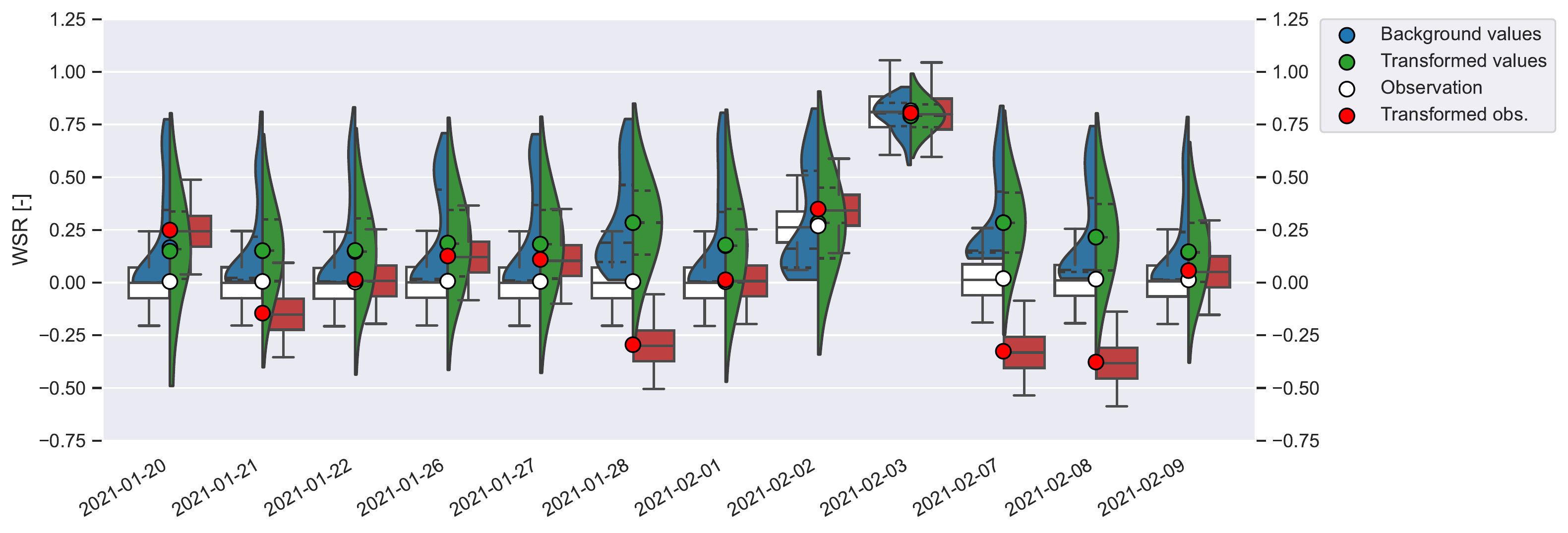}\caption{Subdomain 1}
    \end{subfigure}\hfill
    \begin{subfigure}[b]{0.495\textwidth}
    \centering\includegraphics[trim=0 0 5.11cm 0,clip,width=\linewidth]{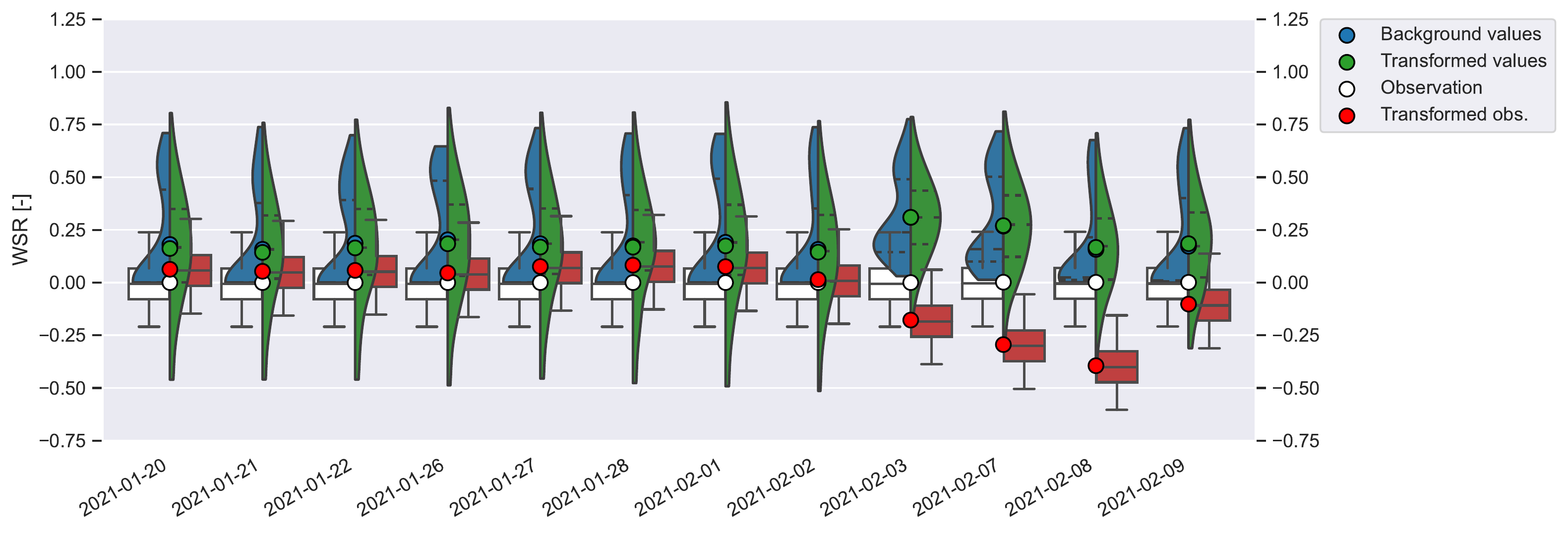}\caption{Subdomain 2}
    \end{subfigure}
    \begin{subfigure}[b]{0.495\textwidth}
    \centering\includegraphics[trim=0 0 5.11cm 0,clip,width=\linewidth]{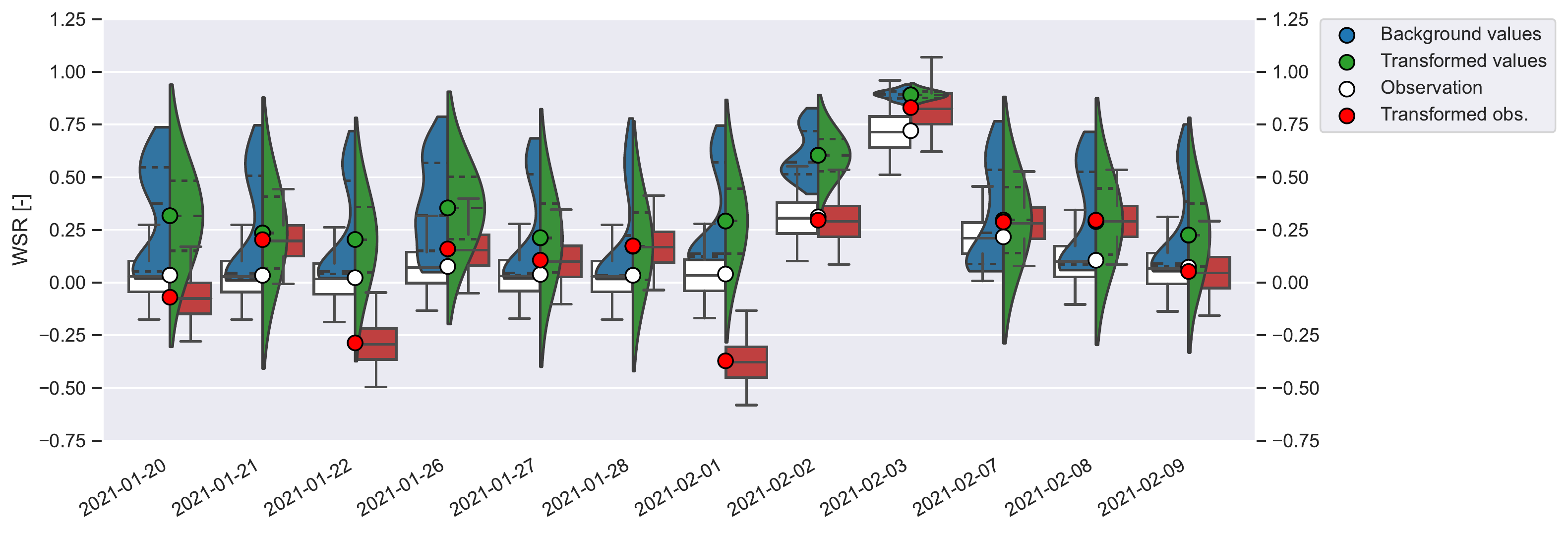}\caption{Subdomain 3}
    \end{subfigure}\hfill
    \begin{subfigure}[b]{0.495\textwidth}
    \centering\includegraphics[trim=0 0 5.11cm 0,clip,width=\linewidth]{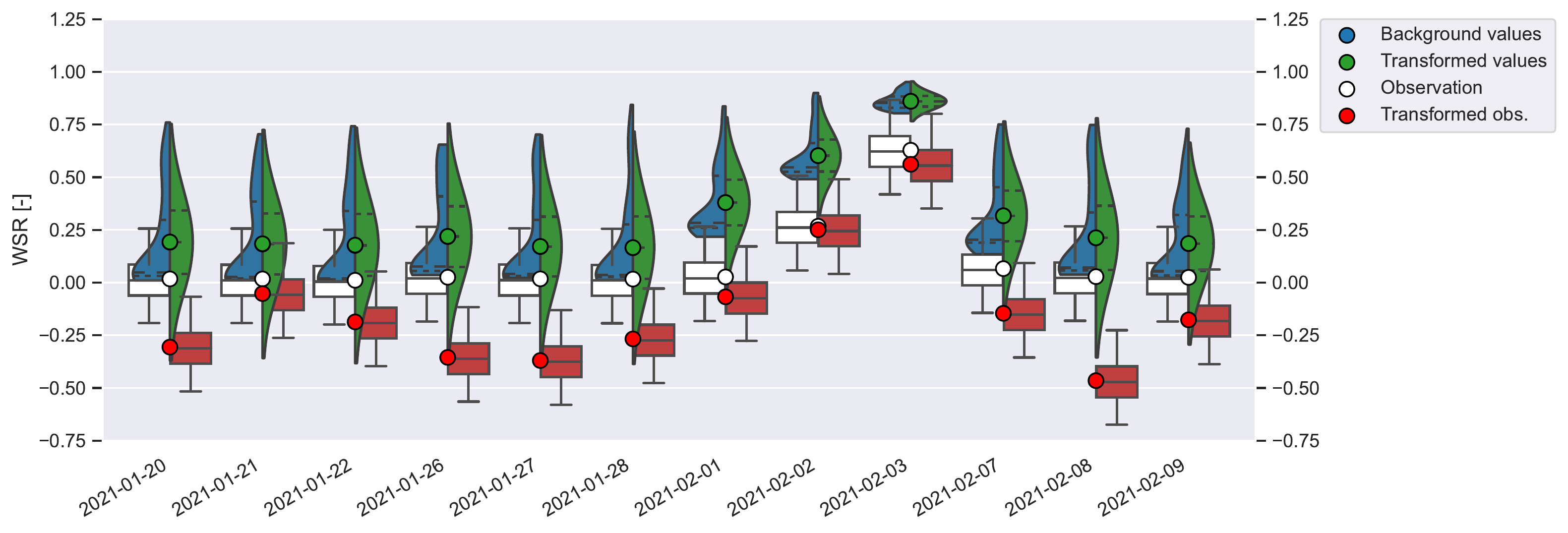}\caption{Subdomain 4}
    \end{subfigure}
% \end{figure*}%
% \begin{figure*}[t]\ContinuedFloat
    % \centering

    \begin{subfigure}[b]{0.59\textwidth}
    \centering\includegraphics[width=\linewidth]{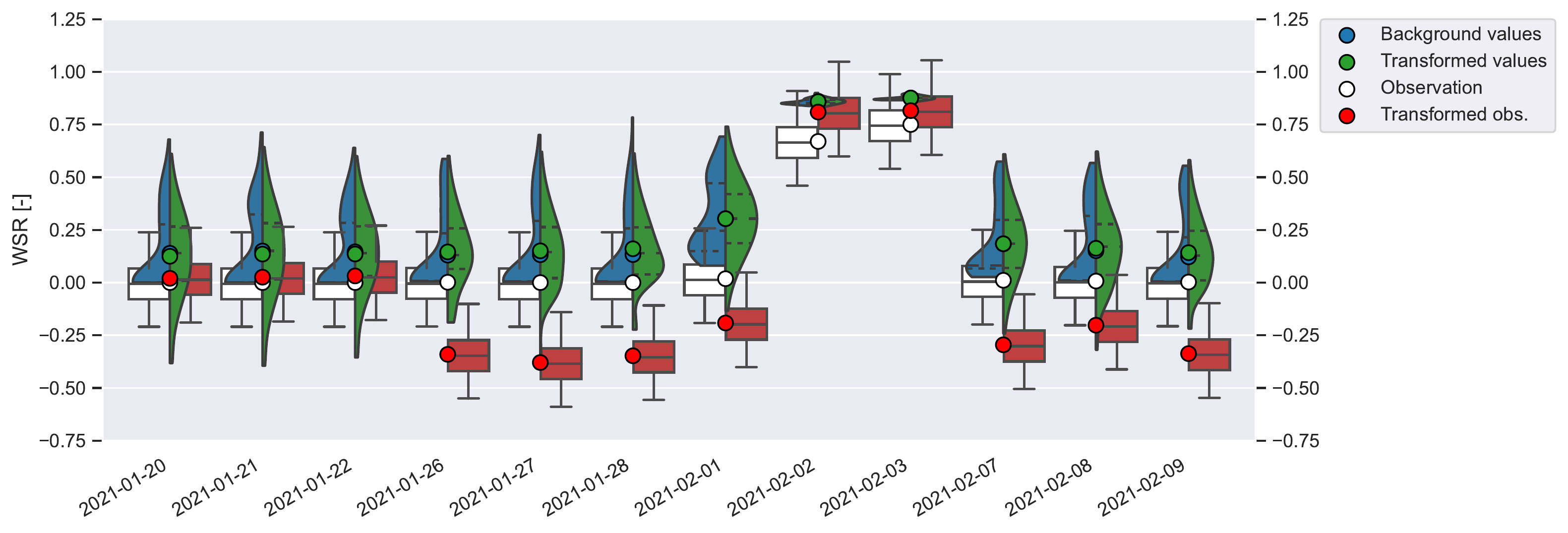}\caption{Subdomain 5}
    \end{subfigure}
    \caption{Split-violin plots computed considering all observations from the real event over each of the five floodplain subdomains for every S1 overpass time.}
    \label{fig:WSR_violinplots_each_zone_each_day}
\end{figure*}

Similarly to Figure~\ref{fig:WSR_violinplots}, the split-violin plots in Figure~\ref{fig:WSR_violinplots_each_zone_each_day} show the PDFs for the model equivalent for the WSR observations for 2021 flood event computed considering all WSR observations over each of the five subdomains \sophie{of the floodplain } (subdomain 1 to 5 illustrated in panel (\textit{a}) to (\textit{e})) for each S1 overpass time (along the $x$-axis), with all 75 members of the ensemble. %\sophie{Je pense qu'on peut supprimer cette phrase car la figure est déjà introduite dans la section 3 - The PDF in the physical space is represented in blue on the left split of the violin plot, whereas the PDF in the transformed space in represented in green on the right split of the violin plot. The observed WSR value in the physical space (respectively, transformed space) is represented with a white (respectively, red) dot. The prescribed standard deviation for the observation error is indicated by the white (respectively, red) box plot in the physical (respectively, transformed) space. }
% \sophie{Je crois que ce n'est pas l'endroit où introduire cette notion - On peut en parler dans la section 3 qd on montre la figure 6, puis en reparler dans la discussion - The observation error standard deviation is preserved in the transformed space. A different approach based on the anamorphosis of each perturbed observations for the stochastic version of the EnKF was tested, but it did not bring any improvement to the DA results.}
% As aforementioned, the tested second approach using the standard deviation in the gaussian space, computed based on the transformed variable samples, did not bring any improvement to the DA results, thus it is not presented in this paper. 
Here again, it is worth-noting that for all subdomains and all dates, the PDFs in the transformed space are well rendered Gaussian, with the mean of the PDF reaching small values before and after the flood peak and large values (close to 1) near the peak (2021-02-03).

\subsection{Results in the control space}
\label{sect:res_control}

\subsubsection{Results for OSSE experiment}
Figure~\ref{fig:control_OSSE} displays the analyzed parameters from the OSSE experiments. 
%%The reference parameter values (\textit{Truth}) are plotted in black and the default values (those of {\it Free run}), which were summarized in Table~\ref{tab:PDFs}, are indicated by horizontal dashed lines. The results for the different DA experiments are plotted with blue lines for IDA, green lines for IHDA and red lines for IGDA. The S1 overpass times over the 2021 event are depicted by vertical dashed lines. The analyzed values for $K_{s_k}$ (with $k\in [0,6]$) are shown on the left column. The analyzed values for the inflow correction $\mu$ is shown in the top panel of the right column while the analyzed values in both IHDA and IGDA experiments for $\delta H_k^a$ with $k\in [1,5]$ are shown on the other panels of the right column, respectively in green and in red. The default value for $\delta H_k$ is $0$. The bottom right panel displays the upstream forcing for reference purpose.
\begin{figure*}[h]
  \centering
  \begin{subfigure}[b]{\textwidth}
  \centering
  \includegraphics[width=0.8\linewidth]{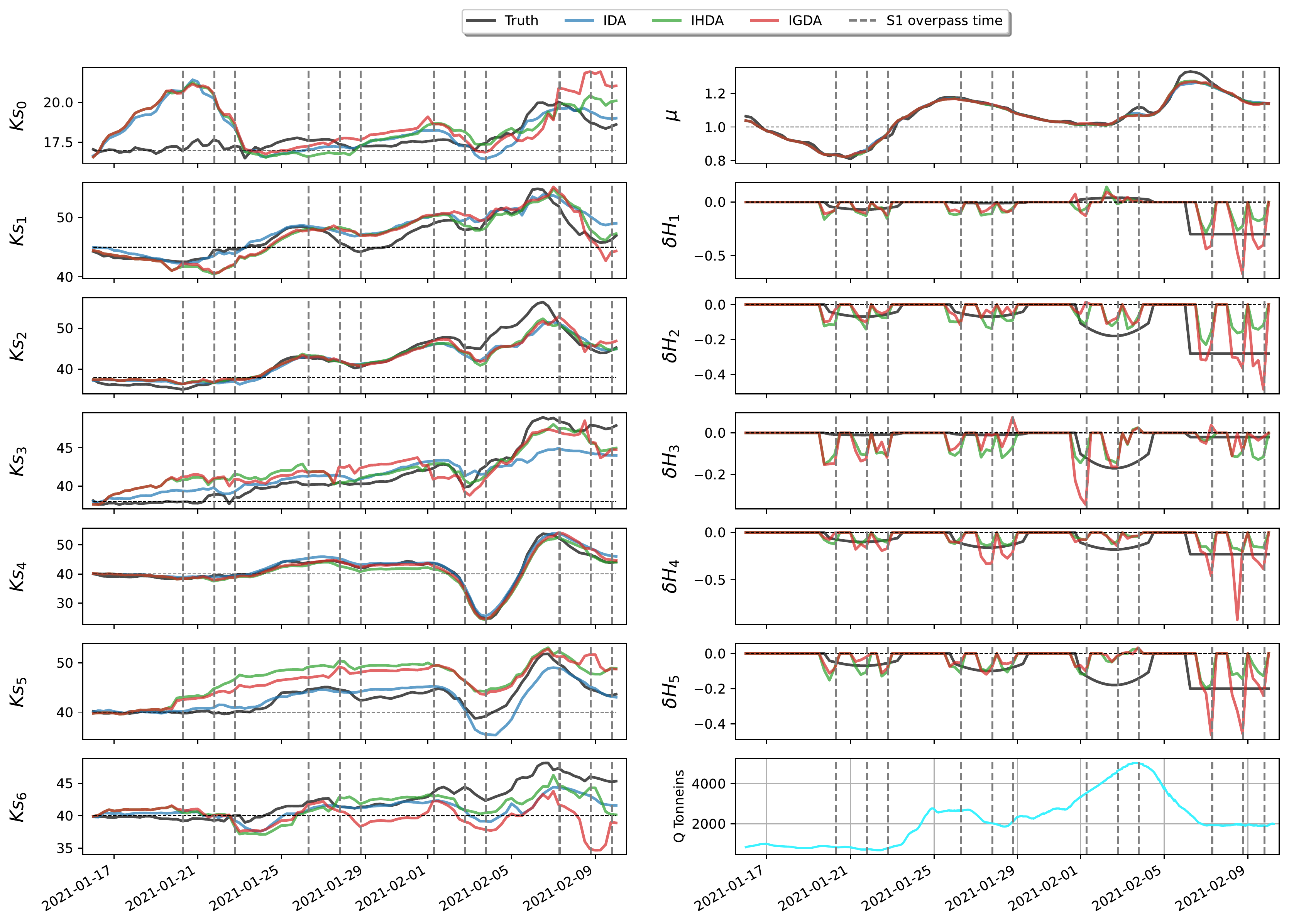}
  \caption{in OSSE}
  \label{fig:control_OSSE}
  \end{subfigure}
  \caption{Analyzed values of the control vector, (a) in OSSE, and (b) in real event 2021. Left: friction coefficients in the floodplain $K_{s_0}$, and in the river bed $K_{s_k}$ (with $k\in [1,6]$); Right: multiplicative correction to the inflow $\mu$, state correction $\delta H_k$ (with $k \in [1,5]$), and upstream forcing $Q(t)$, resulting from DA for IDA (blue), IHDA (green), and IGDA (red). Horizontal dashed lines indicate the a priori values from calibration. Vertical dashed lines indicate S1 overpass times.}
\end{figure*}%
\begin{figure*}[t]\ContinuedFloat
  \captionsetup{list=off,format=cont}
  \begin{subfigure}[b]{\textwidth}
  \centering
  \includegraphics[width=0.8\linewidth]{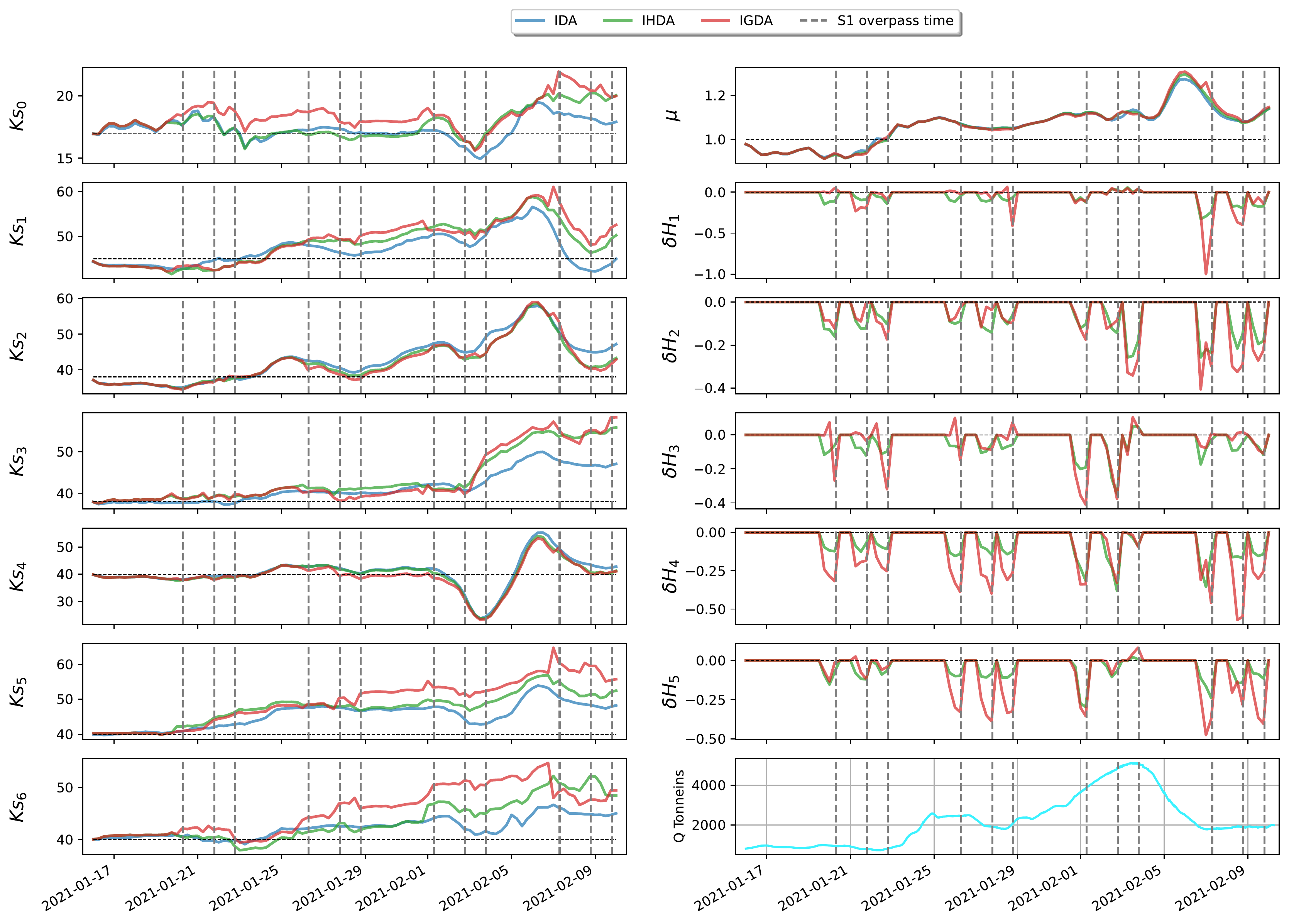}
  \caption{in real event}
  \label{fig:control_real}
  \end{subfigure}
  \caption{Analyzed values of the control vector (continued).}
\end{figure*}

In the river bed, the assimilation of in-situ WL observations with a classical EnKF in IDA suffices to \sophie{successfully} retrieve the true friction coefficients, however with a lesser success \sophie{for} $K_{s_5}$ and $K_{s_6}$. The addition of WSR data in IHDA and IGDA does not bring much of an improvement, except for a slight improvement for $K_{s_3}$ near Marmande during the flood recess. For $K_{s_0}$ in the floodplain, the dispersion of the results is more visible at the flood peak and afterwards, when the sensitivity of the flow to the floodplain friction is most important. Yet, as the velocity of the flow is relatively small, this sensitivity remains weak. 
The dispersion of the analysis for $K_{s_5}$ and $K_{s_6}$ is most likely due to equifinality issues, as the downstream part of the flow is also influenced by previous river segments with $K_{s_3}$ and $K_{s_4}$. As previously noted in \cite{nguyenagu2022}, the in-situ WL observation at Marmande is perfectly efficient at constraining the friction coefficient $K_{s_4}$ downstream of the gauging station and the analysis for the multiplicative factor $\mu$ is near to perfect for all three DA experiments. 
When WSR observations are assimilated at S1 overpass times and the correction for the WL is included in the control vector (for IHDA and IGDA experiments),  the analysis for $\delta H_k$ $(k \in [1,5])$ is close to the prescribed values for the reference run. The GA in experiment IGDA tends to yield larger increment\sophie{s} (negative values) during the flood recess. 

% \clearpage
\subsubsection{Results for real experiment}
The analyzed parameters from the different DA experiments in real 2021 flood event are shown in Figure~\ref{fig:control_real}. %The layout and color code for  Figure~\ref{fig:control_real} are similar to that of  Figure~\ref{fig:control_OSSE}. 
For all DA experiments, the analyzed values for the friction coefficients in the river bed and the floodplain remain within physical ranges, even though with more dispersion at the flood peak and during the flood recess where the innovations in real event are larger. 
The analysis for IDA visibly differs from the results of the other DA experiments that assimilated WSR observations with smaller corrections added to the default values, yet all DA analyses follow similar trends for all the components of the control vector. The analyses on $\mu$ are similar for IDA, IHDA, and IGDA for the whole event. This suggests that the in-situ WLs observed at Tonneins are enough to constraint the multiplicative correction to the inflow and that the use of additional data in the floodplain is not necessary. The GA leads to \sophie{a} slightly different analysis compared to IHDA, with larger WL correction (more negative) than IHDA. 
In contrast to the OSSE, this assessment of the results in the control space does not allow to quantitatively evaluate the DA experiments, since the true values of the controlled parameters are not known.
% \begin{figure*}[!htpb]
%   \centering
%   \includegraphics[width=0.8\linewidth]{fig/control_real.png}
%   \caption{Analyzed values of the control vector for IDA (green), IWDA (red),  IHDA (violet), and IGDA (brown) in real event. The default values are represented with horizontal dashed lines, whereas the S1 overpass times are shown with vertical dashed lines. Left column: friction coefficients in the floodplain $K_{s_0}$, and in the river bed $K_{s_k}$ (with $k\in [1,6]$). Right column, from top to bottom: multiplicative correction to the inflow $\mu$, state correction  $\delta H_k$ (with $k \in [1,5]$), and upstream forcing $Q(t)$.}
%   \label{fig:control_real}
% \end{figure*}

\subsection{Results in the observation space}
\label{sect:res_observation_space}

\subsubsection{Water levels at observing stations (for OSSE and for real experiment)}
% \begin{comment}
% \sophie{Resultats dans l'observation space : pour l'instant, on a le tableau des RMSE sur H aux stations d'observations calculée parrapport à la référence à Tonneins, Marmande, La réole. Pour faire un peu différent du papier WRR, on pourrait montrer une time series de H aux stations pour Truth, IDA, IWDA, IHDA et IGDA.}
% \end{comment}

\begin{figure*}[!]
    \centering
    \begin{subfigure}[b]{\textwidth}
    \centering
    \includegraphics[width=0.32\linewidth]{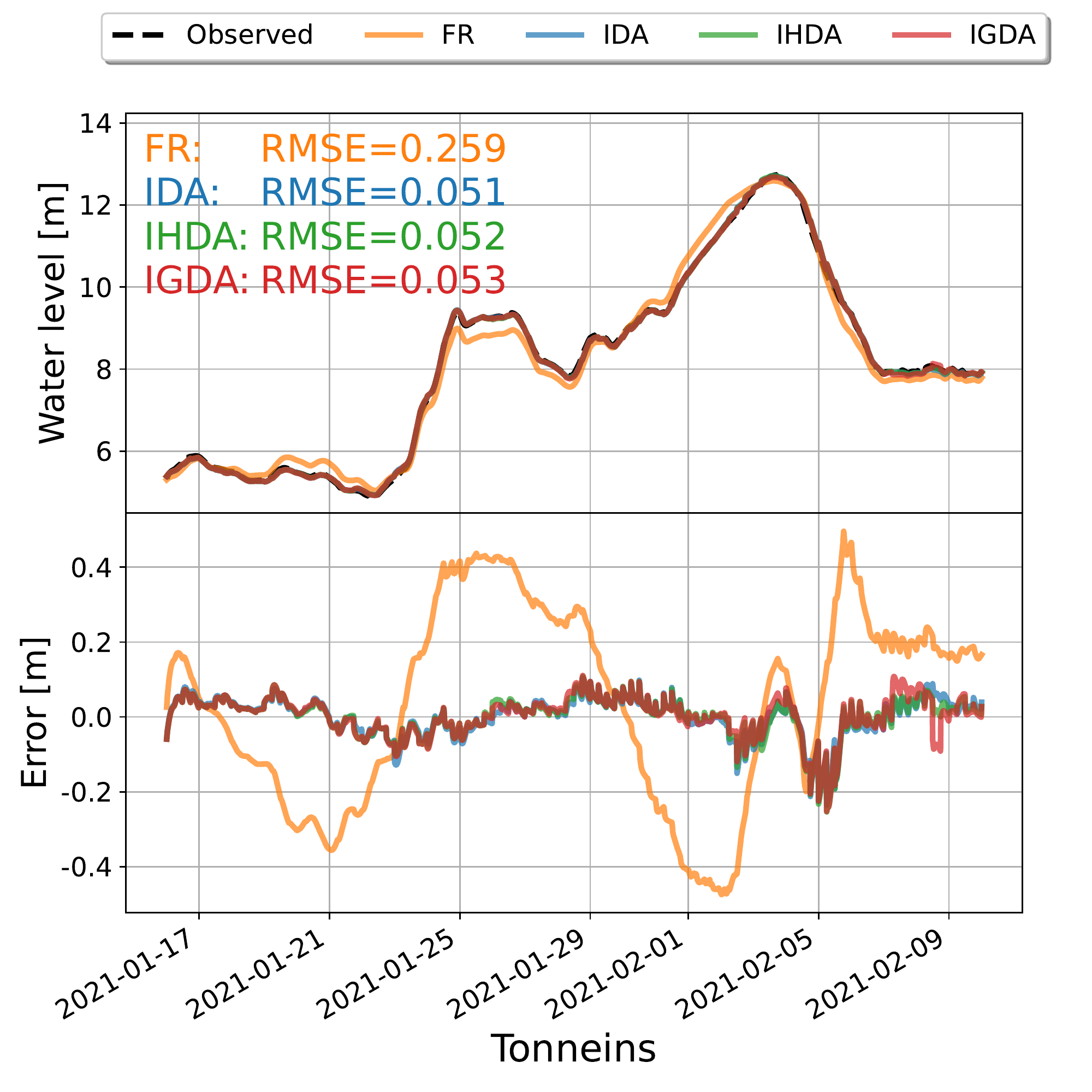}
    \hfill
    \includegraphics[width=0.32\linewidth]{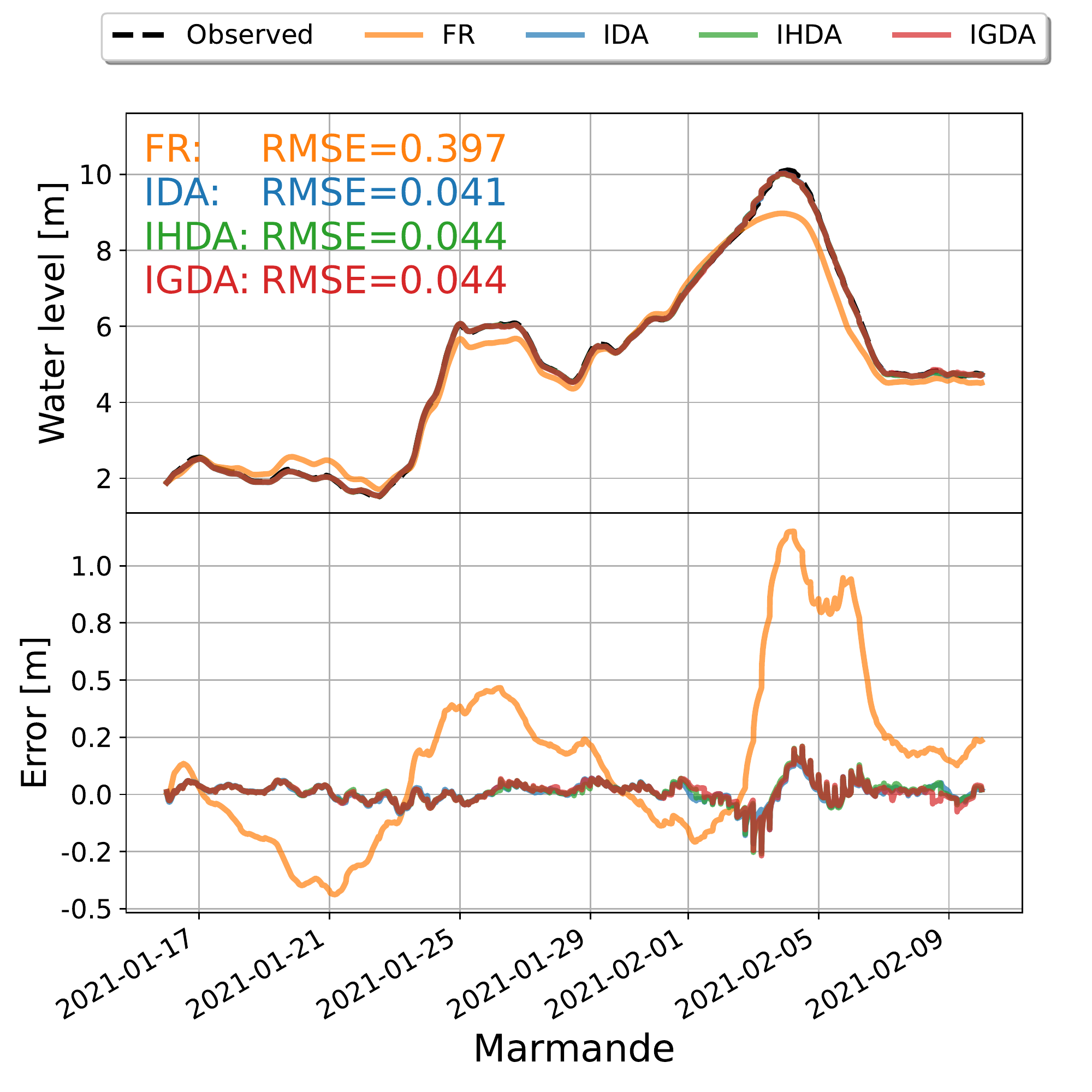}
    \hfill
    \includegraphics[width=0.32\linewidth]{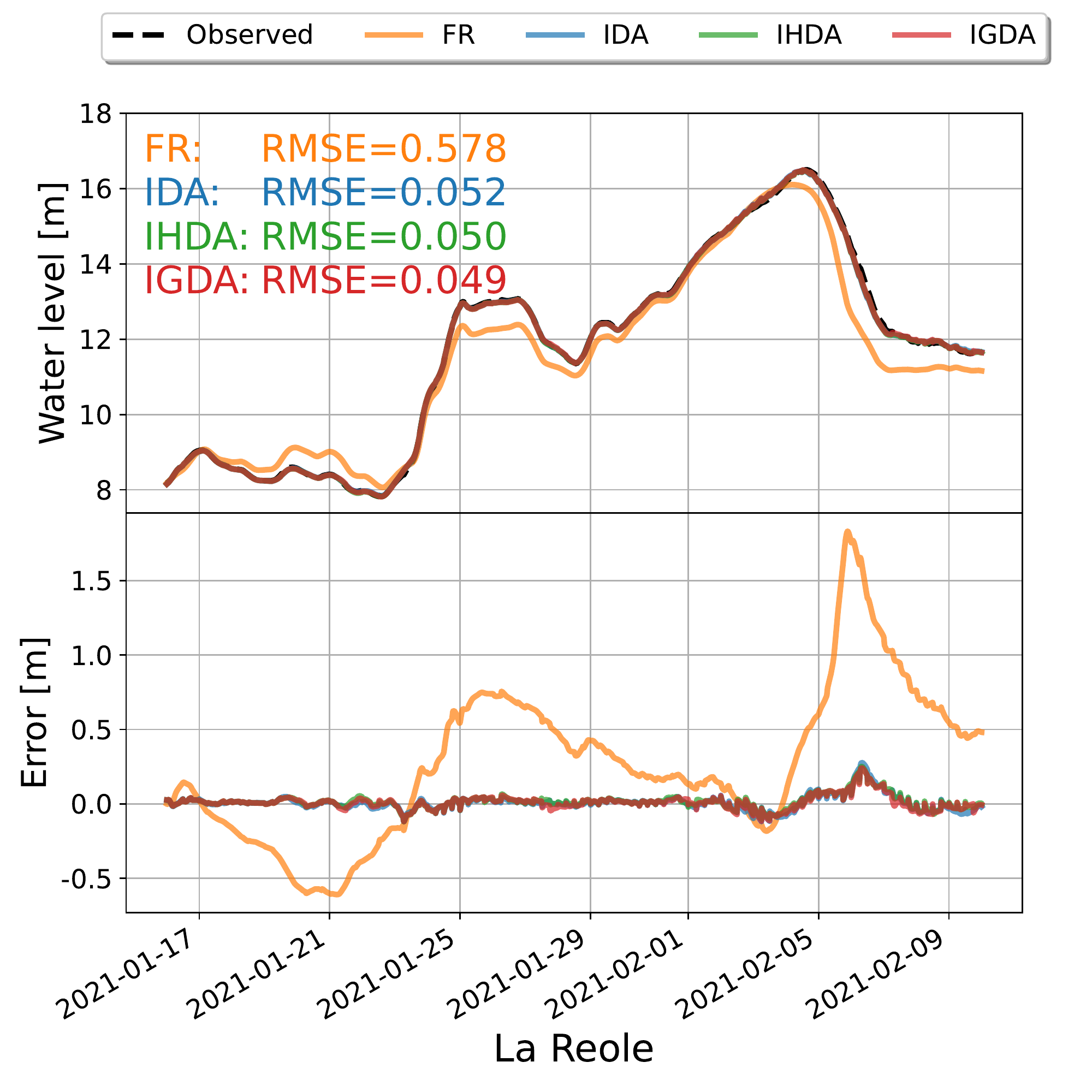}
    \caption{in OSSE}
    \label{fig:WL_OSSE}
    \end{subfigure}\\[10pt]
    
    \begin{subfigure}[b]{\textwidth}
    \centering
    \includegraphics[width=0.32\linewidth]{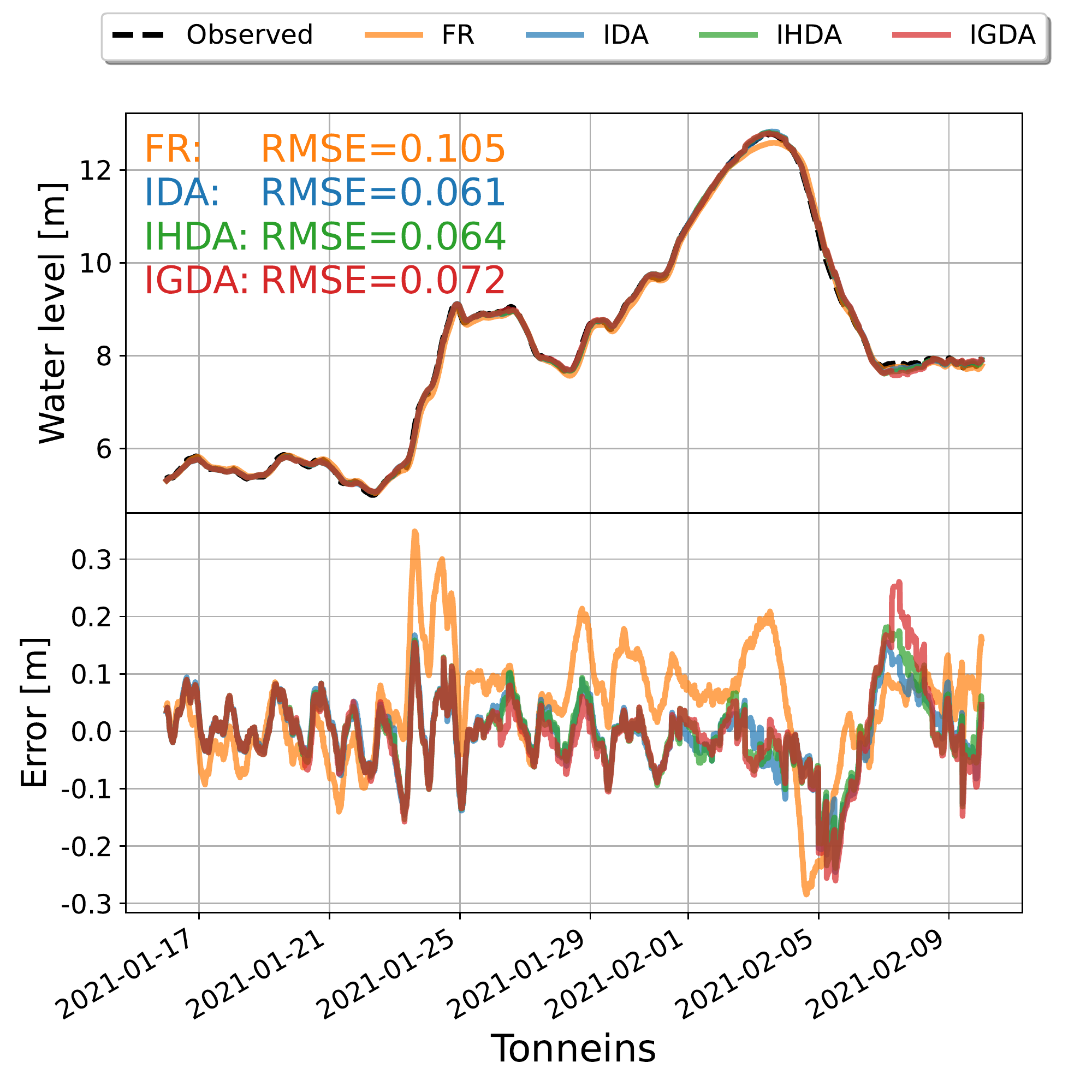}
    \hfill
    \includegraphics[width=0.32\linewidth]{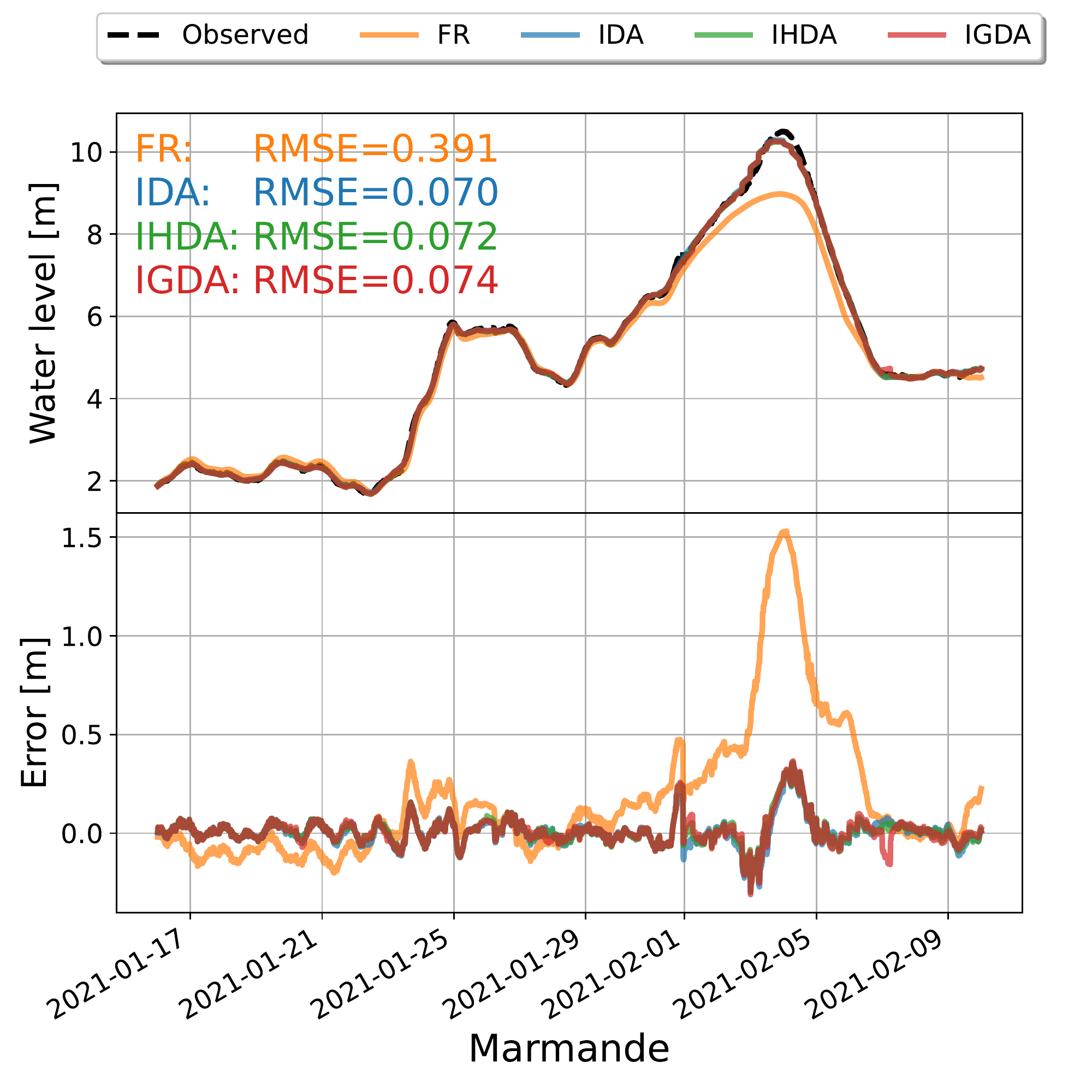}
    \hfill
    \includegraphics[width=0.32\linewidth]{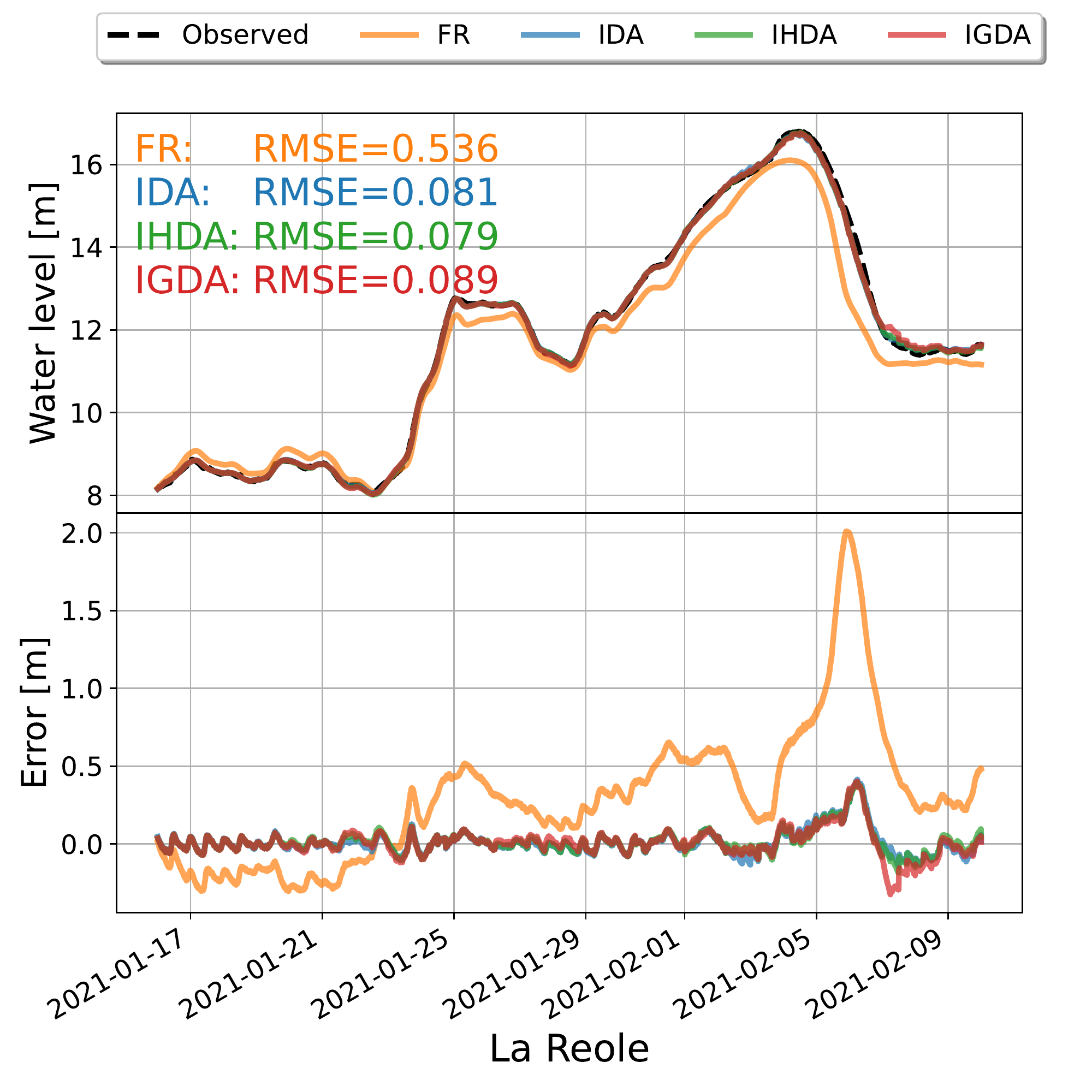}
    \caption{in real event}
    \label{fig:WL_real}
    \end{subfigure}
    \caption{Water level $H$ (upper plots) for all experiments (FR in orange, IDA in blue, IHDA in green and IGDA in red), and their respective error (lower plots) with respect to the (a) reference WL in OSSE, or (b) observed WL in real event. From left to right: results at three VigiCrue observing stations Tonneins (left), Marmande (middle), La Réole (right). Resulting WL RMSE computed over time is indicated with corresponding color, for each station}.
    % \label{}
\end{figure*}

%\paragraph{Results for OSSE experiment}

The comparison of WL resulting from the different FR and DA experiments over the entire simulation duration are shown in Figure~\ref{fig:WL_OSSE} for the OSSE, and in Figure~\ref{fig:WL_real} for the real flood event at the three observing stations, namely Tonneins (left panel), Marmande (middle panel) and La Réole (right panel). For each subfigure, the top panel represents the WLs---reference WL in black-dashed lines overlapped by simulated WLs in color solid lines---while the bottom panel represents the misfits between the observed and simulated WL by each experiment (FR in orange, IDA in blue, IHDA in green, and IGDA in red). The RMSE computed over the entire event, for the WL from FR simulation, as well as from IDA, IHDA and IGDA analyses, with respect to the reference WL (or real in-situ WL observations) at the three observing stations is indicated in respective subfigures.
% Table~\ref{tab:RMSEOSSE} for the OSSE and in Table~\ref{tab:RMSE_real} for the real event, with respect to the real in-situ WL observations. For each observing station, the lowest RMSE values are boldfaced.
%Table~\ref{tab:RMSEOSSE}
It can be noted that all DA experiments succeed in significantly reducing the WL errors compared to those of FR.
For all three observing stations, the reduction in RMSE with respect to FR reaches 80-90\% with slightly different values between IDA, IHDA, and IGDA. This validates the performance of the DA strategy. Most importantly, it demonstrates that the assimilation of in-situ WL in the river bed suffices to constraint the simulated hydraulic state close to the WL measurements at these gauge stations. The merits of assimilating WSR observation is \sophie{noticeable}, on the other hand, when assessing the dynamics of the floodplain. Similar conclusions are drawn from WL timeseries and RMSE assessment for the 2021 real event presented in Figure~\ref{fig:WL_real} where RMSEs are computed with respect to gauge stations measurements. As expected, for real event, the DA experiments yield a lesser improvement compared to the OSSE, yet it remains very significant. For instance, the WL RMSE at Marmande is reduced from 39.1 cm (FR) to 7.4 cm (IGDA).

\subsubsection{WSR in the floodplain}

\paragraph{Results for OSSE experiment}

\begin{figure*}[h]
    \centering
    \begin{subfigure}[b]{0.49\textwidth}
    \centering
    % \includegraphics[width=0.49\linewidth]{fig/plot_H5cm_OSSE_w_IGDA.pdf}
    % \hfill
    \includegraphics[width=\linewidth]{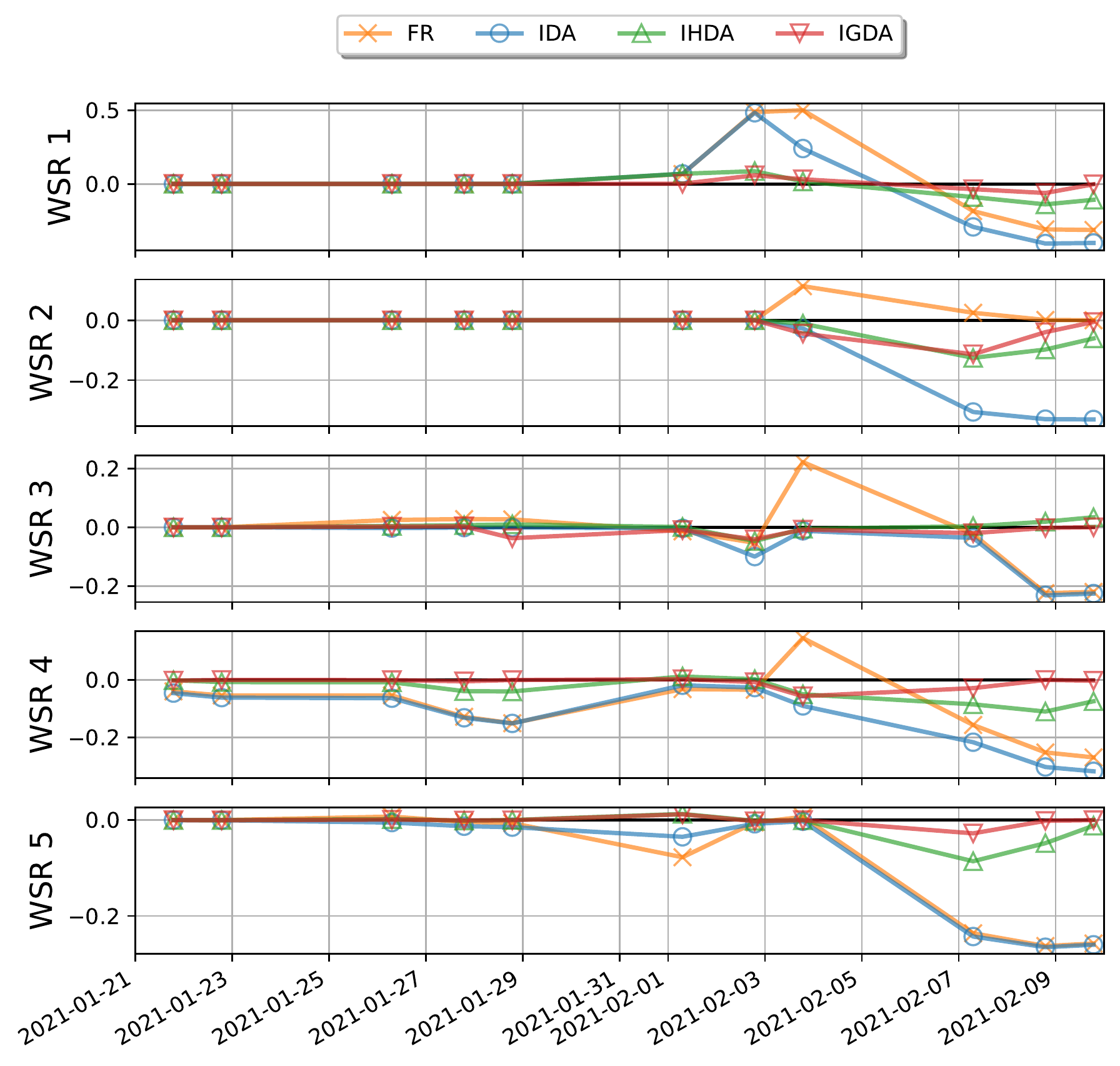}
    \caption{in OSSE}
    \label{fig:WSR_OSSE}
    \end{subfigure}
% \end{figure*}%
\hfill
% \begin{figure*}[t]\ContinuedFloat   
%     \captionsetup{list=off,format=cont}
    \begin{subfigure}[b]{0.49\textwidth}
    \centering
    % \includegraphics[width=0.49\linewidth]{fig/plot_H5cm_2021_w_IGDA.pdf}
    % \hfill
    \includegraphics[width=\linewidth]{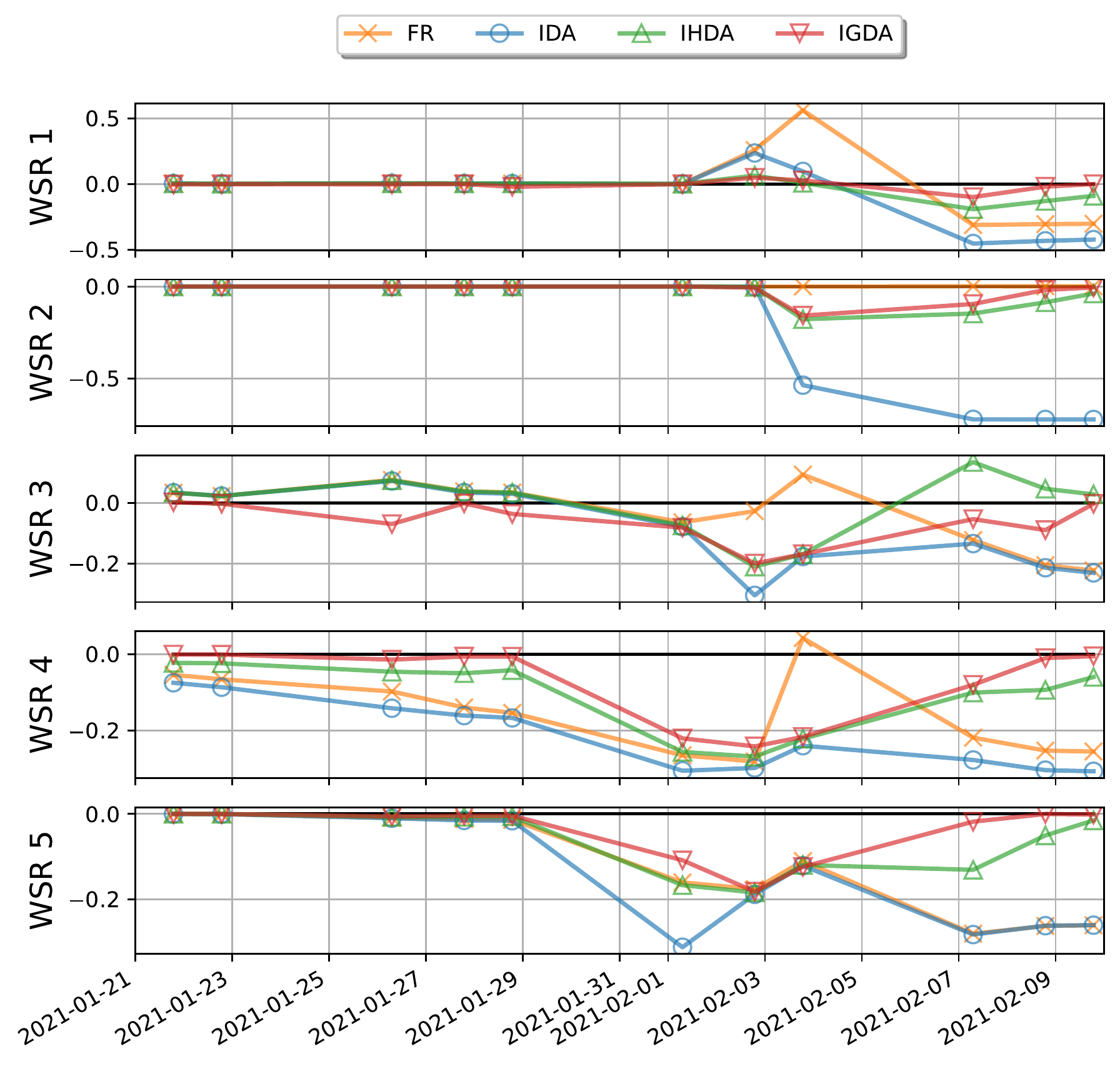}
    \caption{in real event}
    \label{fig:WSR_real}
    \end{subfigure}
    \caption{Misfits of the truth/observed WSR minus the simulated WSR values computed for FR (orange), IDA (blue), IHDA (green), and IGDA (red) over the five subdomains of the floodplain.}
\end{figure*}

Figure~\ref{fig:WSR_OSSE} displays the misfit between the synthetical WSR values computed from the reference simulation (black line) and the WSR computed for FR and DA experiments over the five subdomains of the floodplain (same color code as in Figure~\ref{fig:control_OSSE}). Between the beginning of the event and the flood rising limb (around 2021-02-01), the assimilation of WSR has virtually no impact as the water has not yet overflowed the floodplain. 
The WSR values in the reference and the experiments are thus null or close to zero.
Globally speaking, FR tends to underestimate the flooding for the OSSE experiment. The assimilation of in-situ WL data (IDA) allows for a significant improvement of WSR near the flood peak but the flood recess suffers from T2D incapacity to \sophie{effectively} empty the floodplains. The impact of assimilating WSR observations is only visible when the flood starts around 2021-02-01 (when reference WSR values are above zero). Analyzed WSR values are further improved when WSR observations are assimilated at S1 overpass times in the two other DA experiments. Indeed, when WSR observations are assimilated and the correction for the WL is included in the control vector (experiments IHDA and IGDA), the analysis succeeds in retrieving WSR values that are close to the truth at the flood peak and \sophie{in} emptying the flood plain after the peak has passed. The results for IHDA and IGDA are quite similar---although IGDA performs slightly better, especially at the flood peak and during the recess period---arguing that when the Gaussian hypothesis is violated, the EnKF analysis may be suboptimal but remains satisfactory in terms of WSR. 

%The WSR in the five floodplain subdomains are obtained using the simulated WL in FR experiment, and using the analyzed WL in the three DA experiments with a threshold of 5 cm for the wet/dry definition. They are compared to the WSR computed from the reference simulation and shown in Figure~\ref{fig:WSR_OSSE}. The WSR values are shown in Figure~\ref{fig:WSR_OSSE} (left panel) and the misfit between the reference and the simulated WSR values are shown in Figure~\ref{fig:WSR_OSSE} (right panel).
%The WSR for the truth are plotted in black, the WSR for FR are in orange. The color code for the DA experiments is the same as in Figure~\ref{fig:control_OSSE}: IDA in blue, IHDA in green and IGDA in red. 
%From the beginning of the event up to the flood rising limb (around 2021-02-01), the impact of assimilating WSR is insignificant as the water has not yet overflowed to the floodplain. The WSR values in the reference and the experiment are thus null or close to zero.

\paragraph{Results for real experiment}

Figure~\ref{fig:WSR_real} displays the misfit between the S1-derived WSR at S1 overpass times (black line) and the WSR computed for FR and DA experiments over the five subdomains of the floodplain. For the 2021 event, FR tends to underestimate the extension of the flooding whereas all the DA experiments tend to overflood, especially during \sophie{the} flood recess period. It should be noted that the performance of IDA, which does not benefit from WSR observations (i.e. not assimilated), is not improved compared to that of FR. Indeed, in contrast to FR, IDA mostly lead to overflooding throughout the event. On the other hand, IHDA and IGDA experiments provide improvements, shown by the misfits of WSR for these DA experiments being smaller than those of FR, especially during the flood recess over all subdomains of the floodplain. IGDA brings further noticeable improvement to IHDA when WSR data are the most informative, at flood peak and during the flood recess.

A considerable overprediction in the subdomain 4 and 5 at the first S1 overpass time during the rising limb (2021-02-01 07:00) \sophie{should} be noted for all experiments, in Figure~\ref{fig:WSR_real}.
At this particular moment (7$^{th}$ vertical line in Figure~\ref{fig:Htvigicrue2021}) when the WLs in the riverbed are relatively high, the SAR BS tends to intensify due to the increased soil moisture in the floodplain due to rainfalls and the soil being dampened but not yet flooded. The BS will however decrease later on when the floodplain is flooded. Such a behavior leads to the resulting SAR-derived flood extent map at this date yielding fewer detected wet pixels than expected. This situation will be investigated in future works.

\begin{figure*}[h]
\centering
\begin{subfigure}[t]{\textwidth}
\centering
\begin{tabular}{l@{\hskip 1mm}c@{\hskip 1mm}c@{\hskip 1mm}c@{\hskip 1mm}c@{\hskip 1mm}c@{\hskip 1mm}c}
% && FR & IDA & IHDA & IGDA&\\
&& Truth & FR & IDA$_{osse}$ & IHDA$_{osse}$ & IGDA$_{osse}$\\\hline
% \rotatebox{90}{2021-02-02 19:00}& \rotatebox{90}{before peak} &
% \includegraphics[width=0.23\linewidth]{fig/FR_osse_1623600.png}&
% \includegraphics[width=0.23\linewidth]{fig/IDA_osse_1623600.png}&
% \includegraphics[width=0.23\linewidth]{fig/IHDA_osse_1623600.png}&
% \includegraphics[width=0.23\linewidth]{fig/IGDA_osse_1623600.png}& \rotatebox{90}{S1}\\
% && $\mathrm{CSI}=71.07\%$ & $\mathrm{CSI}=71.14\%$ & $\mathrm{CSI}=93.63\%$ & $\mathrm{CSI}=94.87\%$&\\\hline
\rotatebox{90}{2021-02-03 19:00}& \rotatebox{90}{flood peak} & \includegraphics[width=0.18\linewidth]{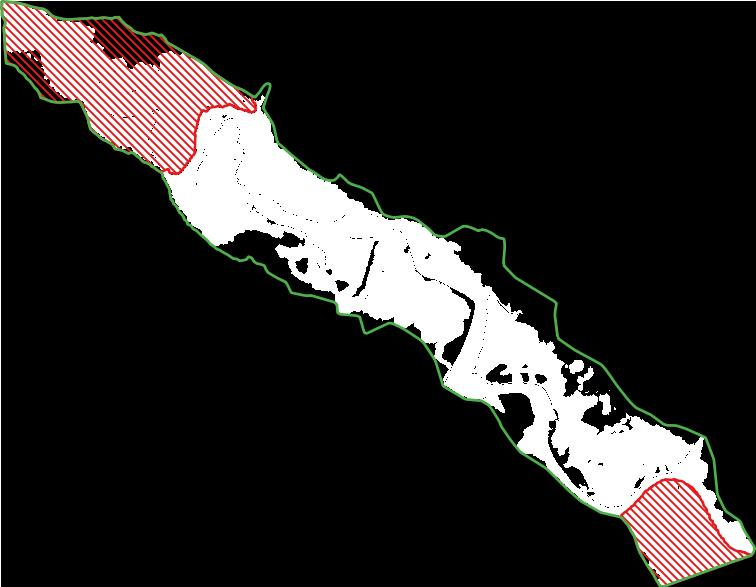} &
\includegraphics[width=0.18\linewidth]{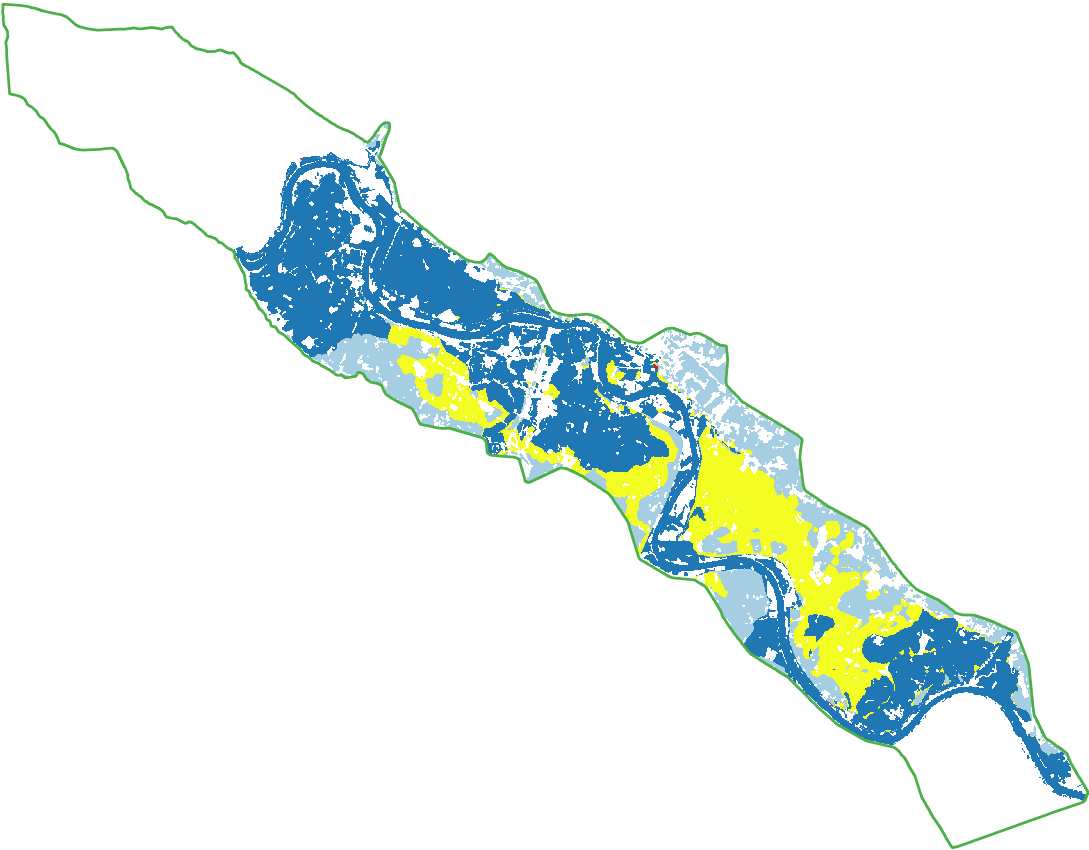}&
\includegraphics[width=0.18\linewidth]{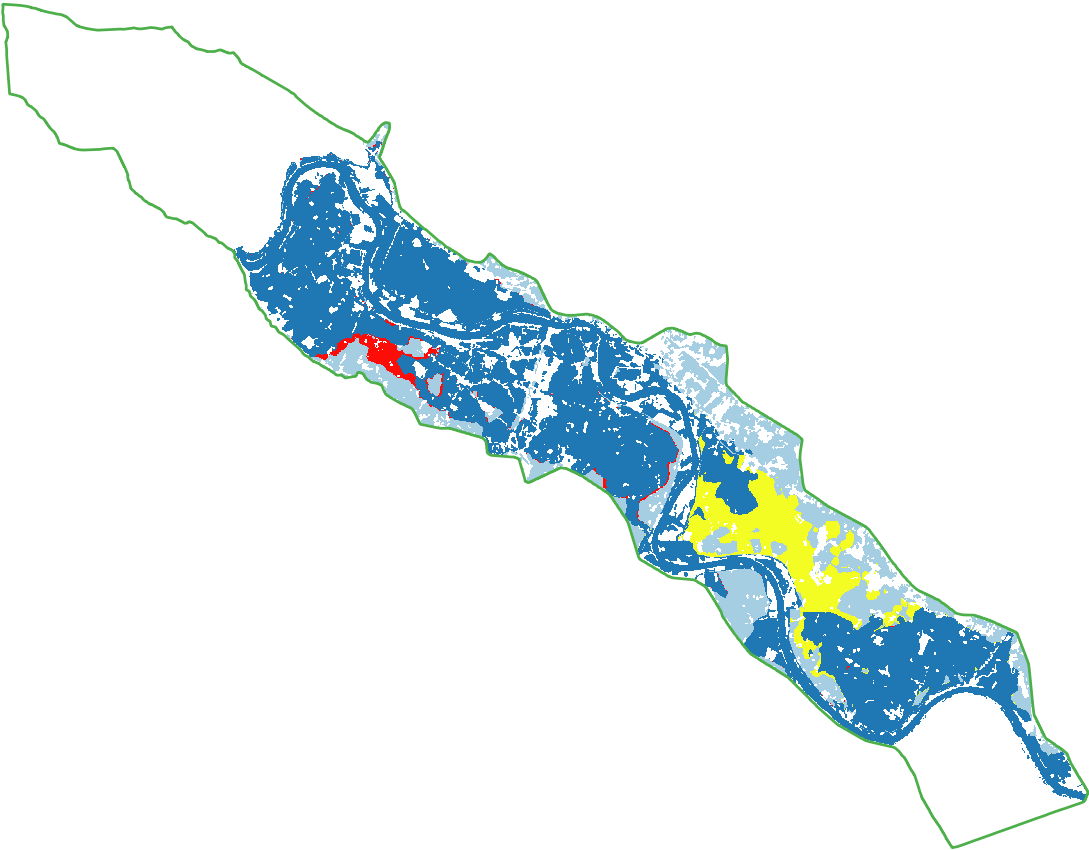}&
\includegraphics[width=0.18\linewidth]{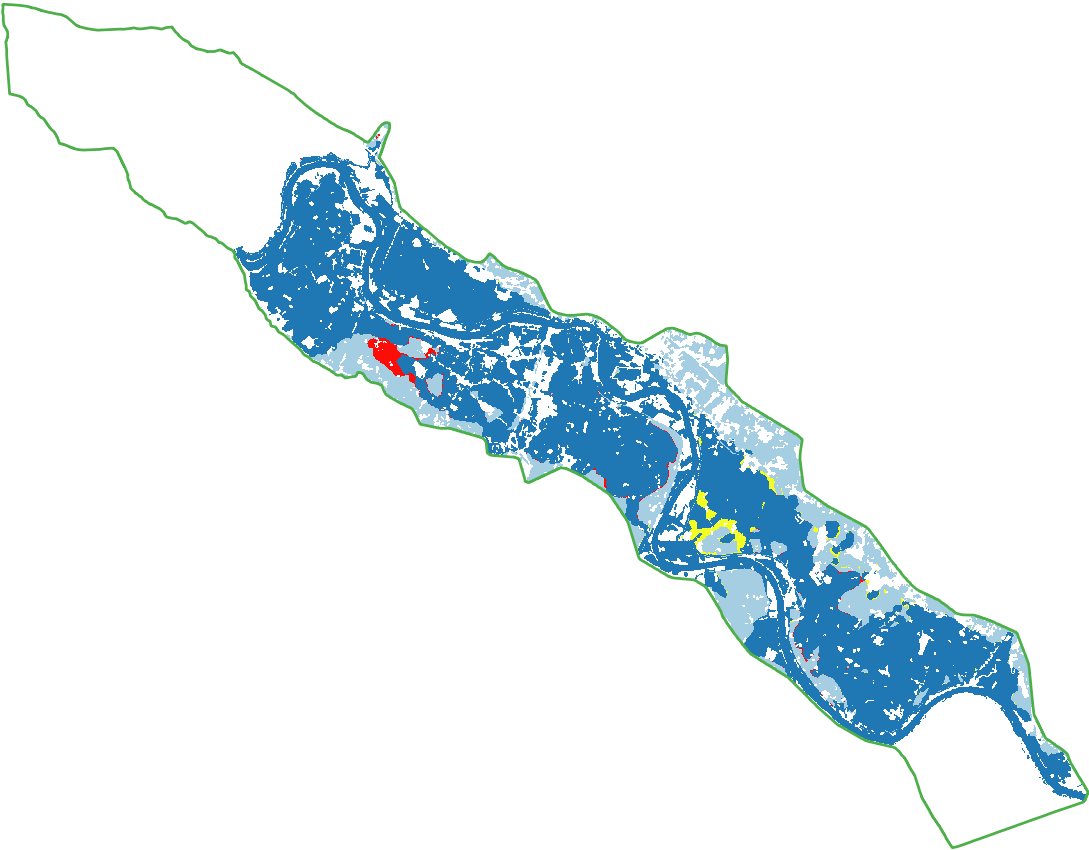}&
\includegraphics[width=0.18\linewidth]{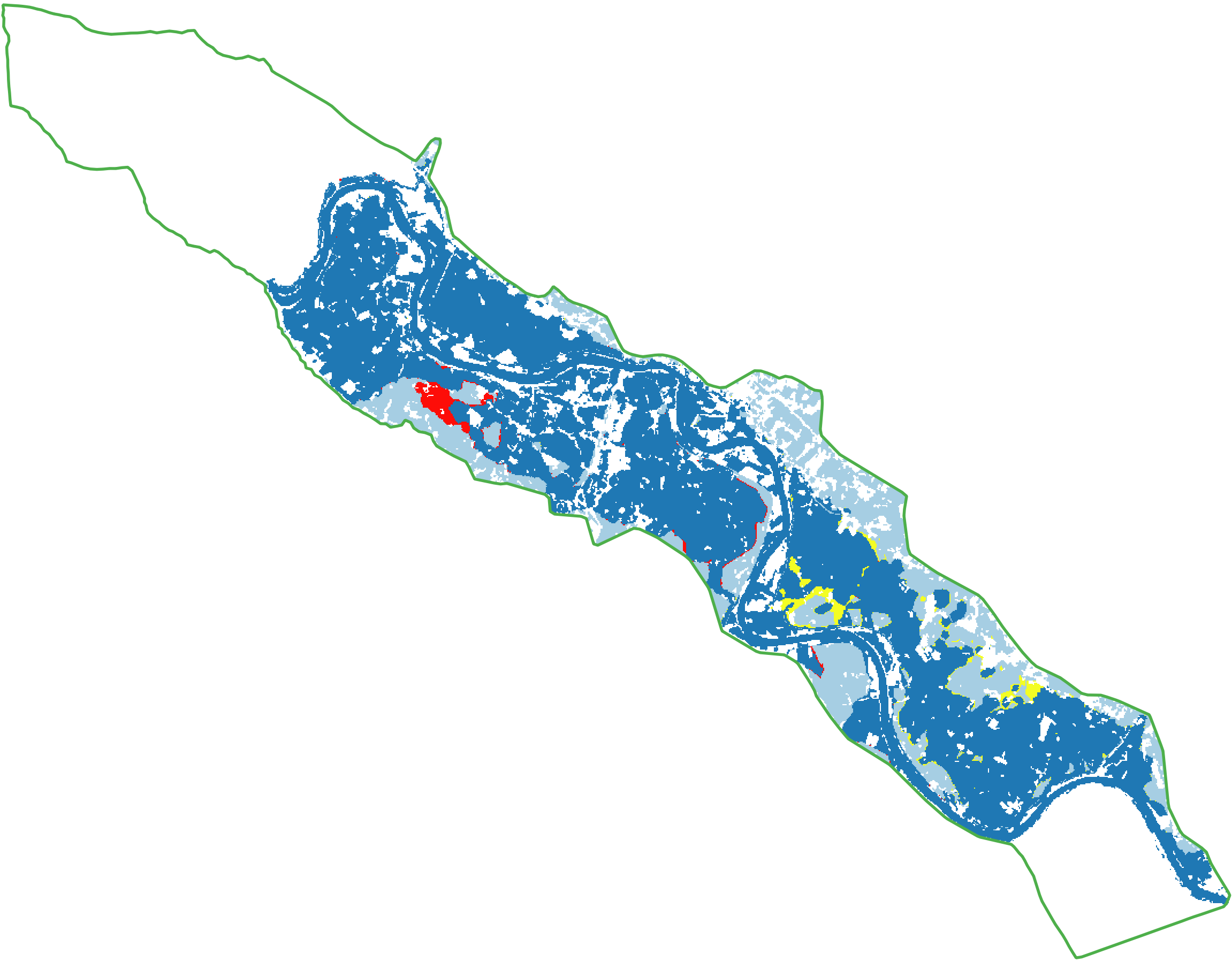}\\
&&& $\mathrm{CSI}=71.41\%$ & $\mathrm{CSI}=86.70\%$ & $\mathrm{CSI}=97.26\%$ & $\mathrm{CSI}=96.72\%$\\\hline
\rotatebox{90}{2021-02-07 07:00}& \rotatebox{90}{falling limb} &
\includegraphics[width=0.18\linewidth]{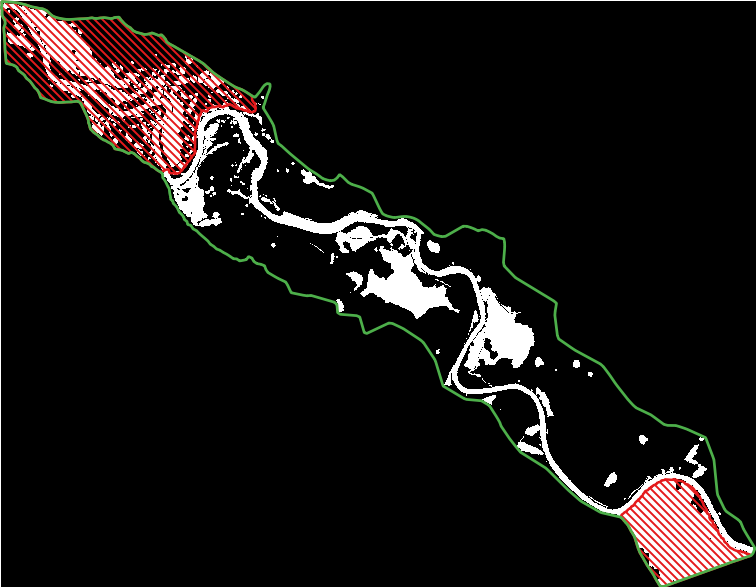}& \includegraphics[width=0.18\linewidth]{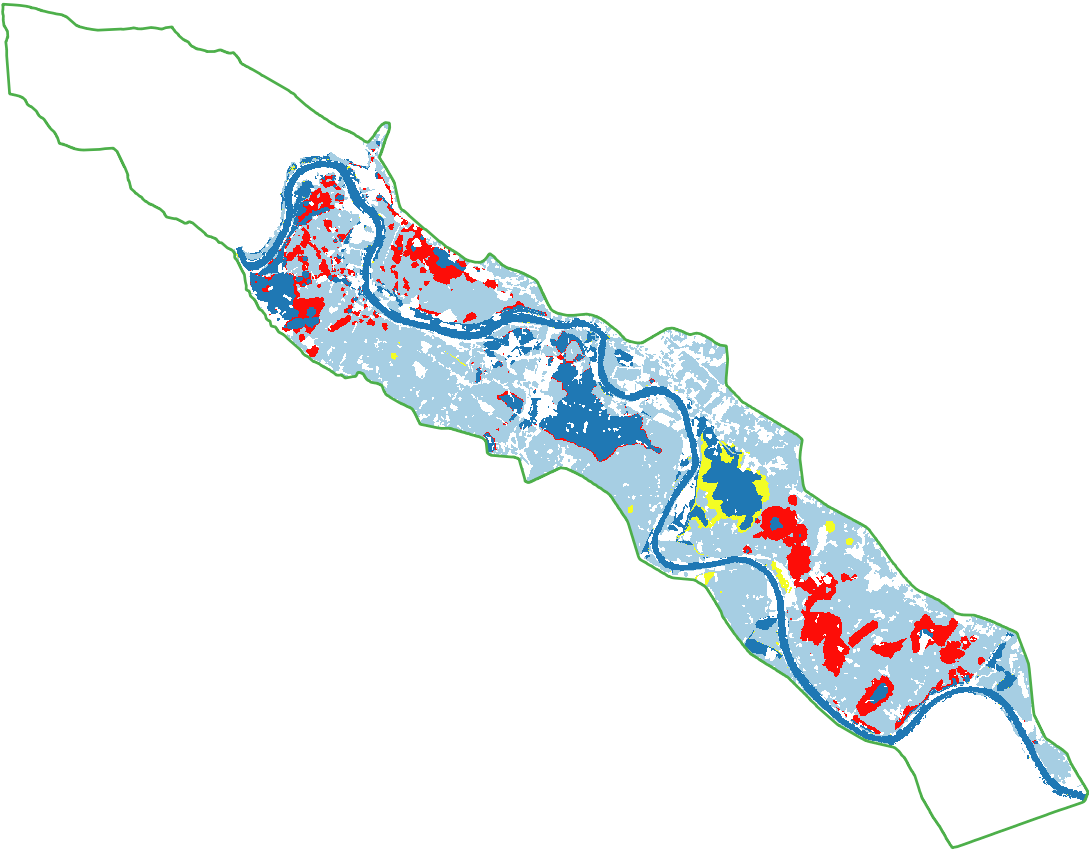}&
\includegraphics[width=0.18\linewidth]{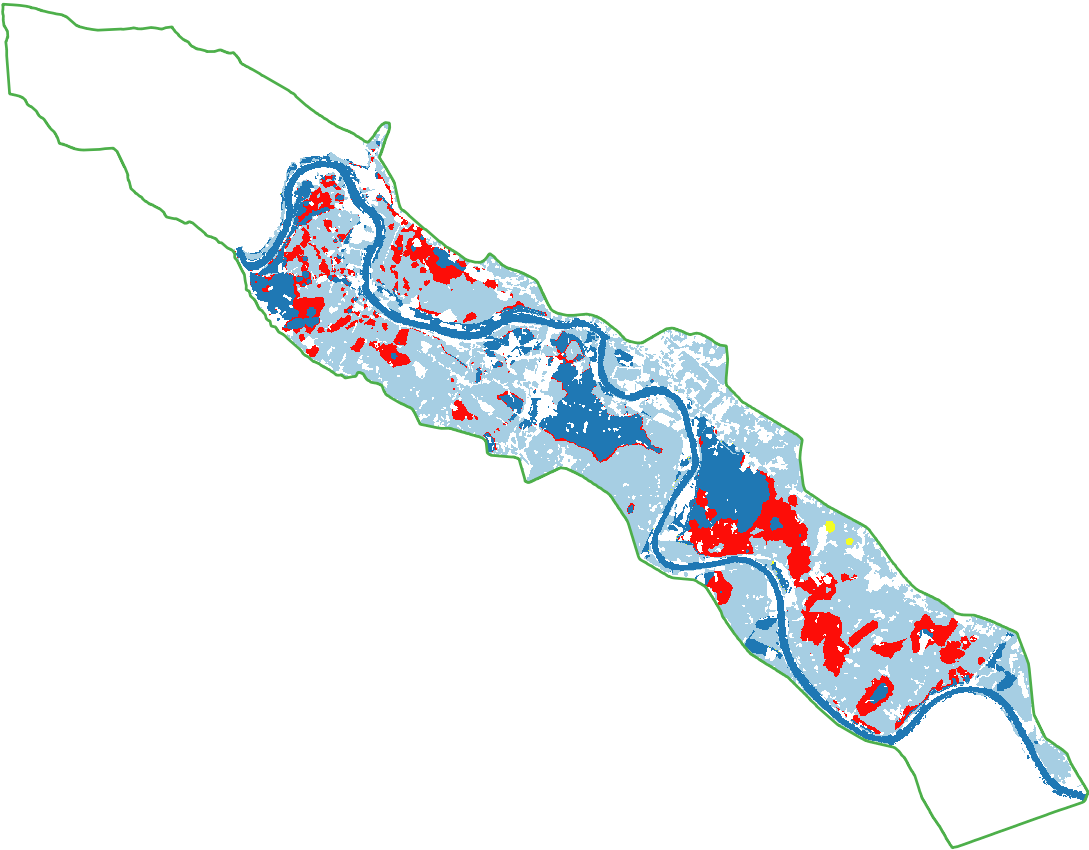}&
\includegraphics[width=0.18\linewidth]{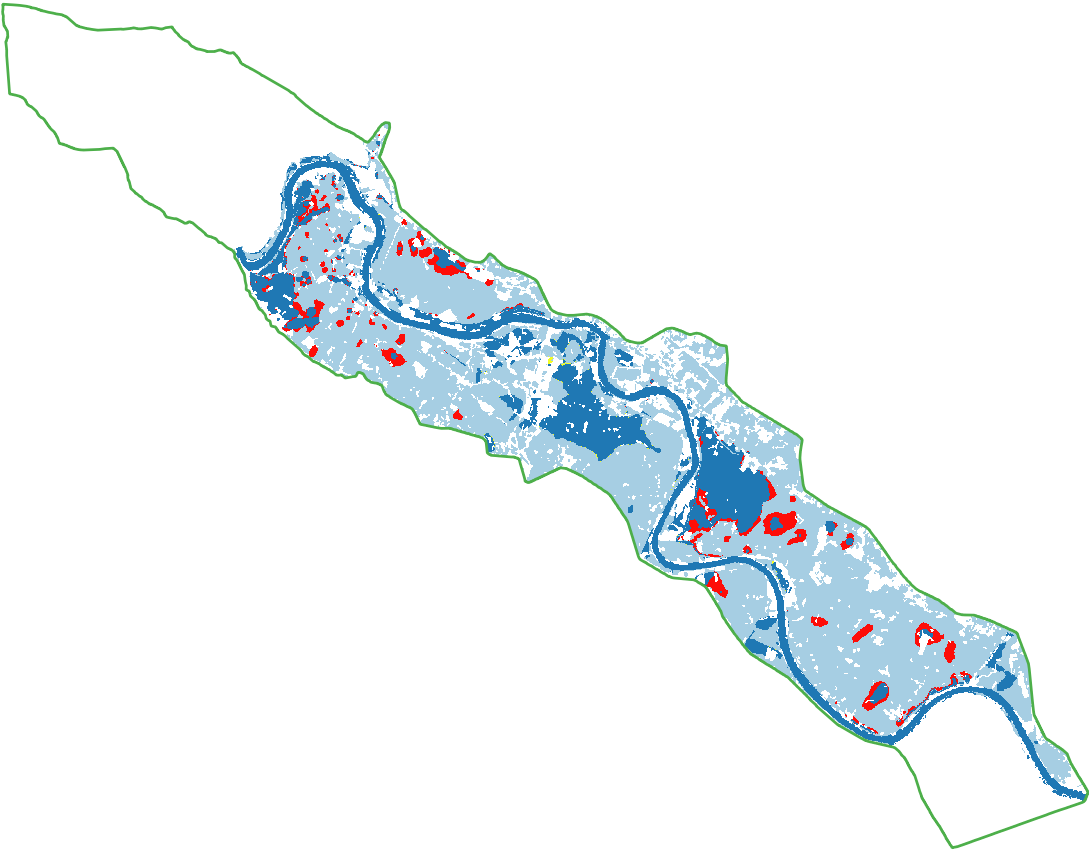}&
\includegraphics[width=0.18\linewidth]{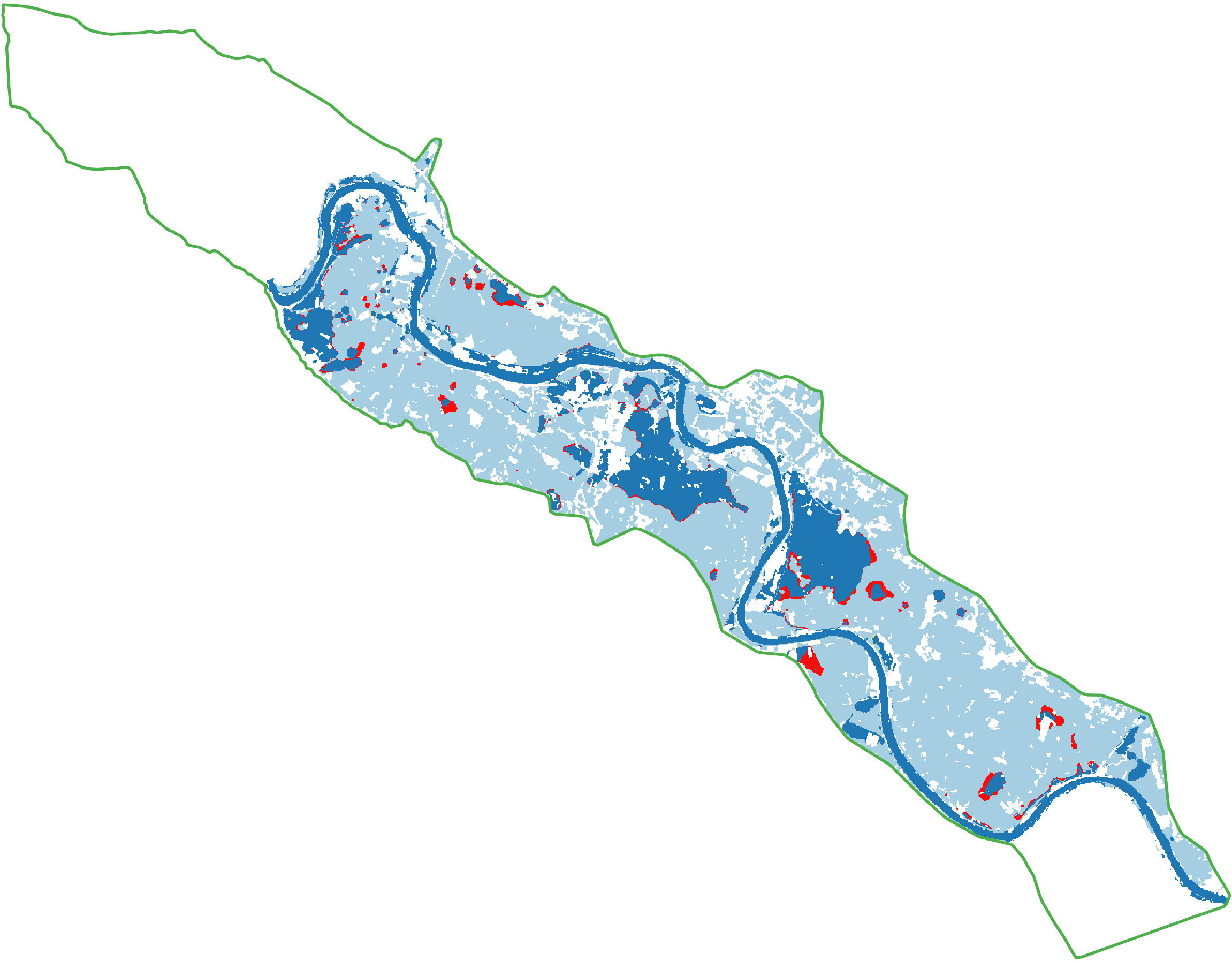}\\
&&& $\mathrm{CSI}=63.64\%$ & $\mathrm{CSI}=62.47\%$ & $\mathrm{CSI}=83.76\%$ & $\mathrm{CSI}=91.89\%$\\ \hline
% && \multicolumn{4}{c}{\includegraphics[trim=0 15cm 17.5cm 0, clip,width=0.35\linewidth]{fig/legend.pdf}}\\
\end{tabular}
\caption{in OSSE}
\label{fig:conti_OSSE}
\end{subfigure}%\\[10pt]
\caption{Contingency maps for FR, IDA, IHDA and IGDA with respect to (a) synthetical flood extent maps in OSSE, and (b) S1-derived flood extents in real event.
%between simulated flood extents (from left to right: FR, IDA, IHDA and IGDA) with respect to (a) synthetical flood extent maps in OSSE, and (b) S1-derived flood extents in real event.
}
\end{figure*}%

\begin{figure*}[t]\ContinuedFloat
\captionsetup{list=off,format=cont}
\centering
\begin{subfigure}[t]{\textwidth}
\centering
\begin{tabular}{l@{\hskip 1mm}l@{\hskip 1mm}c@{\hskip 1mm}c@{\hskip 1mm}c@{\hskip 1mm}c@{\hskip 1mm}c}
&& Observed & FR & IDA & IHDA & IGDA\\\hline
% \rotatebox{90}{2021-02-02 19:00}& \rotatebox{90}{before peak} &
% \includegraphics[width=0.23\linewidth]{fig/FR_osse_1623600.png}&
% \includegraphics[width=0.23\linewidth]{fig/IDA_osse_1623600.png}&
% \includegraphics[width=0.23\linewidth]{fig/IHDA_osse_1623600.png}&
% \includegraphics[width=0.23\linewidth]{fig/IGDA_osse_1623600.png}& \rotatebox{90}{S1}\\
% && $\mathrm{CSI}=42.45\%$ & $\mathrm{CSI}=43.40\%$ & $\mathrm{CSI}=44.55\%$ & $\mathrm{CSI}=44.82\%$&\\\hline
\rotatebox{90}{2021-02-03 19:00}& \rotatebox{90}{flood peak} &
\includegraphics[width=0.18\linewidth]{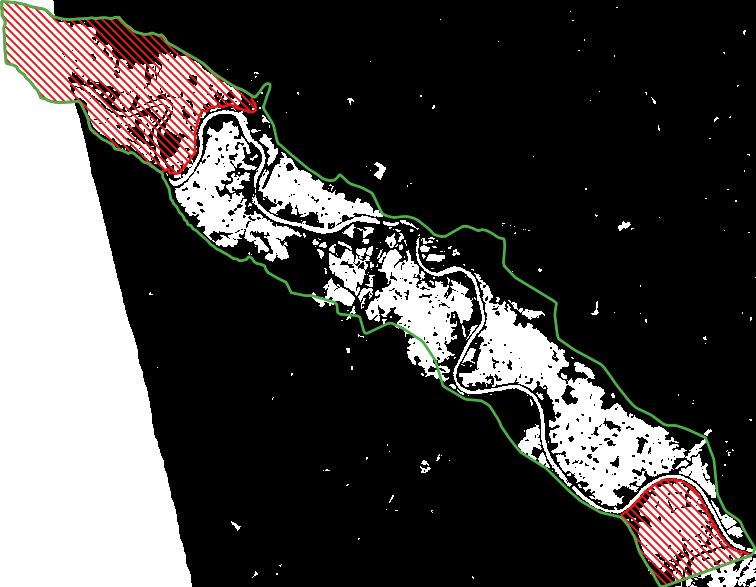}&\includegraphics[width=0.18\linewidth]{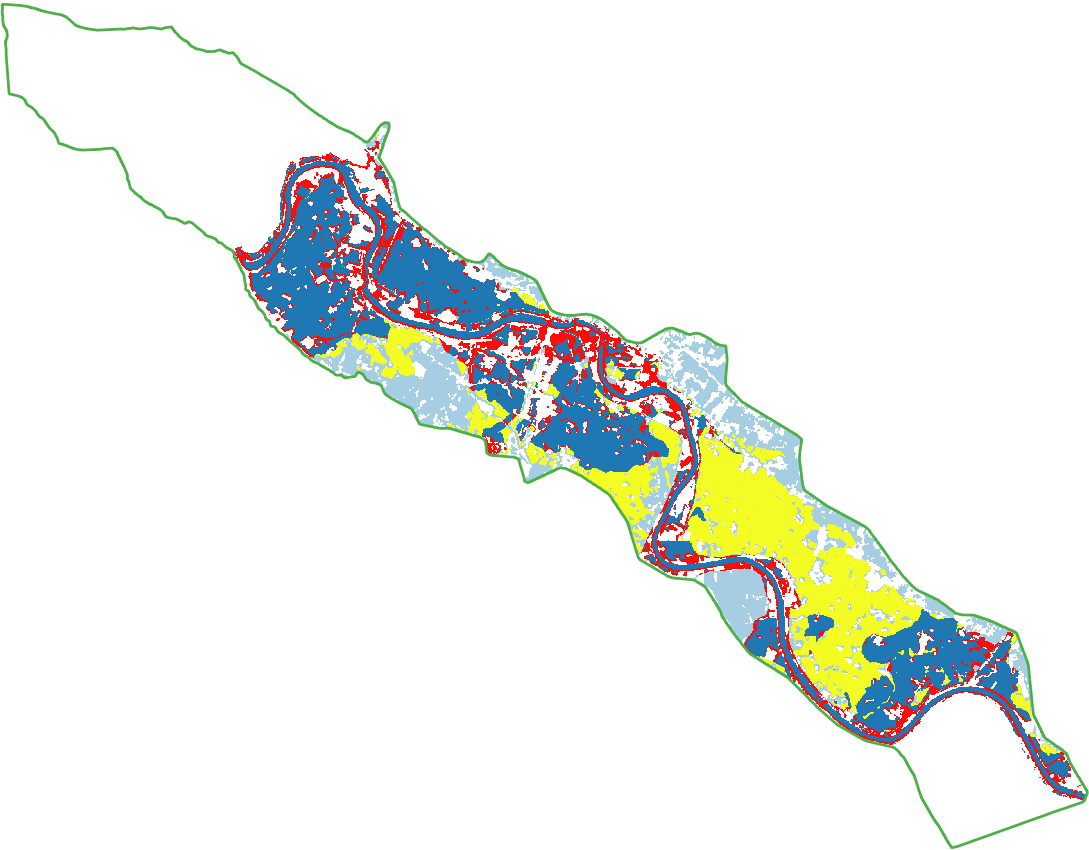}&
\includegraphics[width=0.18\linewidth]{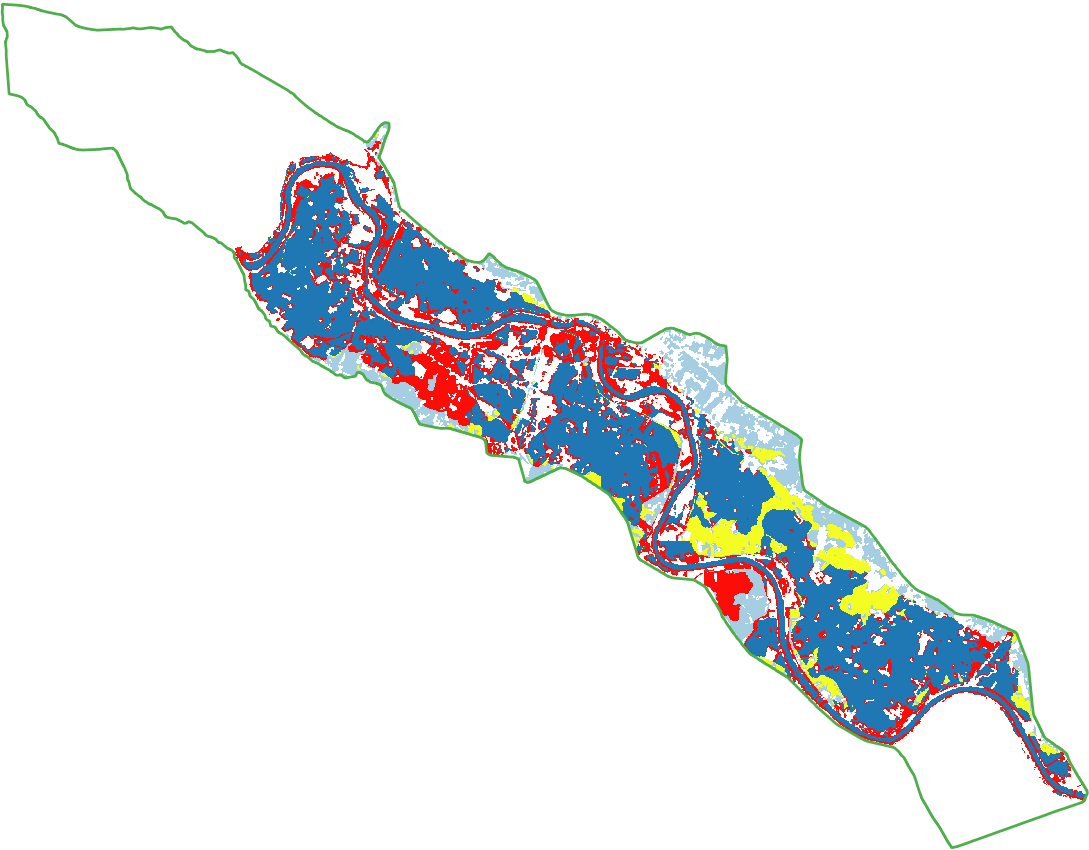}&
\includegraphics[width=0.18\linewidth]{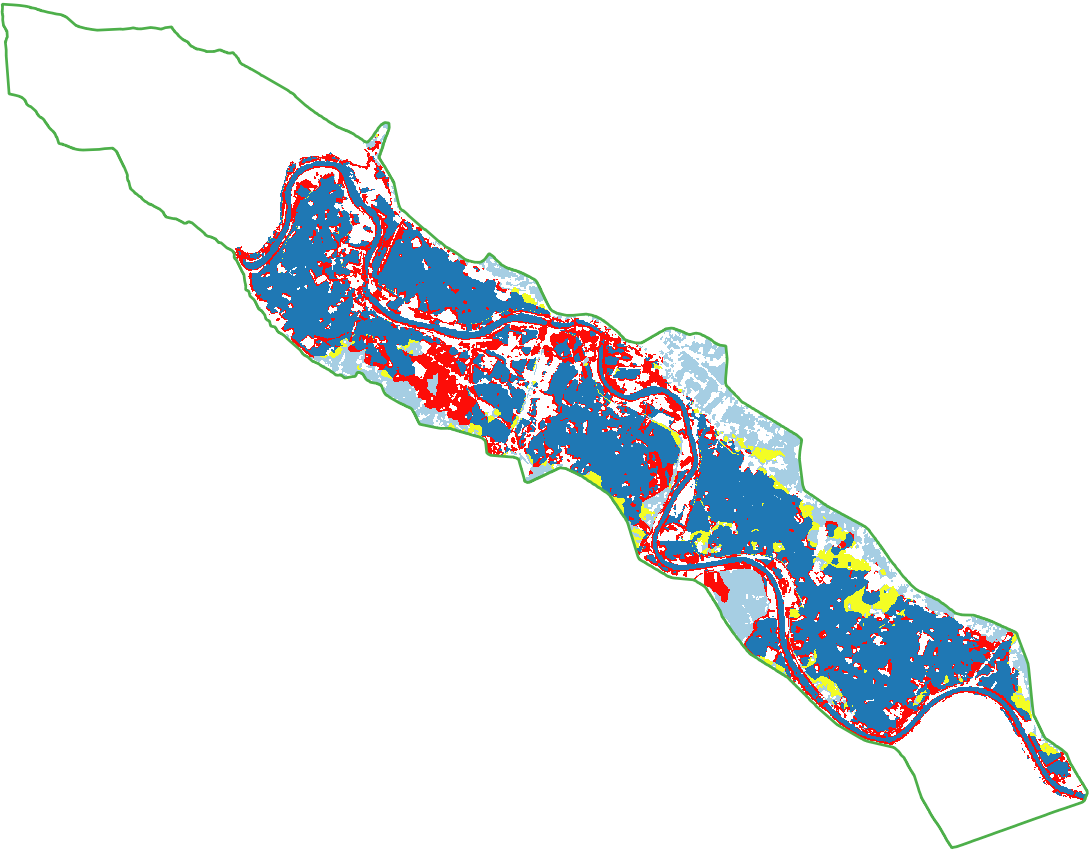}&
\includegraphics[width=0.18\linewidth]{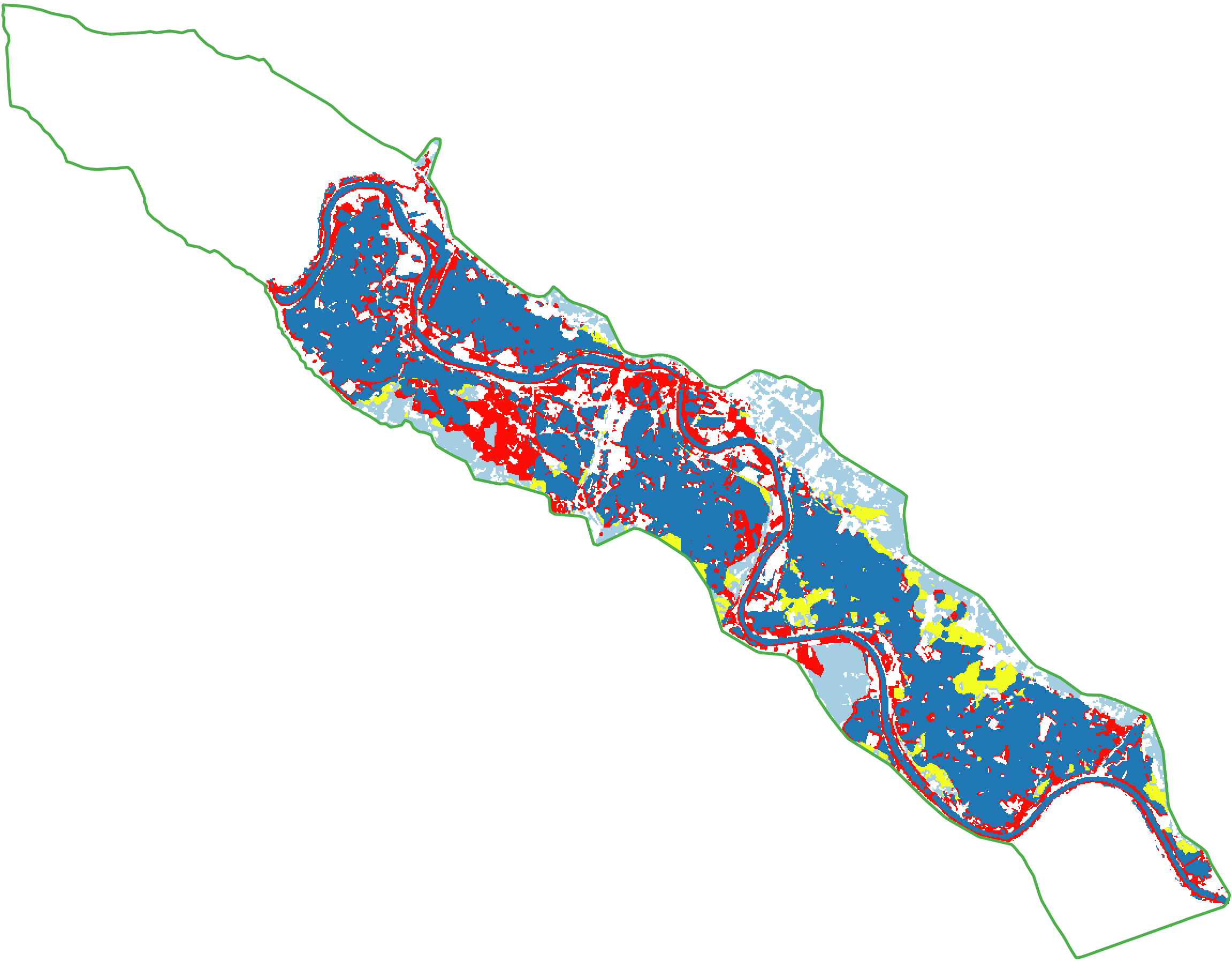}\\
&&& $\mathrm{CSI}=50.38\%$ & $\mathrm{CSI}=66.02\%$ & $\mathrm{CSI}=69.71\%$ & $\mathrm{CSI}=68.96\%$\\ \hline
\rotatebox{90}{2021-02-07 07:00}& \rotatebox{90}{falling limb} &
\includegraphics[width=0.18\linewidth]{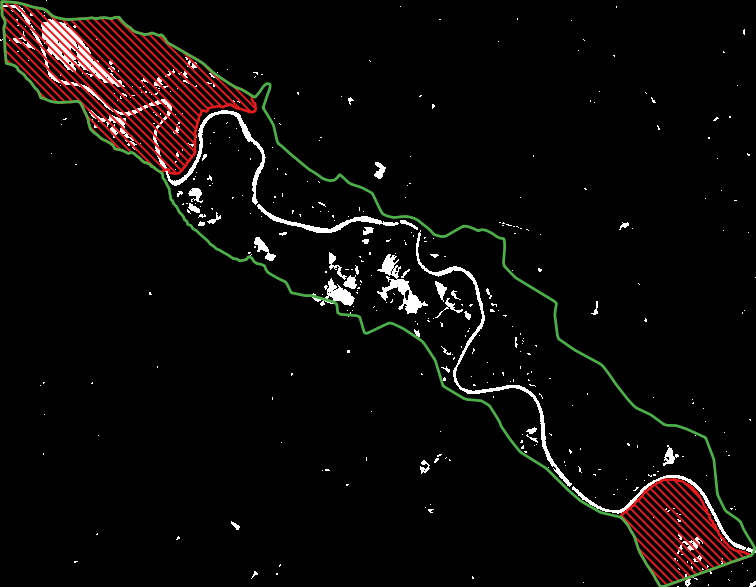} &
\includegraphics[width=0.18\linewidth]{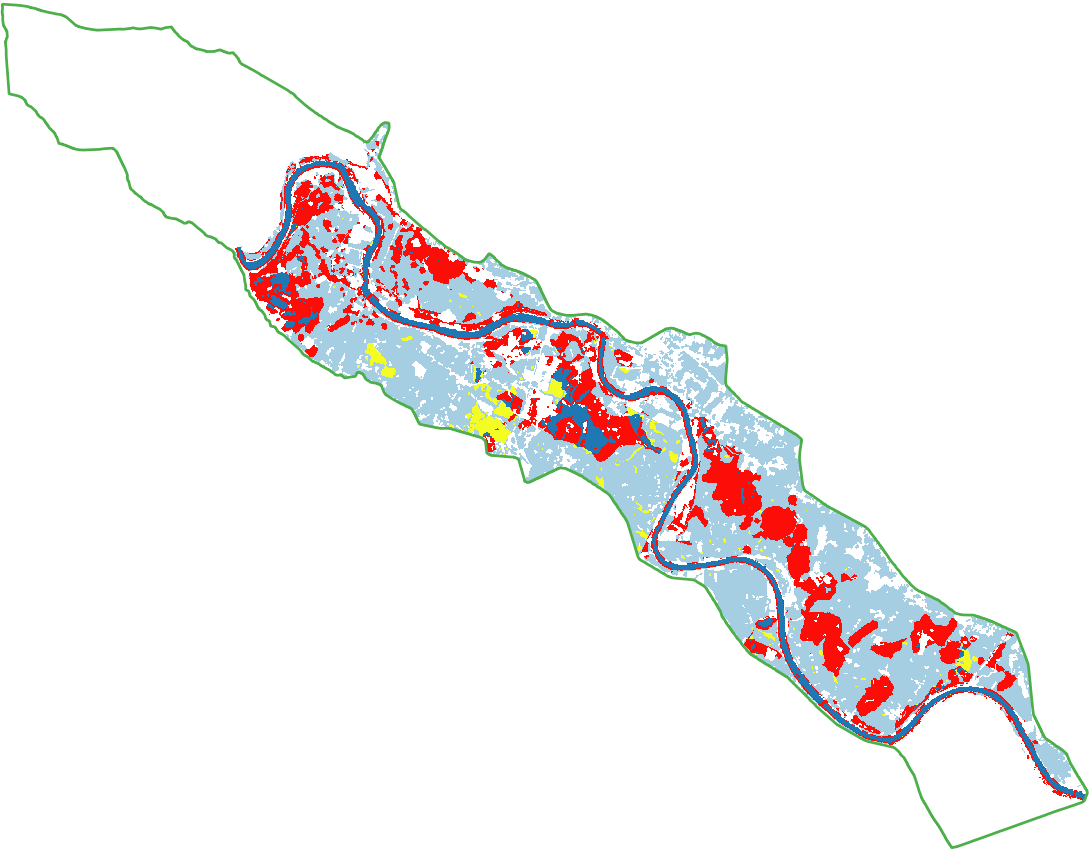}&
\includegraphics[width=0.18\linewidth]{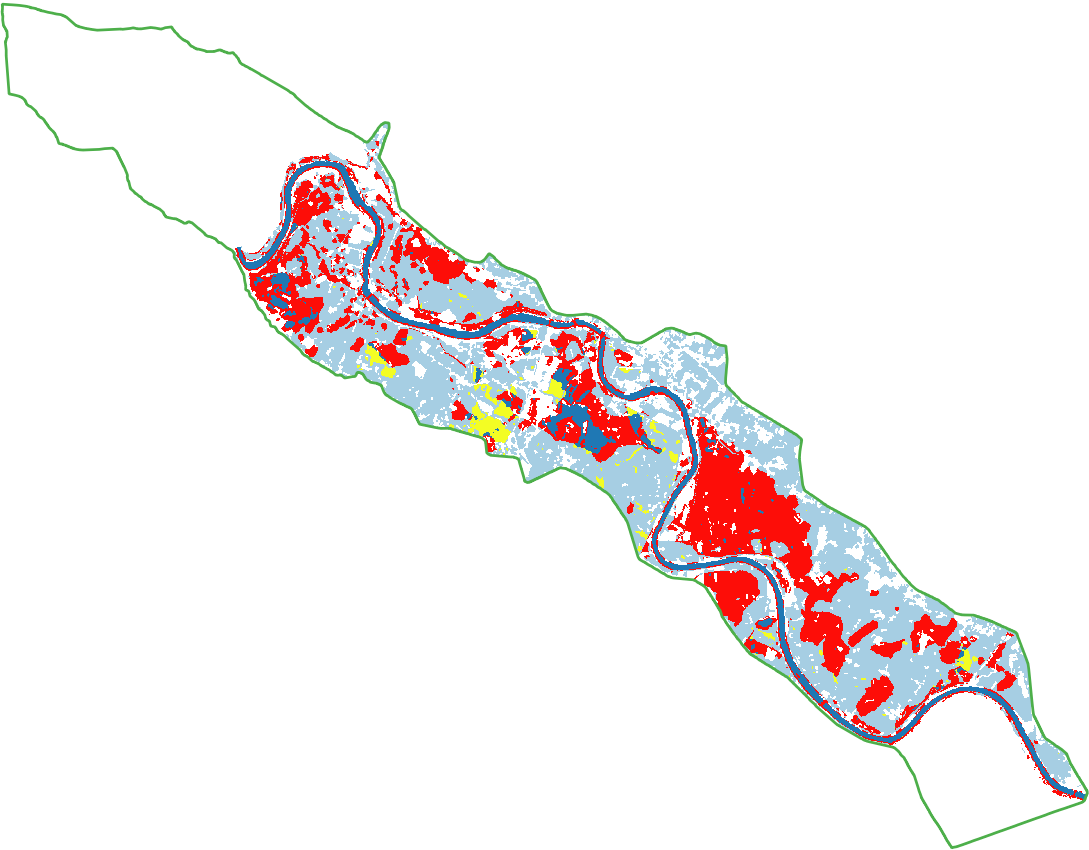}&
\includegraphics[width=0.18\linewidth]{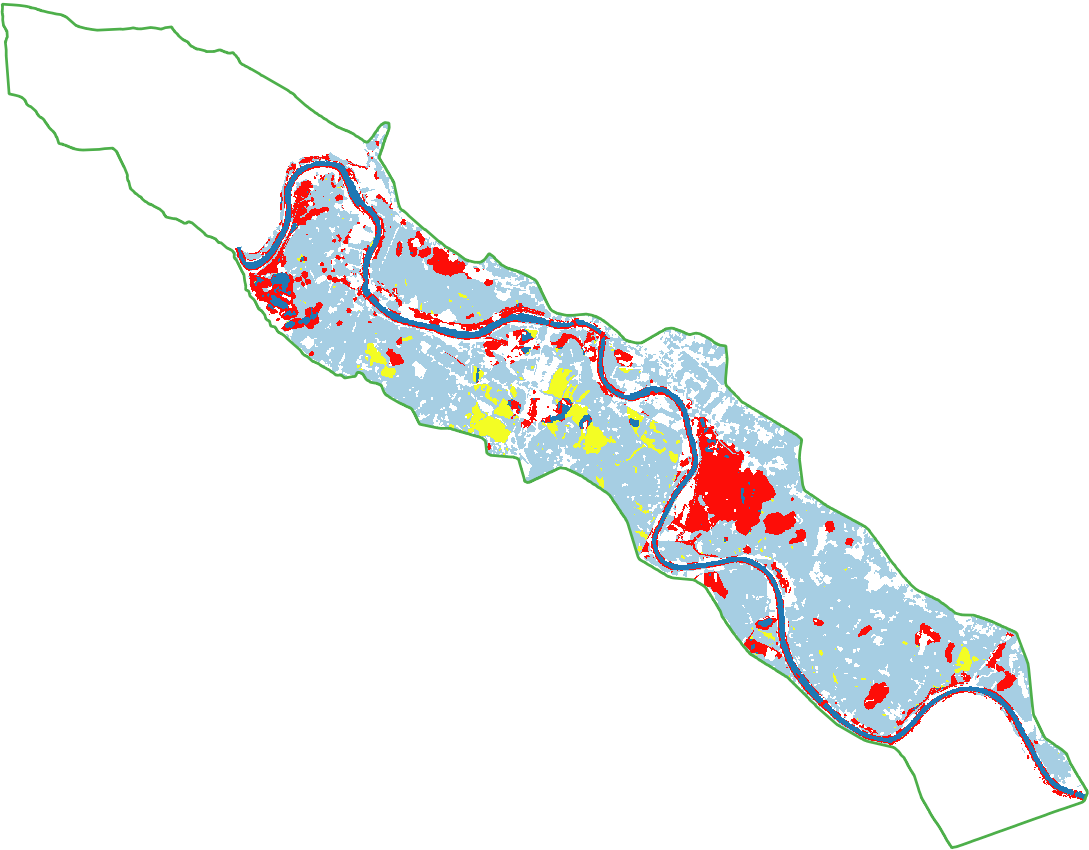}&
\includegraphics[width=0.18\linewidth]{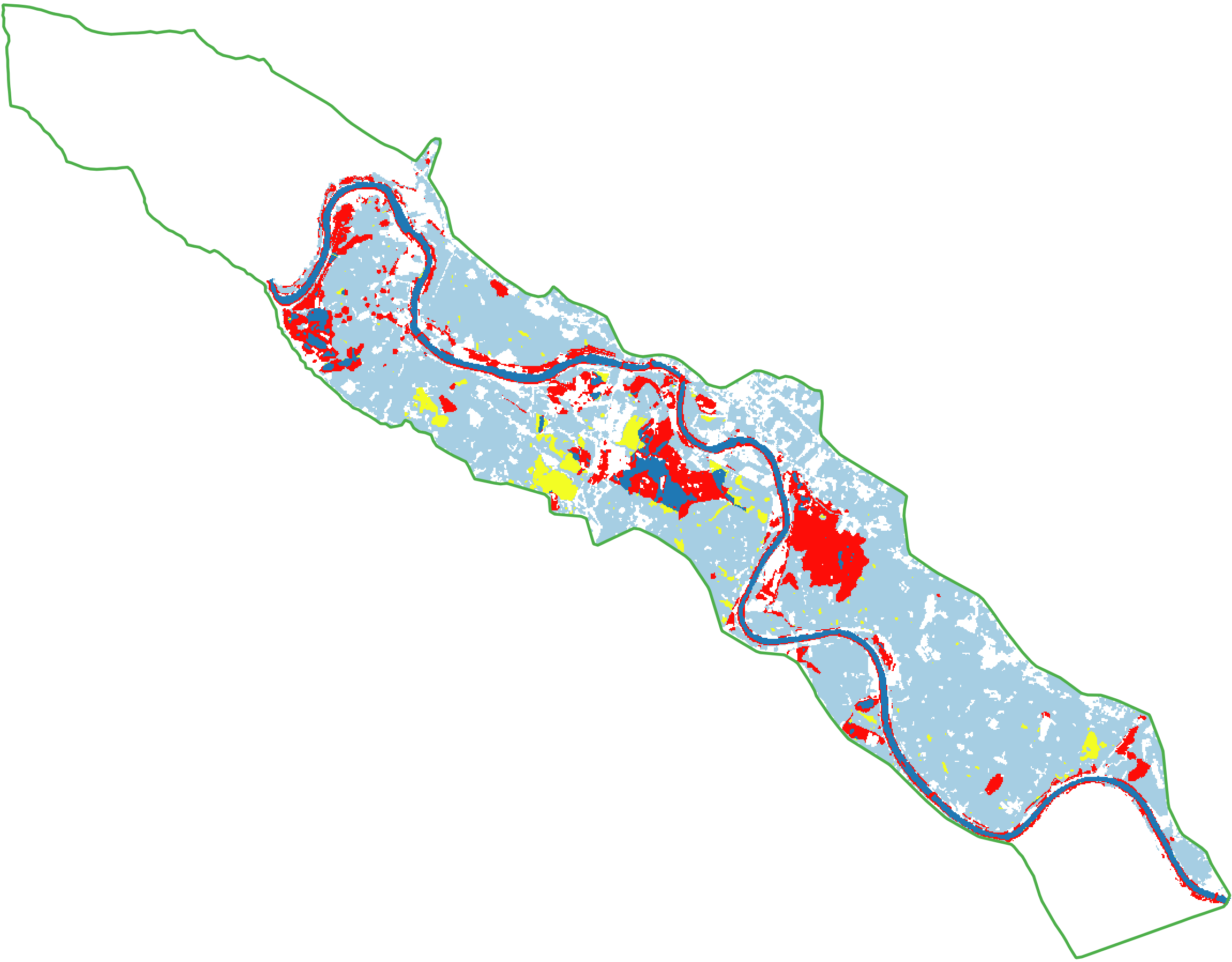}\\
&&& $\mathrm{CSI}=24.83\%$ & $\mathrm{CSI}=21.31\%$ & $\mathrm{CSI}=27.00\%$ & $\mathrm{CSI}=32.60\%$\\\hline
% & \multicolumn{4}{c}{\centering\includegraphics[trim=0 15cm 17.5cm 0, clip,width=0.35\linewidth]{fig/legend.pdf}}\\
\end{tabular}
\caption{in real event}
\label{fig:conti_2021}
\end{subfigure}

\centering\includegraphics[trim=0 14cm 17.5cm 0, clip,width=0.3\linewidth]{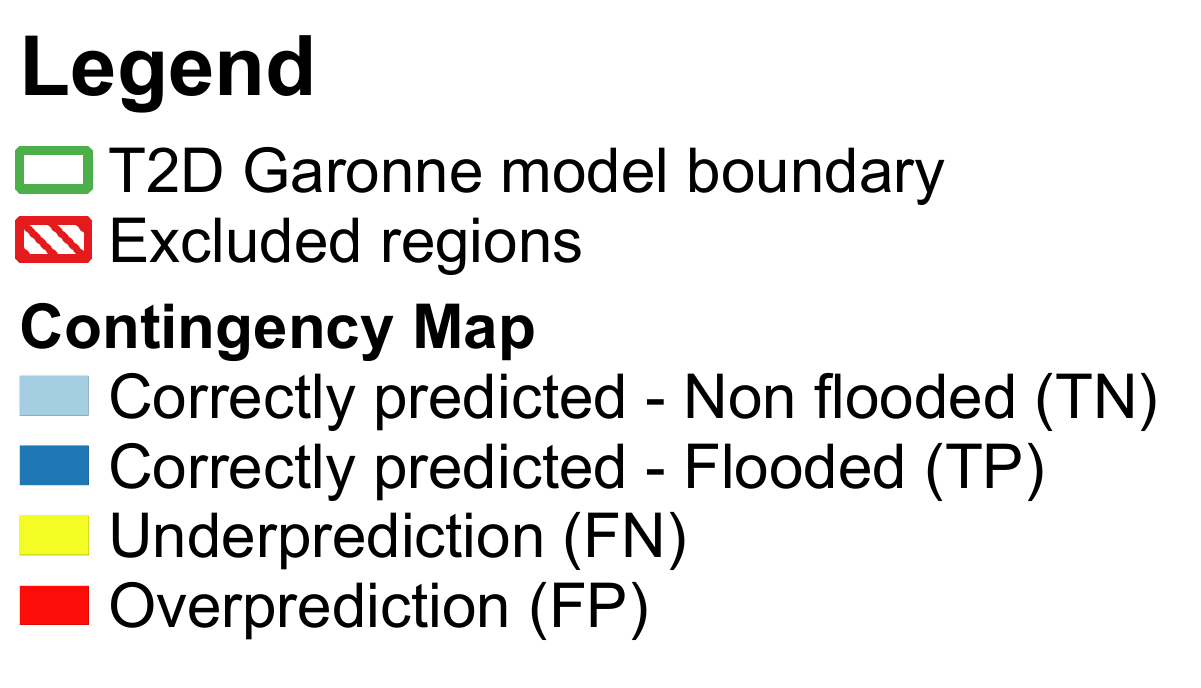}

\caption{Contingency maps computed between simulated flood extents (continued).}
% \label{}
\end{figure*}

% The computation of the CSI over the 5 subdomains of the flood plain confirms the previous remarks. \sophie{As shown in Table~\ref{tab:CSI_OSSE}, the CSI is significantly improved when WSR are assimilated in IHDA and IGDA,} and the water level is corrected with a negative increment in the flood plain. Nevertheless, the GA does not bring any significant improvement. 

% \clearpage
\subsubsection{2D validation with contingency maps and $\mathrm{CSI}$ scores}
\paragraph{Results for OSSE experiment}
Figure~\ref{fig:conti_OSSE} shows the reference flood extent maps (first column) and then the resulting contingency maps and $\mathrm{CSI}$ score for FR and DA experiments computed with respect to the reference simulation at the time of the flood peak (2021-02-03 19:00) and during water recess (2021-02-07 07:00). 
\offline{The two red-hatched regions are excluded from the assessments due  artificial flooding of the first meander \sophie{in the Garonne T2D model} and unreliable topography near \sophie{La Réole}.}
In the contingency maps, the correctly predicted pixels as non-flooded and flooded are represented in light blue and in dark blue, respectively. The wet pixels incorrectly predicted as non-flooded (or underprediction) are shown in yellow, whereas the dry pixels incorrectly predicted as flooded (or overprediction) are indicated in red. 
%Contingency maps are shown for the synthetical 2021 event at the time of the flood peak (2021-02-03 19:00) and during water recess (2021-02-07 07:00). The resulting $\mathrm{CSI}$ scores are also included. 

The flooding is significantly underestimated by FR at the flood peak, as reflected by the large number of yellow underprediction pixels in the floodplain. The assimilation of in-situ WL data in IDA ($\mathrm{CSI}=86.70\%$) improves the dynamics of the flow compared to FR  ($\mathrm{CSI}=71.41\%$). Yet, the assimilation of WSR observations in IHDA and IGDA yields much higher CSI and better flood extent representation, thanks to the extended control vector involving the associated hydraulic state correction. Indeed, the underprediction and overprediction areas in FR are significantly reduced when WSR are assimilated for both illustrated times of the flood. IHDA and IGDA results are quite close at the flood peak (respectively, 97.26\% and 96.72\%). During the flood recess, as both IHDA and IGDA allow for an effective emptying of the floodplain, the flood extents are significantly reduced and are in better agreement with the synthetical flood extents, whereas IDA struggles to reduce the overprediction areas exhibited in FR (shown by their CSI = 62.47\% for FR and CSI = 63.64\% for IDA). IHDA and IGDA's resulting CSIs are improved with respect to that of FR: from $\mathrm{CSI}=62.47\%$ for FR to $\mathrm{CSI}=83.76\%$ for IHDA and $\mathrm{CSI}=91.89\%$ for IGDA.  This results confirm that, when observations are assimilated in the floodplain, the GA strategy brings an improvement with respect to the classical EnKF, especially when the WSR data are most informative with respect to the imperfect dynamics of the numerical model, namely during the flood recess.

%However, the correction of the hydraulic state in the subdomains of the floodplain associated with the assimilation of WSR in IHDA and IGDA leads to an effective drying of the floodplain that is in good agreement with the synthetical observations. The amount of overprediction areas in the floodplain is significantly reduced for IHDA ($\mathrm{CSI}=83.76\%$), and even more so for IGDA ($\mathrm{CSI}=91.89\%$), showing the efficiency of the GA.

%, with a slightly better score for advantage of IHDA ($\mathrm{CSI}=97.26\%$) over IGDA ($\mathrm{CSI}=96.72\%$) in terms of CSI score.
%During the water recess, IDA fails to bring any improvement with respect to FR, with both experiments exhibiting overprediction areas for the most part ($\mathrm{CSI}=62.47\%$ versus $\mathrm{CSI}=63.64\%$). 

% \clearpage
\paragraph{Results for real experiment}

Figure~\ref{fig:conti_2021} displays the observed flood extent maps derived from S1 images (first column) and the contingency maps for the 2021 flood event at the flood peak (2021-02-03 19:00) and during recess (2021-02-07 07:00) for every experiment. Similarly to what was observed for OSSE,  the assimilation of WSR data brings a significant improvement for the representation of the flood extent with respect to FR and also to IDA. It should first be noted that, as expected, for the real experiment, the CSI scores for all experiments remain smaller than those of the OSSE experiments. Indeed, the numerical model struggles to simulate a flow that is in agreement with the observations in both the river bed and the floodplain. At the flood peak, the assimilation of WSR data in IHDA and IGDA brings a significant improvement over all subdomains with respect to FR, shown by fewer underpredicted pixels. Same conclusions are drawn during the flood recess. Indeed, IGDA outperforms IHDA even though some overpredicted areas still remain. These results confirm the merits of assimilating WSR observations, along with the GA step in the DA analysis, even in real event configuration that is more challenging the OSSE configuration.

\subsubsection{Post-event measure validation with High Water Marks observations (only for real experiment)}

% \begin{figure*}[h]
%     \centering
%     \includegraphics[trim=0 3.5cm 0 0, clip, width=0.85\linewidth]{fig/2021_HWM_AGU_IGDA.pdf}
%     \caption{Post-event HWM validations over the 2021 flood event. Yellow triangle indicates an underprediction by the simulation whereas red triangle indicates an overprediction.}
%     \label{fig:HWM}
% \end{figure*}

\begin{figure}[h]
    \centering
    \includegraphics[width=0.85\linewidth]{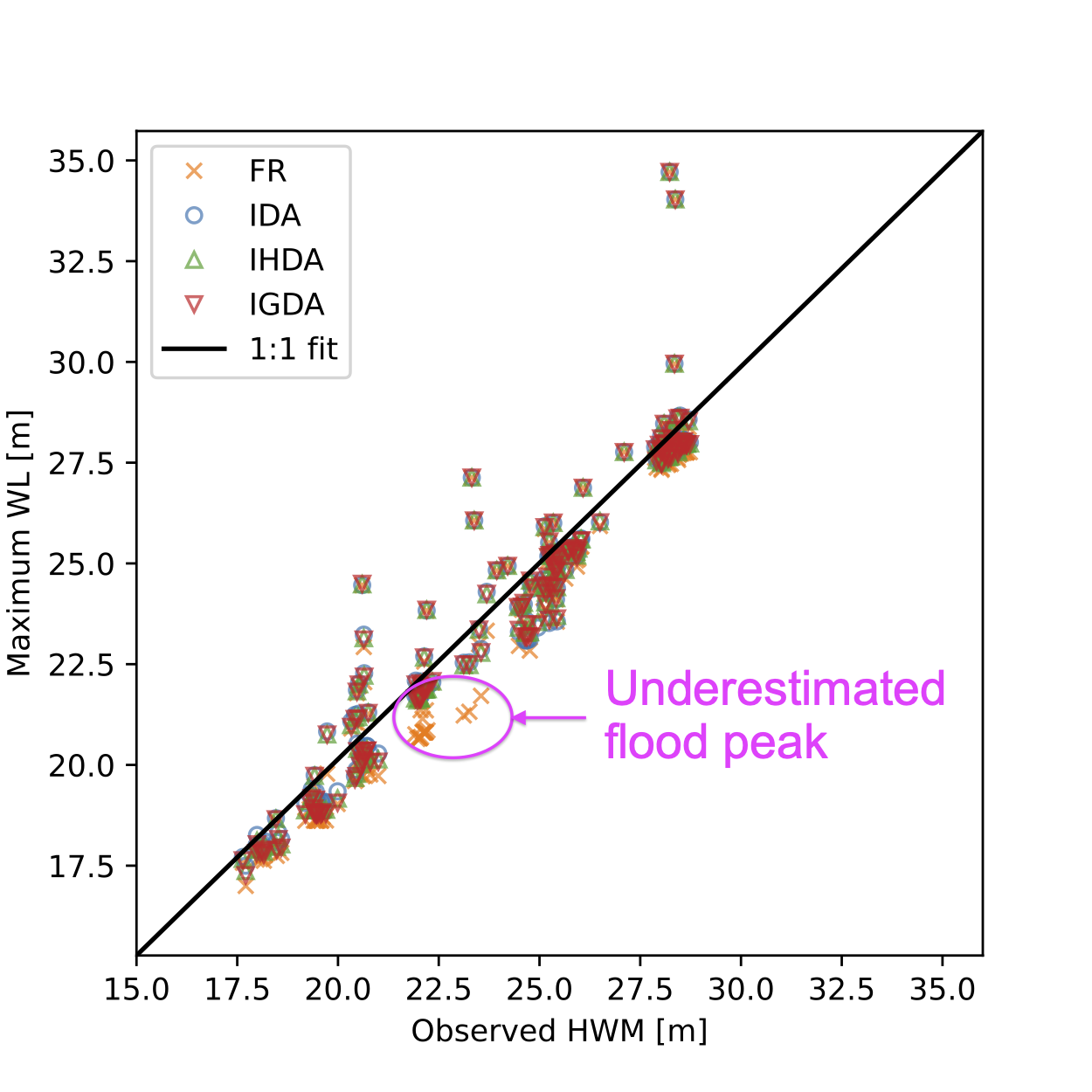}
    \caption{Comparison of HWMs between experiments.}
    \label{fig:HWM_pointclouds}
\end{figure}

A validation of the DA strategies with \sophie{respect to } independent data was \sophie{finally} carried out using the collective public datasets of HWM for the 2021 flood event. This allows to evaluate the highest simulated WLs spatially distributed at various points on the river banks and within the floodplain at the flood peak. 
Figure~\ref{fig:HWM_pointclouds} gathers the scatter plot of observed highest WL (in $x$-axis) and its simulated values (in $y$-axis) from the different experiments. The diagonal 1:1 line represents the perfect simulation scenario.
Compared to FR results (shown by orange crosses that underestimated highest WL values at many locations, encircled in violet), DA allows to significantly reduce the misfits between the observed and simulated highest WL. However, similar results are found between IDA, IHDA, and IGDA. Indeed, since this validation only concerns the highest WL throughout the event, the merits of IGDA demonstrated mostly over the flood recess (Figure~\ref{fig:conti_2021}) remain imperceptible with respect to other DA experiments.
%(shown in our previous work \cite[Fig. 14]{nguyenagu2022}).
% The HWM errors between the simulated WL and the observed WL are classified into four ranges, taking $\pm$ 1 meter as a baseline for small errors.
% % and more than those as large errors. 
% It appears that the assimilation of WSR significantly reduces , 

\section{Conclusion and perspectives}
\label{ConclusionPerspectives}

This \offline{article presents a follow-up of our previous study \cite{nguyenagu2022} \sophie{where a Gaussian anamorphosis is proposed to deal with the non-gaussiannity of RS observations' errors}. It highlights the merits of a dual state-parameter EnKF, involving the assimilation of SAR-derived flood extents jointly with gauge water-level data. The proposed method has been validated with OSSE experiments and then assessed in hindcast mode with a real flood event for significant flooding over the Garonne Marmandaise catchment.}
It was shown that the assimilation of in-situ WL data improves the simulation in the river bed but the complementary assimilation of WSR is key to improve the dynamics in the floodplain, especially at the flood peak and during the flood recess when the T2D model alone struggles to dry out the floodplain. The present work emphasizes on the non-Gaussian properties of observations errors associated with the WSR measurements computed over the subdomains of the floodplain.  A Gaussian anamorphosis (GA) strategy, previously proposed in the literature to deal with ecological and hydrogeological variables, was adapted here. The GA maps the observation vector and is model equivalent onto a transformed space where the Gaussianity assumption \offline{becomes} valid. The anamorphosis function is computed based on the ensemble of simulations that yield the model equivalents of the WSR observations. In our case study, the bijectivity of the GA function is not guaranteed as there are many occurrences of either strictly null or unity values of WSR, respectively when the subdomain is entirely dry or flooded. \offline{Therefore, the} GA function was made bijective by separating the similar values close to the $[0,1]$ bounds with a negligible random noise, and then by extrapolating the tails of the function between the leftmost or rightmost values to the unreachable bounds when needed. The major findings from our numerical experiments are summarized as follows:
\begin{itemize}
\item GA succeeds in transforming the non-Gaussian distribution of  observation errors into a Gaussian distribution, for each Sentinel-1 overpass time and for each relevant subdomain of the floodplain. 
\item GA, \offline{applied} in IGDA experiment, leads to slightly better results than the classical EnKF in IHDA, when assimilating both in-situ WL and WSR observations \offline{with a dual state-parameter estimation}.
\end{itemize}
 These conclusions advocate for using a GA step when possible, but it also demonstrates that while the classical EnKF is sub-optimal in the presence of non-Gaussianity, its analysis still remains valid and reliable.

As a perspective for this work, one may revisit the assumption of a uniform correction of WL within the floodplain subdomains, as well as the definition of these subdomains, in order to allow for a finer correction of the hydraulic state. Thus far, the DA strategy was implemented with a cycled DA algorithm, only issuing forecast for the following DA cycle. The impact of the DA correction should further be investigated in full forecast mode, considering various lead times that exceed the propagation time of the hydraulic network. For that purpose, the characteristics of the time-varying errors in the control vector should be investigated, and the strategy for prescribing the DA correction beyond the end of the assimilation cycle shall be proposed. By default, a persistent correction could be applied, yet it is expected that such a simple approach could struggle to follow the dynamics of the floods, especially when rapidly changing. In order to tackle longer forecast lead times, chaining the hydrodynamics model with a large-scale hydrologic model could be considered. This approach is currently being investigated. In this perspective, the hypothesis on the stationarity of the errors in the elements of the control vector becomes all the more important, yet it may not stand between the assimilation and the forecast periods. Finally, a major perspective for this research work stands in the assimilation of various RS-derived data from SAR (Sentinel-1, TerraSAR-X), optical (Sentinel-2) and altimetry (Sentinel-6, and the recently launched SWOT) images in order to enrich the observation network and better cover different phases of flood events.

\section*{Acknowledgment}

% Funding for this work was provided by CNES and CERFACS.
% Funding for this work was provided by CNES and CERFACS within the framework of the Space for Climate Observatory (SCO). 
The authors gratefully thank the  Electricité de France (EDF) for providing the TELEMAC-2D model on the Garonne Downstream catchment, and the SCHAPI, SPCs Garonne-Tarn-Lot and Gironde-Adour-Dordogne for providing in-situ data. They also would like to thank R. Hostache (IRD), R. Fjortoft (CNES), and T. Koleck (CNES) for fruitful discussions and advices, and Q. Bonassies (CERFACS) for helping with the manuscript revision. 
% Lastly, the authors would like to thank the anonymous reviewers whose comments and suggestions helped improve this manuscript.
%Ehouarn Simon from IRIT

% \section{Bibliographie}
% \label{sec:Bibliographie}	
{\small
\bibliographystyle{IEEEtranN}
\bibliography{./ref.bib} 
}

% Fin du document
\end{document}